\documentclass[aps,prx,twocolumn,superscriptaddress,10pt]{revtex4-2}

\usepackage{graphicx}
\usepackage{dcolumn}
\usepackage{bm}
\usepackage{braket}
\usepackage{bbold}
\usepackage{dsfont}
\usepackage{amsmath}
\usepackage{mathtools}
\usepackage{subfigure}
\usepackage{extarrows}
\usepackage{amssymb}
\usepackage{esint}
\usepackage{qcircuit}
\usepackage{tikz}
\usepackage{xcolor}
\usepackage{dirtytalk}
\usetikzlibrary{trees,arrows}
\usepackage{listings}
\usepackage{comment}
\usepackage{makecell}
\usepackage{soul}

\begin{document}

\newcommand{\redmark}[1] {\color{red}\textbf{#1}\color{black}\normalsize}
\newcommand{\bluemark}[1] {\color{blue}\textbf{#1}\color{black}\normalsize}
\newcommand{\brownmark}[1] {\color{purple}\textbf{#1}\color{black}\normalsize}
\newcommand{\uvec}[1]{\boldsymbol{\hat{\textbf{#1}}}}

\lstset{%
  language=C,
  basicstyle=\ttfamily\tiny,
  frame=lines,
  numbers=none,
  breaklines=True             
}

\title{Preparing Valence-Bond-Solid states on noisy intermediate-scale quantum computers}

\author{Bruno Murta}
\email{bpmurta@gmail.com}
\affiliation{Theory of Quantum Nanostructures Group, International Iberian Nanotechnology Laboratory (INL), 4715-330 Braga, Portugal}
\affiliation{Departamento de F\'{i}sica, Universidade do Minho, 4710-057 Braga, Portugal}
\author{Pedro M. Q. Cruz}
\affiliation{Theory of Quantum Nanostructures Group, International Iberian Nanotechnology Laboratory (INL), 4715-330 Braga, Portugal}
\affiliation{ICFO - Institut de Ci\`{e}ncies Fot\`{o}niques, The Barcelona Institute of Science and Technology, 08860 Castelldefels, Spain}
\affiliation{Faculdade de Ci\^{e}ncias, Universidade do Porto, 4169-007 Porto, Portugal}
\author{J. Fern\'{a}ndez-Rossier}
\affiliation{Theory of Quantum Nanostructures Group, International Iberian Nanotechnology Laboratory (INL), 4715-330 Braga, Portugal}

\date{\today}

\begin{abstract}
Quantum state preparation is a key step in all digital quantum simulation algorithms. Here we propose methods to initialize on a gate-based quantum computer a general class of quantum spin wave functions, the so-called Valence-Bond-Solid (VBS) states, that are important for two reasons. First, VBS states are the exact ground states of a class of interacting quantum spin models introduced by Affleck, Kennedy, Lieb and Tasaki (AKLT). Second, the two-dimensional VBS states are universal resource states for measurement-based quantum computing. We find that schemes to prepare VBS states based on their tensor-network representations yield quantum circuits that are too deep to be within reach of noisy intermediate-scale quantum (NISQ) computers. We then apply the general non-deterministic method herein proposed to the preparation of the spin-$1$ and spin-$\frac{3}{2}$ VBS states, the ground states of the AKLT models defined in one dimension and in the honeycomb lattice, respectively. Shallow quantum circuits of depth independent of the lattice size are explicitly derived for both cases, making use of optimization schemes that outperform standard basis gate decomposition methods. Given the probabilistic nature of the proposed routine, two strategies that achieve a quadratic reduction of the repetition overhead for any VBS state defined on a bipartite lattice are devised. Our approach should permit to use NISQ processors to explore the AKLT model and variants thereof, outperforming conventional numerical methods in the near future.

\end{abstract}

\pacs{Valid PACS appear here}

\maketitle

\section{Introduction}\label{Section1}

Quantum many-body phenomena are ubiquitous in nature, but their study has been hampered by the difficulty \cite{Laughlin00} in carrying out large-scale numerical simulations to probe their defining emergent features \cite{Anderson72}. Digital quantum computers hold the promise of a more scalable simulation of quantum many-body phenomena than what is possible with conventional numerical methods by exploiting the principle of superposition and the natural encoding of entanglement \cite{Nielsen00}. Indeed, some of the key limitations faced by leading numerical methods, notably the sign problem \cite{Troyer05} in quantum Monte Carlo methods, the exponential wall problem \cite{Kohn99} faced by exact diagonalization, and the entanglement area laws \cite{Eisert10} that hinder the viability of tensor networks beyond one dimension, are overcome by gate-based quantum computation \cite{Ortiz01, Preskill12}.

However, digital quantum simulation is yet to become a standard method in the study of quantum many-body systems. This is due to the limitations \cite{Stolze04} of the current noisy intermediate-scale quantum (NISQ) computers \cite{Preskill18}, namely the accumulation of both coherent and incoherent errors due to faulty gate operations \cite{cruz2021testing}, and the limited coherence times of qubits that restrict the depth of circuits that can be faithfully executed. Although these issues will eventually be overcome through quantum error correction \cite{Lidar13}, meanwhile can the strengths of NISQ devices be exploited to solve practical problems beyond what is possible via conventional numerical methods? 

Two complementary approaches have been followed towards this goal. The first involves the development of hybrid variational algorithms (cf. \cite{Cerezo21} and references therein) that trade circuit depth for the parallel execution of independent circuits and delegate part of the computational task to conventional processors. The second consists of identifying problems for which the prospect of a quantum speed-up \cite{Jozsa03} is plausible, either because conventional methods are not scalable or because the problem itself is especially suited for NISQ devices.

In the vein of the second approach, several proposals \cite{Verstraete09, Schmoll17, CerveraLierta18, Xiao21, Cervia21, VanDyke21, VanDyke211, Bespalova21} involving the preparation of exact eigenstates of integrable quantum many-body models on digital quantum computers have arisen in the literature recently. Even though analytical methods allow to probe many features of these models, the computation of other properties remains challenging. For example, although the spectrum of the quantum XXZ model was determined exactly over sixty years ago \cite{Orbach58}, calculating some of its correlation functions remains an active area of research \cite{Babenko21}, which motivated the development of a routine to prepare the corresponding Bethe ansatz states on quantum hardware \cite{VanDyke21, VanDyke211}. Likewise, the addition of a sufficiently strong magnetic field to the Kitaev model \cite{Kitaev06} disrupts the fractionalization upon which its integrability is based, in which case a hybrid variational scheme on a NISQ device \cite{Bespalova21} may constitute a competitive strategy.

The class of quantum many-body states that will be the focus of this article are the so-called \textit{Valence-Bond-Solid} (VBS) states, which are the exact ground states of Affleck-Kennedy-Lieb-Tasaki (AKLT) models \cite{AKLT87, AKLT88}. The ground state of the bilinear-biquadratic spin-$1$ AKLT model is the landmark VBS state, because it provided the first strong piece of evidence to support Haldane's conjecture \cite{Haldane83} that the integer-spin Heisenberg models are gapped in one dimension. Nevertheless, the VBS construction can be generalized to arbitrary lattices in higher dimensions \cite{AKLT87, AKLT88}. One that will deserve our special attention is the spin-$\frac{3}{2}$ VBS state on the honeycomb lattice, as this is a resource state for universal quantum computation \cite{Wei11, Miyake11} within the measurement-based paradigm of Raussendorf and Briegel \cite{Raussendorf01, Raussendorf03, Briegel09}. 

This paper is organized as follows. Section \ref{Section2} introduces the VBS states and the respective AKLT models. Section \ref{Section3} outlines the challenge of preparing VBS states on digital quantum computers and discusses the methods currently available in the literature to accomplish it. This discussion motivates the development of a novel NISQ-friendly scheme, which is detailed in Section \ref{Section4}. Section \ref{Section5} presents a shallow basis gate decomposition of the key building block of the proposed NISQ-friendly scheme for the spin-$1$ and spin-$\frac{3}{2}$ VBS states. In light of the probabilistic nature of our method, Sections \ref{Section6} and \ref{Section7} present two alternative ways of achieving a quadratic reduction of the resulting repetition overhead. In order to bypass the basis gate decomposition of a large unitary matrix, notably for high-spin VBS states, Section \ref{Section8} presents a systematic method to find a quantum circuit to implement the key element in the preparation of VBS states. Section \ref{Section9} includes a detailed discussion of the resources required to prepare the spin-$1$ and spin-$\frac{3}{2}$ VBS states in near-term quantum hardware, taking qubit connectivity constraints into account. In addition, two applications, one for spin-$1$ and another for spin-$\frac{3}{2}$, for which quantum advantage is attainable are introduced. The final section highlights the main conclusions of this paper.

\section{AKLT Models. VBS States}\label{Section2}

In one dimension, the spin-$1$ AKLT model \cite{AKLT87, AKLT88} is a special case of the bilinear-biquadratic Hamiltonian (cf. \cite{Trebst06} and references therein)
\begin{equation}
    \mathcal{H}_{\textrm{BLBQ}}^{S = 1}(\beta) = \sum_{n = 1}^{N-1} \vec{S}_n \cdot \vec{S}_{n+1} + \beta (\vec{S}_n \cdot \vec{S}_{n+1})^2,  
\label{eq_BLBQ_Ham}
\end{equation}
with $\beta = \frac{1}{3}$ and $\vec{S}_{n}=\left(S_{n}^{x},S_{n}^{y},S_{n}^{z}\right)$, where $S_n^i$ are the spin-1 matrices (cf. Appendix \ref{AppB}). The phase diagram of $\mathcal{H}_{\textrm{BLBQ}}^{S = 1}(\beta)$ comprises a gapless phase for $\beta > 1$, a dimerized phase for $\beta < -1$ and the gapped Haldane phase for $\beta \in (-1,1)$. The latter includes both the spin-$1$ Heisenberg model \cite{Haldane83} ($\beta = 0$) and the spin-$1$ AKLT model \cite{AKLT87, AKLT88} ($\beta = \frac{1}{3}$).

The spin-1 AKLT model, $\mathcal{H}_{\textrm{AKLT}}^{S = 1} \equiv \mathcal{H}_{\textrm{BLBQ}}^{S = 1}(\beta = \frac{1}{3})$, can be expressed (up to a factor and an additive constant) in terms of a sum of the local projectors $P_{n, n+1}^{S=2}$ that map neighboring spins to the subspace of total spin $|\vec{S}_n + \vec{S}_{n+1}|^2 = 2(2+1)$ (cf. Appendix \ref{AppA} for the derivation):
\begin{equation}
    \mathcal{H}_{\textrm{AKLT}}^{S = 1} = 2 \sum_{n = 1}^{N-1} \Big( P_{n, n+1}^{S = 2}(\vec{S}_n, \vec{S}_{n+1}) - \frac{1}{3} \Big).
\label{eq_AKLT_S_1}
\end{equation}
Ignoring the constants, all eigenenergies must be non-negative because each term is a projector. As a result, a state satisfying $P_{n, n+1}^{S=2} \ket{\psi_0} = 0$ for all pairs of neighboring spins must be a ground state of the AKLT model.

\begin{figure}[h]
\includegraphics[width=0.9\linewidth]{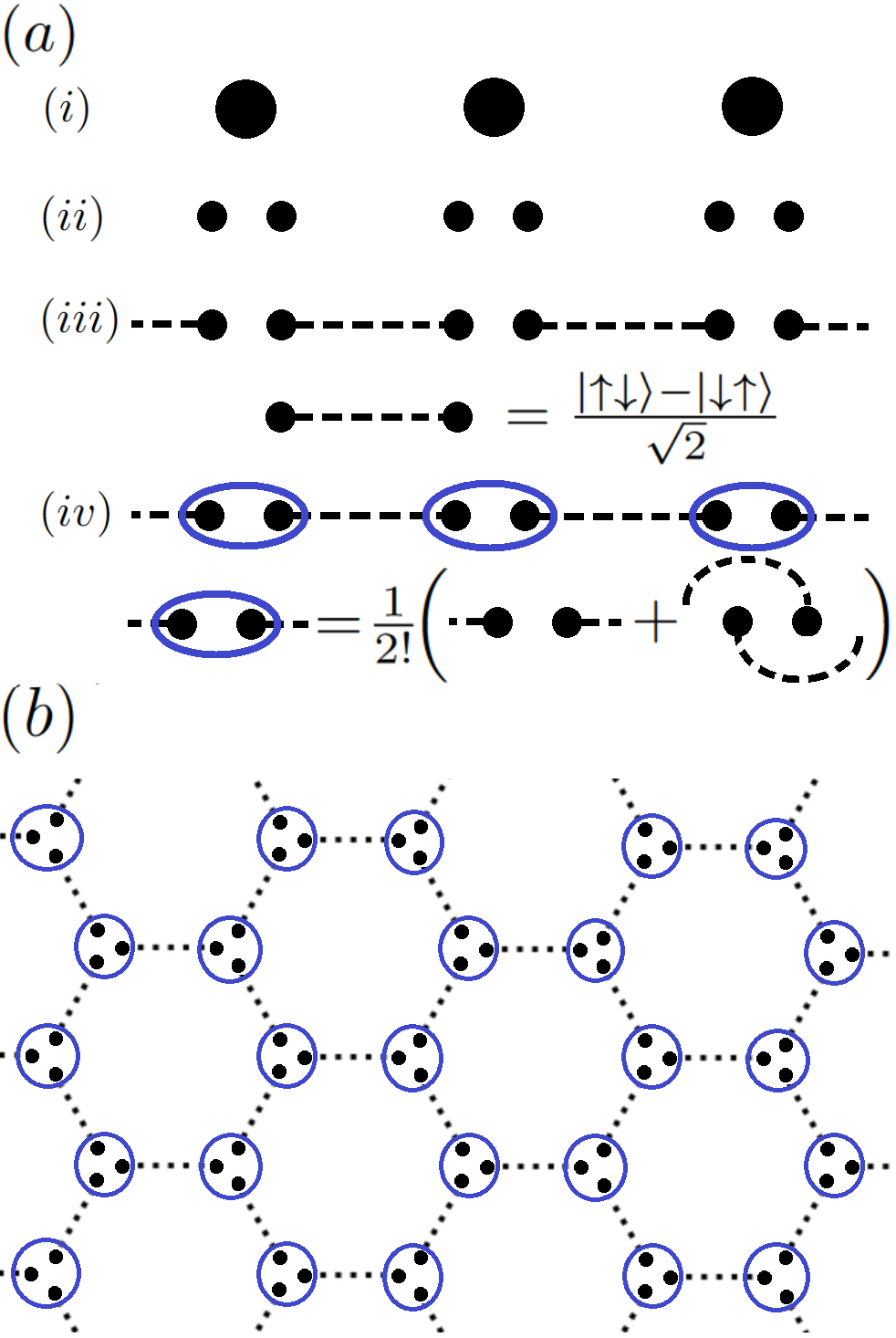}
\caption{(a) Illustrated description of construction of spin-$1$ Valence-Bond-Solid (VBS) states in one-dimensional lattice. (i) Local degree of freedom at each lattice site is spin-$1$. (ii) Each spin-$1$ is decomposed into two spins-$\frac{1}{2}$. (iii) A valence bond $\frac{\ket{\uparrow \downarrow} - \ket{\downarrow \uparrow}}{\sqrt{2}}$ is prepared at every pair of nearest-neighboring sites, each site contributing a spin-$\frac{1}{2}$. This ensures the ground state condition for the AKLT Hamiltonian (cf. Eq. (\ref{eq_AKLT_S_1})) is satisfied. (iv) The local symmetrization operator $\mathcal{S}$ is applied at every pair of qubits that encode a single site in order to retrieve the original spin-$1$ local degree of freedom. (b) Extension of VBS construction scheme to honeycomb lattice. Since coordination number is $3$, there are $3$ spins-$\frac{1}{2}$ per lattice site, each being involved in a single valence bond. Application of $\mathcal{S}$ at every site produces a spin-$\frac{3}{2}$. This spin-$\frac{3}{2}$ VBS state is the ground state of the AKLT Hamiltonian stated in Eq. (\ref{eq_AKLT_S_3_2}).}
\label{Scheme1}
\end{figure}

The VBS construction scheme ensures this ground state condition is satisfied as follows. Originally, the total spin of any pair of neighboring spins-$1$, $\vec{S}_{\textrm{total}}^{n,n+1} = \vec{S}_n + \vec{S}_{n+1}$, can take the values $1 \oplus 1 = 0, 1, 2$, where $\oplus$ denotes angular momentum addition. To dispose of ${S}_{\textrm{total}}^{n,n+1} = 2$, each local spin-$1$ is decomposed into two spins-$\frac{1}{2}$, so that a valence bond $\frac{1}{\sqrt{2}}(\ket{\uparrow \downarrow} - \ket{\downarrow \uparrow})$ can be created between the two adjacent spins-$\frac{1}{2}$ for each pair of neighboring sites (cf. Fig. \ref{Scheme1}(a)). Since the valence bond is a (spin-$0$) singlet, now $S_{\textrm{total}}^{n,n+1}$ can only take values $\frac{1}{2} \oplus 0 \oplus \frac{1}{2} = 0, 1$, as desired. Finally, to ensure that the physical degrees of freedom at each site are the expected spins-$1$, the respective pair of spins-$\frac{1}{2}$ must be symmetrized,
\begin{equation}
    \mathcal{S}(\ket{\sigma}_L \ket{\sigma'}_R) = \frac{1}{2} \Big( \ket{\sigma}_L \ket{\sigma'}_R + \ket{\sigma'}_L \ket{\sigma}_R \Big),
\label{eq_Symm_S_1}
\end{equation}
with $\sigma, \sigma' \in \{ \uparrow, \downarrow \}$. 

This VBS construction can be straightforwardly extended to arbitrary lattices in any dimensions \cite{AKLT87, AKLT88,Tasaki20}. Given some lattice $(\Lambda, \mathcal{L})$, where $\Lambda$ is the set of sites and $\mathcal{L}$ the set of links, every site $n \in \Lambda$ with coordination number $\mathcal{N}_n$ is associated with a local spin-$\frac{\mathcal{N}_n}{2}$ \footnote{Our notation includes the most general case of a graph with locally varying coordination number, in which case the local spin varies from site to site. However, the particular cases considered in this paper will correspond to regular lattices where every site is equivalent, and therefore all local spins are the same.}. Each local spin is decomposed into $\mathcal{N}_n$ spins-$\frac{1}{2}$, one for each lattice link $(n,n') \in \mathcal{L}$ emanating from site $n$, with $n' \in \Lambda$ a nearest-neighboring site of $n$. In the first step, a valence bond is created along each link, such that each spin-$\frac{1}{2}$ is involved in one valence bond. Henceforth, this product state of valence bonds will be referred to as
\begin{equation*}
    \ket{\psi_{\textrm{pre-VBS}}} = \bigotimes_{(n,n') \in \mathcal{L}} \frac{\ket{\uparrow}_{(n;n')} \ket{\downarrow}_{(n';n)} - \ket{\downarrow}_{(n;n')} \ket{\uparrow}_{(n';n)}}{\sqrt{2}},
\end{equation*}
where $\ket{\sigma}_{(n;n')}$ denotes the state of the spin-$\frac{1}{2}$ from site $n$ that is associated with neighboring site $n'$. The second and final step amounts to encoding a spin-$\frac{\mathcal{N}_n}{2}$ by applying the local symmetrization operator,
\begin{equation}
    \mathcal{S}_n \Bigg(\bigotimes_{i = 1}^{\mathcal{N}_n} \ket{\sigma_i}\Bigg) = \frac{1}{\mathcal{N}_n!} \sum_{\mathcal{P}} \Bigg(\bigotimes_{i = 1}^{\mathcal{N}_n} \ket{\mathcal{P}(\sigma_i)}\Bigg),
\label{eq_Symm}
\end{equation}
at every site $n$. $\mathcal{P}$ is the generator of all $\mathcal{N}_n!$ permutations of the $\mathcal{N}_n$ spins-$\frac{1}{2}$ belonging to site $n$. Appendix \ref{AppB} includes the derivation of the matrix representations of $\mathcal{S}$ for the spin-$1$ and spin-$\frac{3}{2}$ cases. 

The combination of these two steps yields the (unnormalized) Valence-Bond-Solid (VBS) state:
\begin{equation}
    \ket{\psi_{\textrm{VBS}}} = \bigotimes_{n \in \Lambda} \mathcal{S}_n \ket{\psi_{\textrm{pre-VBS}}}.
\end{equation}

Fig. \ref{Scheme1}(b) shows a schematic representation of the VBS state associated with the honeycomb lattice. Each dashed black line illustrates a valence bond, while the blue loops represent the local symmetrization. For both sublattices, every site has coordination number $\mathcal{N}_n = 3$, so the local physical degree of freedom corresponds to a spin-$\frac{3}{2}$, encoded by $3$ spins-$\frac{1}{2}$. As in one dimension, the valence bonds guarantee the total spin $\vec{S}_{\textrm{total}}^{n,n'} = \vec{S}_n + \vec{S}_{n'}$ resulting from the sum of the two neighboring spins-$\frac{3}{2}$ $\vec{S}_n$ and $\vec{S}_{n'}$ does not take its maximum value $S_{\textrm{total}}^{n,n'} = 3$, thus ensuring this VBS state is the ground state of the Hamiltonian (cf. Appendix \ref{AppA} for the derivation)
\begin{equation}
\begin{aligned}
    & \mathcal{H}_{\textrm{AKLT}}^{S = \frac{3}{2}} = \sum_{(n,n') \in \mathcal{L}} P_{n,n'}^{S = 3}(\vec{S}_n, \vec{S}_{n'}) = \\
    & = \sum_{(n,n') \in \mathcal{L}} \Big( \vec{S}_n \cdot \vec{S}_{n'} + \frac{116}{243} (\vec{S}_n \cdot \vec{S}_{n'})^2 + \frac{16}{243} (\vec{S}_n \cdot \vec{S}_{n'})^3 \Big).
\end{aligned}
\label{eq_AKLT_S_3_2}
\end{equation}

Despite the historic significance of the spin-$1$ VBS state in one dimension (1D), both for its support to Haldane's conjecture \cite{Haldane83} and for its pioneering role in the development of matrix product states (MPS) \cite{Fannes92, Klumper93}, the spin-$\frac{3}{2}$ VBS state on the honeycomb lattice is a potentially more relevant trial state for digital quantum simulation. Indeed, while most properties of the 1D spin-$1$ AKLT model can be computed analytically or, if needed, numerically (e.g., via DMRG \cite{White92, Rommer97, Schollwock05, Schollwock11}), analytical results for the spin-$\frac{3}{2}$ AKLT model on the honeycomb lattice are far scarcer and conventional numerical simulations less scalable. In fact, in 1D, the proof of the existence of the spectral gap appeared already in the original AKLT papers \cite{AKLT87, AKLT88} and even exact results for some excited states have been derived \cite{Arovas89, Moudgalya18, Moudgalya181}. For the honeycomb lattice, in turn, the nonzero gap of the AKLT model was only recently established \cite{Pomata20, Lemm20} and no exact excited states are known.

There are two noteworthy differences between the spin-$1$ and the spin-$\frac{3}{2}$ VBS states. The first concerns their computational power within the context of measurement-based quantum computation \cite{Raussendorf01, Raussendorf03, Briegel09}: While the spin-$1$ VBS state can only simulate restricted computations involving arbitrary single-qubit gates \cite{Gross07, Brennen08}, the spin-$\frac{3}{2}$ VBS state has been shown to be a resource state for the simulation of universal quantum circuits \cite{Wei11, Miyake11}. The second difference lies in the relation to the corresponding Heisenberg model: In 1D the spin-$1$ AKLT and Heisenberg models are both in the Haldane phase \cite{Trebst06}, whereas in the honeycomb lattice the ground state of the spin-$\frac{3}{2}$ Heisenberg model is N\'eel-ordered \cite{AKLT88}, thus implying a phase transition separating it from the spin-$\frac{3}{2}$ AKLT model, which has a unique disordered ground state \cite{AKLT88}.

In the remainder of this paper, the methods will first be presented for the general case of a spin-$S$ VBS state, with $S \in \{1, \frac{3}{2}, 2, \frac{5}{2}, 3, ...\}$. In addition, the two important examples highlighted above, the spin-$1$ VBS state (the simplest and canonical example of a VBS state) and the spin-$\frac{3}{2}$ VBS state (the natural candidate VBS state to achieve quantum advantage), will be discussed in detail.

\section{VBS States on Quantum Hardware: Outline of Problem}\label{Section3}

\subsection{Failure of Standard Quantum Simulation}

The VBS construction scheme involves two steps:
\begin{enumerate}
    \item{Preparing $\ket{\psi_{\textrm{pre-VBS}}}$, the product state of valence bonds, one for each lattice link.}
    \item{Applying the local symmetrization operator $\mathcal{S}$ at each lattice site of $\ket{\psi_{\textrm{pre-VBS}}}$, yielding $\ket{\psi_{\textrm{VBS}}}$.}
\end{enumerate}

In the language of quantum information theory, a valence bond is nothing more than the Bell state $\ket{\Psi^{-}} = \frac{1}{\sqrt{2}} (\ket{01} -\ket{10})$, where the mapping $\{ \ket{\uparrow} \longleftrightarrow \ket{0}, \ket{\downarrow} \longleftrightarrow \ket{1}\}$ is implied. Hence, the first step can be straightforwardly implemented by applying the two-qubit subcircuit shown in Fig. \ref{Circ1}(a) at each pair of qubits $\ket{n;n'}$ and $\ket{n';n}$ representing the two spins-$\frac{1}{2}$ along a lattice link $(n,n') \in \mathcal{L}$. All such subcircuits can be applied in parallel, thus resulting in a layer of depth $1$ $\textsc{cnot}$ \footnote{Since the execution time and error rates of two-qubit gates are significantly greater than those of single-qubit gates (especially standard ones like $Z$, $X$ or $H$), we ignore the latter in the calculation of the depth of this circuit.}, assuming merely connectivity between two such qubits.

The challenge lies in the second part of the VBS construction. This is due to the non-unitarity of the local symmetrization operator $\mathcal{S}$, which has two consequences. First, its implementation on quantum hardware is non-trivial, since quantum circuits are inherently unitary, as they ultimately amount to time-evolution operators of closed quantum systems (at least, within the gate-based paradigm) \cite{Stolze04}. Second, applying $\mathcal{S}$ amounts to projecting $\ket{\psi_{\textrm{pre-VBS}}}$ onto the local subspace of exchange-symmetric states. As a result, ignoring finite-size effects, the overlap between the easy-to-prepare $\ket{\psi_{\textrm{pre-VBS}}}$ and the final (normalized, hence the tilde) $\ket{\tilde{\psi}_{\textrm{VBS}}}$ decreases exponentially with the number of lattice sites $N$,
\begin{equation}
    \braket{\tilde{\psi}_{\textrm{VBS}} | \psi_{\textrm{pre-VBS}}} = p^{N/2},
\label{eq_overlap_main}
\end{equation}
where $p \in (0,1)$ is a constant set by the local spin considered. In general, $p$ is given by the fraction of symmetric spin states selected by $\mathcal{S}$. In particular, $p = \frac{3}{4}$ for the spin-$1$ VBS state and $p = \frac{1}{2}$ for the spin-$\frac{3}{2}$ VBS state. Appendix \ref{AppC} presents an explicit derivation of these two results and explains how the general approach is applied.

This exponentially vanishing overlap implies that attempting to prepare $\ket{\psi_{\textrm{VBS}}}$ starting from the easy-to-prepare $\ket{\psi_{\textrm{pre-VBS}}}$ via a standard digital quantum simulation strategy (e.g., quantum phase estimation \cite{Kitaev95, Nielsen00}, quantum cooling methods \cite{Motta20,McArdle19, Wang17}) becomes impractical for sufficiently large systems. Instead, Wang \cite{Wang17} started from a computational basis state, a good enough input state for the proof-of-concept simulation of spin-$1$ AKLT chain with just $N = 3$ sites but, given the exponential scaling of the size of the Hilbert space, certainly not for large enough systems to probe the thermodynamic limit.

\subsection{Matrix Product States in 1D}

\begin{figure*}
\includegraphics[width=\linewidth]{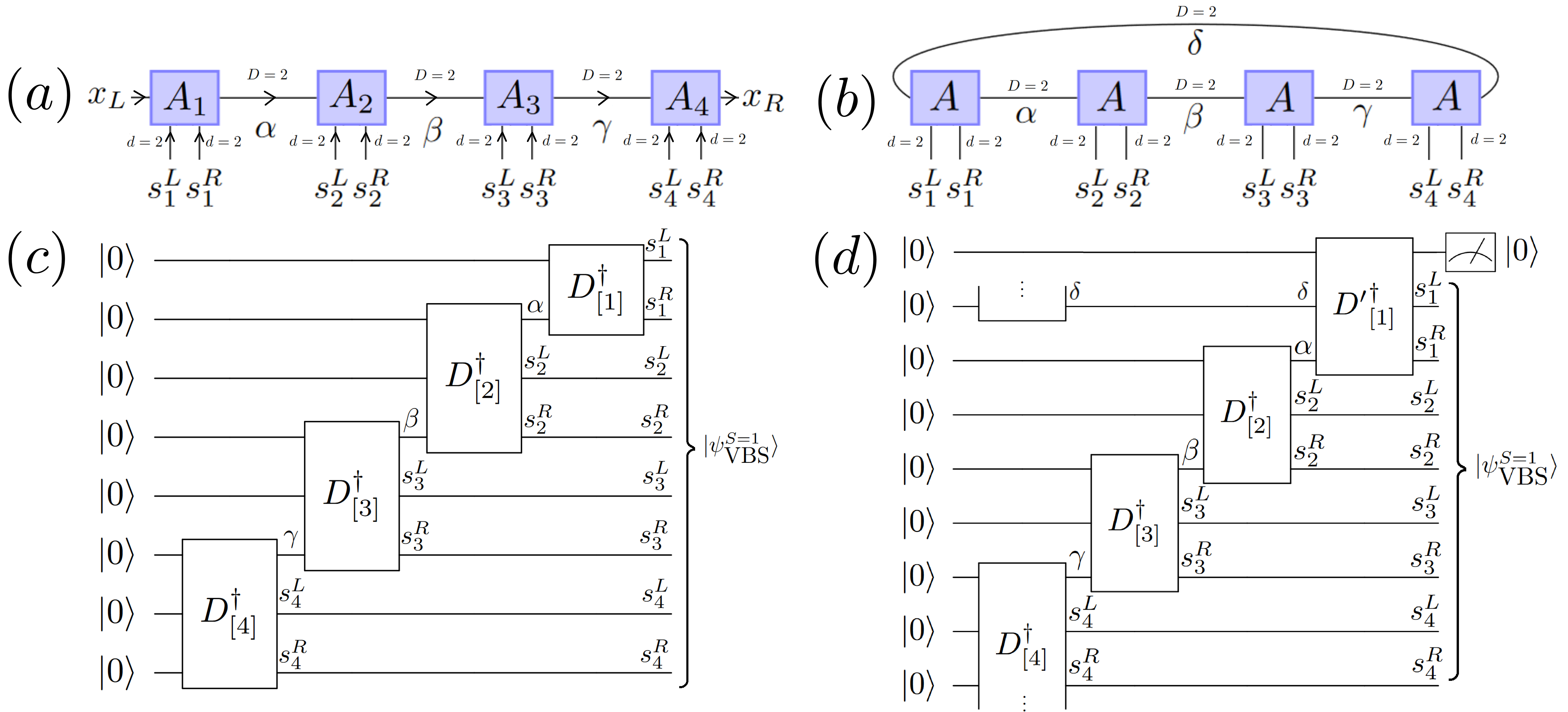}
\caption{(a,b) Matrix product state (MPS) representation of a one-dimensional spin-$1$ Valence-Bond-Solid (VBS) state with open (a) and periodic (b) boundary conditions for $N=4$ sites. Due to the translational invariance, all tensors in (b) are given by Eq. (\ref{local_tensor_two_qubits}), where two physical indices, each of dimension $d = 2$, are used instead of the conventional single index with $d = 3$. The MPS with open boundary conditions in (a) can be obtained by opening the bond corresponding to the sum over $\delta$ in (b) and setting the two resulting free indices $x_L$ and $x_R$ to one of the two possible values, thus producing one of the four degenerate VBS states for a spin-$1$ AKLT chain. The arrows in (a) allude to the fact that the MPS is in left-canonical form, in which case the local tensors vary from site to site, hence the site subscript used. (c,d) High-level scheme of quantum circuits that initialize spin-$1$ VBS states with open (c) and periodic (d) boundary conditions for $N=4$ sites by exploiting their MPS representation. $D^{\dagger}_{[i]}$ is the inverse of the unitary matrix product operator \cite{Cirac17, Chen18} $D_{[i]}$ that disentangles site $i$ from the remainder of the MPS. The explanation of how to obtain $D_{[i]}$ from the corresponding local tensor $A_i$ in the MPS can be found in Appendix \ref{AppTrueD}. For $N$ sites and open boundary conditions, as in (c), the circuit involves the sequential application of $N-1$ three-qubit operations $D_{[i]}^{\dagger}, i = N, N-1, ..., 2$, followed by the execution of the two-qubit operation $D_{[1]}^{\dagger}$, so the depth is $\mathcal{O}(N)$. The case of periodic boundary conditions introduces two differences. First, the final two-qubit operation, $D_{[1]}^{\dagger}$, is non-unitary, so it has to be embedded in a three-qubit unitary operation ${D'}_{[1]}^{\dagger}$ to be executed on quantum hardware. The additional ancilla must be initialized in $\ket{0}$ and measured in $\ket{0}$ (with probability of $50\%$, regardless of $N$) for the method to be successful. Second, the first operation, $D_{[N]}^{\dagger}$, acts on four qubits instead of three.
\label{fig_mps_vbs}}
\end{figure*}

An alternative approach to the preparation of VBS states on quantum hardware involves exploiting its exact representation in terms of tensor networks of bond dimension $D = 2$. In particular, in one dimension, the spin-$1$ VBS state with periodic boundary conditions can be expressed as a MPS \cite{Schollwock11,Tasaki20} (cf. Fig. \ref{fig_mps_vbs}(b)) 
\begin{equation}
\ket{\psi_{\textrm{VBS}}^{S=1}} = \sum_{\vec{\sigma}} \textrm{Tr}(A^{\sigma_1} A^{\sigma_2} ... A^{\sigma_N}) \ket{\vec{\sigma}},
\end{equation}
where $\ket{\vec{\sigma}} \equiv \bigotimes_{n=1}^{N} \ket{\sigma_i}$ represents the $N$ physical spin-$1$ degrees of freedom (i.e., $\ket{\sigma_i} = (\sigma_i^{+1}, \sigma_i^{0}, \sigma_i^{-1})^{T}$) and the (left- and right-normalized) local rank-$3$ tensors $A^{\sigma}$ are of the form
\begin{equation}
    \scriptstyle{
    A^{+1} = \begin{pmatrix}
     0 & \sqrt{\frac{2}{3}} \\
     0 & 0 \\
    \end{pmatrix}, \; 
    A^{0} = \begin{pmatrix}
     -\frac{1}{\sqrt{3}} & 0 \\
     0 & \frac{1}{\sqrt{3}} \\
    \end{pmatrix}, \;
    A^{-1} = \begin{pmatrix}
     0 & 0 \\
     -\sqrt{\frac{2}{3}} & 0 \\
    \end{pmatrix}.
    }
\end{equation}
On a digital quantum computer, these spins-$1$, $\ket{\sigma_i}$, have to be encoded (with redundancy) in terms of two qubits each, $(\ket{s_{i}^{L}}, \ket{s_{i}^{R}})$. Hence, for our purposes, it is more useful to express the local tensor in the form
\begin{equation}
\begin{aligned}
    & A^{\uparrow \uparrow} = \begin{pmatrix}
     0 & \sqrt{\frac{2}{3}} \\
     0 & 0 \\
    \end{pmatrix},
    A^{\uparrow \downarrow} = \begin{pmatrix}
     -\frac{1}{\sqrt{6}} & 0 \\
     0 & \frac{1}{\sqrt{6}} \\
    \end{pmatrix}, \\
    & A^{\downarrow \uparrow} = \begin{pmatrix}
     -\frac{1}{\sqrt{6}} & 0 \\
     0 & \frac{1}{\sqrt{6}} \\
    \end{pmatrix},
    A^{\downarrow \downarrow} = \begin{pmatrix}
     0 & 0 \\
     -\sqrt{\frac{2}{3}} & 0 \\
    \end{pmatrix},
\end{aligned}
\label{local_tensor_two_qubits}
\end{equation}
where the left- and right-normalization conditions are still satisfied. To obtain the MPS for the spin-$1$ VBS state with open boundary conditions, one can simply remove the virtual index corresponding to the bond linking the two ends of the chain (i.e., the bond between $A^{\sigma_1}$ and $A^{\sigma_N}$) and set each of the resulting free indices $x_L$ and $x_R$ to one of two possible values (cf. Fig. \ref{fig_mps_vbs}(a)), thus yielding the expected four-fold degeneracy of the spin-$1$ AKLT chain. At this point, the two tensors at the ends of the chain, $A^{\sigma_1}$ and $A^{\sigma_N}$, are only right- and left-normalized, respectively, so the MPS has to be brought into left-canonical form to apply the scheme discussed in the following paragraphs and explained in detail in Appendix \ref{AppTrueD}. In the left-canonical MPS, the tensors $A^{\sigma_i}$ at different sites are no longer equal, which justifies the addition of a site subscript, as shown in Fig. \ref{fig_mps_vbs}(a).

A MPS with open boundary conditions, virtual index dimension $D = 2$ and physical index dimension $d = 2$ (e.g., for spins-$\frac{1}{2}$ or spinless fermions) can be initialized exactly on quantum hardware following the single-layer scheme discussed by Ran \cite{Ran20}, which translated an earlier method by Sch\"{o}n et al. \cite{Schoen05} to the language of gate-based quantum computing. As explained in Appendix \ref{AppTrueD}, it is straightforward to extend such method to $d = 4$ (i.e., two qubits at each site, so that a spin-$1$ can be encoded locally), which results in the quantum circuit schematically shown in Fig. \ref{fig_mps_vbs}(c) for the particular case of $N=4$ sites. Apart from the last step, which amounts to a two-qubit operation, for a chain with $N$ sites, this method corresponds to the sequential application of $N-1$ three-qubit operations, thus allowing to prepare a 1D spin-$1$ VBS state with $\mathcal{O}(N)$ depth. 

The prefactor of such linear scaling can be prohibitively large for NISQ hardware, though. Using the state-of-the-art quantum Shannon decomposition \cite{Shende06}, which takes at most $20$ $\textsc{cnot}$ gates to decompose a three-qubit operation, the circuit depth in $\textsc{cnot}$ gates of this MPS-based method should scale as $\mathcal{O}(\sim 20 N)$ for $N$ lattice sites. For concreteness, as discussed in Appendix \ref{AppTrueD}, using the Cirq \texttt{three\_qubit\_matrix\_to\_operations} method \cite{Cirq3QDecomp}, the quantum Shannon decomposition was applied explicitly for the case of periodic boundary conditions described below (cf. Fig. \ref{fig_mps_vbs}(d)) to the three-qubit operation $D^{\dagger}_{[i]}, i = 2, 3, ..., N-1$ \footnote{Thanks to the translational invariance, the unitary disentangler $D_{[i]}$ happens to be the same for $i = 2, 3, ..., N -1$.}. This decomposition resulted in the maximum count of $20$ $\textsc{cnot}$ gates.

It is also possible to adapt this scheme to the preparation of a 1D spin-1 VBS state with periodic boundary conditions (cf. Fig. \ref{fig_mps_vbs}(d)). There are two main differences relative to the previous case. The first difference lies in the fact that the first operation, $D_{[N]}^{\dagger}$, acts on four qubits instead of three. However, since it is the first operation of the circuit, it is not necessary to find a circuit that acts on an arbitrary initial state, but only one that produces the desired effect on the fiducial state $\ket{0}^{\otimes 4}$. In other words, the basis gate decomposition of a $16 \times 16$ unitary matrix is replaced by the determination of the circuit that initializes a four-qubit state, which can be accomplished straightforwardly by exploiting its Schmidt decomposition, as discussed in Section \ref{Section7}. 

The second difference appears in the final operation $D_{[1]}^{\dagger}$, which is non-unitary in this case. As discussed in Appendix \ref{AppTrueD}, this two-qubit non-unitary operation can be embedded in a three-qubit unitary one via the generic method presented in Appendix C of \cite{Lin21}. This involves the addition of an ancillary qubit initialized in state $\ket{0}$, which is measured in the computational basis at the end of the scheme. If the ancilla is found in state $\ket{0}$, which occurs with a probability of $50 \%$ (irrespective of $N$), the VBS state is prepared in the remaining $2N$ qubits.

\subsection{Projected Entangled-Pair States in 2D}

Beyond one dimension, VBS states can no longer be expressed as a MPS with constant bond dimension $D$, but they can nonetheless be cast exactly in the form of a Projected Entangled-Pair State (PEPS) \cite{Cirac21} (a high-dimensional generalization of MPS) with bond dimension $D = 2$ as well. However, to the best of our knowledge, there is no method in the literature to initialize PEPSs on quantum hardware by deriving a quantum circuit from the respective local tensors, as described above for MPSs.

Nevertheless, Schwarz et al. \cite{Schwarz12} developed a quantum routine to prepare injective PEPS on digital quantum computers with polynomial resources, provided that the parent Hamiltonian of which the injective PEPS is the unique ground state is gapped \cite{Perez-Garcia08} and sequences of partial sums of local terms of the parent Hamiltonian can be defined such that each partial sum has a unique ground state of its own. The key idea \cite{Schwarz12} is to perform quantum phase estimation (QPE) \cite{Kitaev95, Nielsen00} with such a partial sum of terms of the parent Hamiltonian, enlarging the number of lattice sites acted upon by one at a time. This therefore gives rise to a sequential application of at least $N$ QPE executions --- possibly more than $N$ since the measurement outcome of QPE may not be the ground state energy, in which case the measurement must be undone via the Marriott-Watrous trick \cite{Jordan1875, Marriott05, Nagaj09}, so that QPE can be repeated until achieving the successful outcome.

For the case of two-dimensional VBS states such as the spin-$\frac{3}{2}$ VBS state on the honeycomb lattice and possibly the spin-$2$ VBS state on the square lattice (given the Tensor Network Renormalization Group (TNRG) results suggesting the existence of a nonzero gap \cite{Garcia-Saez13}), both these conditions are satisfied: The parent Hamiltonian (i.e., the respective AKLT model) is gapped, and the sequences of partial sums, which include only the projectors $P_{n,n'}^{S = S_{\textrm{max}}}(\vec{S}_n, \vec{S}_{n'})$ acting on the sites considered up to that point, also have a unique ground state. Hence, starting from the product state of valence bonds, $\ket{\psi_{\textrm{pre-VBS}}}$, the VBS state can be grown one site at a time via QPE using the partial sums of terms of the AKLT Hamiltonian, thus effectively resulting in a $\mathcal{O}(N)$ circuit depth.

\subsection{Implementation in NISQ Hardware}

In summary, the use of standard digital quantum simulation algorithms such as QPE \cite{Kitaev95, Nielsen00} or quantum cooling methods \cite{Motta20,McArdle19, Wang17} to prepare VBS states on quantum hardware using their parent AKLT Hamiltonians is hampered by the exponentially vanishing overlap with the easy-to-prepare initial state $\ket{\psi_{\textrm{pre-VBS}}}$. As an alternative, VBS states can be constructed one site at a time by exploiting their MPS representation in one dimension or the fact that the AKLT Hamiltonian is a sum of local projectors in any number of dimensions. In any case, despite the $\mathcal{O}(N)$ depth of these sitewise approaches, even for a very small number of lattice sites $N$, the circuit depth is bound to be prohibitively large for NISQ hardware as the MPS construction \cite{Ran20,Schoen05} involves $N-1$ sequential three-qubit operations and the injective PEPS preparation method by Schwarz et al. \cite{Schwarz12} requires at least $N$ sequential QPE executions. In the next section, we take the opposite approach by devising a method that implements the symmetrization at all sites in parallel, resulting in a constant-depth circuit at the cost of requiring the post-selection of the measurement outcomes of ancillary qubits. In a sense, circuit depth is traded for the repetition of the same shallow circuit, in the spirit of hybrid variational algorithms \cite{Cerezo21}. Thus, the resulting method, though probabilistic, is suitable for NISQ devices.

\section{Probabilistic Preparation of Valence-Bond-Solid States}\label{Section4}

The fact that the local symmetrization operator $\mathcal{S}$ is non-unitary means it cannot be expressed as a product of unitary operations (i.e., quantum gates). In any case, $\mathcal{S}$ is Hermitian, so $e^{-i\theta \mathcal{S}}$ is unitary for any $\theta \in \mathbb{R}$ and we can, in principle, find the corresponding quantum circuit. 

\begin{figure}[h]
    \centering{
    \small
    \Qcircuit @C=0.4em @R=0.4em {
    \quad & \quad & \quad & \mbox{\large\ensuremath{(c)}} \quad & \quad & \quad & \quad & \quad & \quad & \quad & \quad & \lstick{|Q_1\rangle} & \qw & {/^{^{2S}}} \qw & \qw & \multigate{3}{\mathcal{V}} & \gate{e^{-i \pi \mathcal{S}_1}} & \qw & \qw & \qw & \qw & \qw & \qw & \qw & \quad & \quad \\
    \quad & \quad & \quad & \quad & \quad & \quad & \quad & \quad & \quad & \quad & \quad & \lstick{|Q_2\rangle} & \qw & {/^{^{2S}}} \qw & \qw & \ghost{\mathcal{V}} & \qw & \qw & \gate{e^{-i \pi \mathcal{S}_2}} & \qw & \qw & \qw & \qw & \qw & \quad & \quad \\
    \quad & \quad & \quad & \quad & \quad & \quad & \quad & \quad & \quad & \quad & \quad & \lstick{\; \; \; \vdots \; \; \;} \\
    \quad & \quad & \quad & \quad & \quad & \quad & \quad & \quad & \quad & \quad & \quad & \lstick{|Q_N\rangle} & \qw & {/^{^{2S}}} \qw & \qw & \ghost{\mathcal{V}} & \qw & \qw & \qw & \gate{e^{-i \pi \mathcal{S}_N}} & \qw & \qw & \qw & \qw & \quad & \quad \\
    \quad & \quad & \quad & \quad & \quad & \quad & \quad & \quad & \quad & \quad & \quad & \lstick{|A_1\rangle} & \qw & \qw & \qw & \gate{H} & \ctrl{-4} & \qw & \qw & \qw & \gate{H} & \meter & \rstick{|1\rangle} \\
    \quad & \quad & \quad & \quad & \quad & \quad & \quad & \quad & \quad & \quad & \quad & \lstick{|A_2\rangle} & \qw & \qw & \qw & \gate{H} & \qw & \qw & \ctrl{-4} & \qw & \gate{H} & \meter & \rstick{|1\rangle} \\
    \quad & \quad & \quad & \quad & \quad & \quad & \quad & \quad & \quad & \quad & \quad & \lstick{\; \; \; \vdots \; \; \;} \\
    \quad & \quad & \quad & \quad & \quad & \quad & \quad & \quad & \quad & \quad & \quad & \lstick{|A_N\rangle} & \qw & \qw & \qw & \gate{H} & \qw & \qw & \qw & \ctrl{-4} & \gate{H} & \meter & \rstick{|1\rangle} \\
    }}
    
    \caption{(a) Two-qubit subcircuit that prepares a single valence bond $\frac{1}{\sqrt{2}}(\ket{\uparrow \downarrow} - \ket{\downarrow \uparrow}) \equiv \frac{1}{\sqrt{2}}(\ket{01} - \ket{10})$. Full circuit $\mathcal{V}$ that prepares $\ket{\psi_{\textrm{pre-VBS}}} = \mathcal{V} \ket{0}^{\otimes 2NS}$ amounts to repetition of this subcircuit across all pairs of qubits representing lattice links. (b) Hadamard test for generic unitary operator $\mathcal{U}$. Setting $\mathcal{U} = e^{-i\pi \mathcal{S}}$ turns input $\ket{0} \otimes \ket{\psi}$ into $\ket{0} \otimes (\mathbb{1} - \mathcal{S}) \ket{\psi} + \ket{1} \otimes \mathcal{S} \ket{\psi}$, so measuring ancilla in $\ket{1}$ ensures symmetrization operator $\mathcal{S}$ is applied at given site in main register. (c) Circuit that prepares spin-$S$ VBS state, $\ket{\psi_{\textrm{VBS}}} = \bigotimes_{n=1}^N \mathcal{S}_n \ket{\psi_{\textrm{pre-VBS}}}$, in $N$-site lattice if all $N$ ancillas are measured in $\ket{1}$. The basis gate decomposition of controlled-$e^{-i \pi \mathcal{S}}$ for $S = 1, \frac{3}{2}$ can be found in Section \ref{Section5}. All $N$ controlled-$e^{-i\pi \mathcal{S}_n}$ are executed in parallel, so the circuit depth is constant, i.e., independent of the system size $N$.}
    \label{Circ1}
\end{figure}
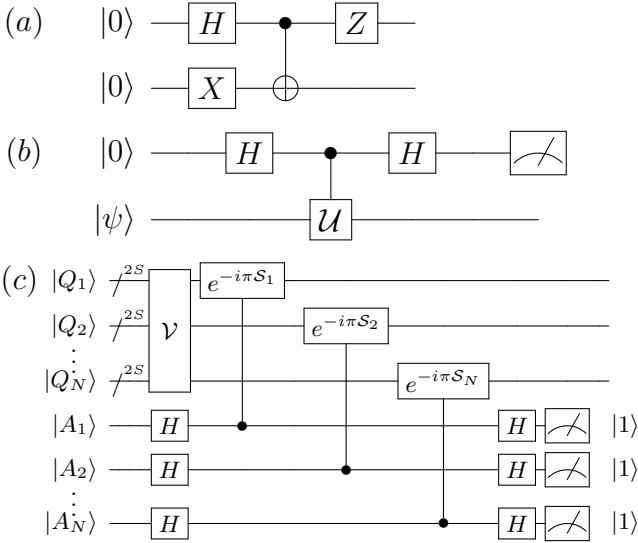

Let us now consider the application of the Hadamard test (cf. Fig. \ref{Circ1}(b)) with $\mathcal{U} = e^{-i \theta \mathcal{S}}$. Labelling the input state on the main register as $\ket{\psi}$, the output of this circuit before the measurement of the ancilla qubit is
\begin{equation}
    \ket{0} \otimes \Bigg[ \frac{\mathbb{1} + e^{-i\theta \mathcal{S}}}{2} \Bigg] \ket{\psi} + \ket{1} \otimes \Bigg[ \frac{\mathbb{1} - e^{-i\theta \mathcal{S}}}{2} \Bigg] \ket{\psi}.
\label{Hadamard_out_1}
\end{equation}
Using the fact that $\mathcal{S}$ is idempotent ($\mathcal{S}^2 = \mathcal{S}$), we have $e^{-i \theta \mathcal{S}} = \mathbb{1} - (1 - e^{-i\theta}) \mathcal{S}$. Replacing in (\ref{Hadamard_out_1}) gives
\begin{equation}
    \ket{0} \otimes \Bigg[\mathbb{1} - \frac{1 - e^{-i\theta}}{2} \mathcal{S} \Bigg] \ket{\psi} + \ket{1} \otimes \Bigg[\frac{1 - e^{-i\theta}}{2} \mathcal{S} \Bigg] \ket{\psi}.
\label{Hadamard_out_2}
\end{equation}
Setting $\theta = \pi$, we obtain
\begin{equation}
    \ket{0} \otimes (\mathbb{1} - \mathcal{S}) \ket{\psi} + \ket{1} \otimes \mathcal{S} \ket{\psi}.
\label{Hadamard_out_3}
\end{equation}
Hence, applying the Hadamard test with $\mathcal{U} = e^{-i\pi \mathcal{S}}$ to an input state $\ket{\psi}$ and retaining only the measurements of the ancilla qubit that yield $\ket{1}$ results in the application of $\mathcal{S}$ to $\ket{\psi}$ at the site in question.

This scheme can be generalized to the symmetrization of the input state $\ket{\psi}$ at all sites $n = 1, 2, ..., N$ by considering $N$ Hadamard tests in parallel, with one ancilla qubit per site. The outcome is of the form
\begin{equation}
\ket{\Phi} \equiv \sum_{m = 0}^{2^N - 1} \ket{m} \otimes \Bigg[ \bigotimes_{n = 1}^N (\mathbb{1} - \mathcal{S}_n)^{1-m_n} \mathcal{S}_n^{m_n} \ket{\psi} \Bigg],
\label{output_eq}
\end{equation}
where $\ket{m}$ is a computational basis state and $m = m_N m_{N-1} ... m_2 m_1$ is the binary representation of $m$. In words, whenever the $n^{\textrm{th}}$ ancilla qubit is measured in $\ket{1}$ (i.e. $m_n = 1$), the $n^{\textrm{th}}$ site of the input state $\ket{\psi}$ is locally symmetrized. As a result, if the $N$-qubit ancilla register is read out in state $\ket{11...11} \equiv \ket{2^N-1}$, the main register is found in the state $\mathcal{S}_1 \mathcal{S}_{2} ... \mathcal{S}_{N-1} \mathcal{S}_N \ket{\psi} \equiv \bigotimes_{n=1}^{N} \mathcal{S}_n \ket{\psi}$.

Fig. \ref{Circ1}(c) shows the quantum circuit that prepares the spin-$S$ VBS state. It consists of three parts: First, the preparation of the product state of valence bonds, $\ket{\psi_{\textrm{pre-VBS}}} = \mathcal{V} \ket{0}$, then the application of the layer of local Hadamard tests, and finally the measurement of the site ancillas in the computational basis, retaining only the trials that yield all site ancillas in state $\ket{1}$. Each local spin-$S$ is encoded by $2S$ qubits (e.g., $3$ qubits for a spin-$\frac{3}{2}$), so the main register has a total of $2NS$ qubits for a $N$-site lattice. For each lattice site, an additional ancilla is required to perform the postprocessing measurement.

\begin{figure}[h]
    \centering{\large \Qcircuit @C=1.2em @R=0.8em { \quad & (a) \quad & \quad & \quad & \lstick{|0\rangle} & \gate{H} & \ctrl{1} & \gate{Z} & \qw \\
    \quad & \quad & \quad & \quad & \lstick{|0\rangle} & \gate{X} & \targ & \qw & \qw \\
    }}
\end{figure}
\vspace{-0.6cm}
\begin{figure}[h]
    \centering{\large \Qcircuit @C=1.2em @R=0.8em {
    \quad & (b) \quad & \quad & \quad & \lstick{|0\rangle} & \qw & \gate{H} & \ctrl{1} & \gate{H} & \qw & \meter \\
    \quad & \quad & \quad & \quad & \lstick{| \psi \rangle} & \qw & \qw & \gate{\mathcal{U}} & \qw & \qw & \qw \\
    }}
\end{figure}
\vspace{-0.6cm}

The fact that $\ket{\psi_{\textrm{VBS}}}$ is only prepared if all $N$ ancillas are measured in $\ket{1}$ after the local Hadamard tests means only a few trials are successful. The average number of trials required to prepare $\ket{\psi_{\textrm{VBS}}}$ corresponds to the inverse of the probability of measuring all $N$ site ancillas in state $\ket{1}$. Using Eq. (\ref{output_eq}) and the orthogonality of the computational basis states, this corresponds to
\begin{equation}
    \begin{aligned}
    P_{11..1} & = \mathrm{Tr}(\ket{11...1} \bra{11...1} \otimes \mathbb{1}_{2^{2NS} \times 2^{2NS}} \ket{\Phi} \bra{\Phi}) \\
    & = |\braket{\psi_{\textrm{VBS}} | \psi_{\textrm{VBS}}}|^2 = p^N,
    \end{aligned}
    \label{eq_prob_Hadamard}
\end{equation}
where in the last step the result derived in Appendix \ref{AppC} was used. The average number of repetitions is thus exponential in the system size $N$, which may become prohibitively large for the intermediate values of $N$ at which quantum advantage may be attainable. Sections \ref{Section6} and \ref{Section7} introduce two methods to achieve a quadratic reduction of this repetition overhead at the cost of a $\mathcal{O}(\log(N))$ and $\mathcal{O}(1)$ overhead in circuit depth, respectively.

\section{Basis Gate Decomposition of $e^{-i \pi \mathcal{S}}$ for Local Spin $S = 1, \frac{3}{2}$}\label{Section5}

We now turn to the problem of compiling the derived algorithm into explicit quantum circuits for the important cases of $S = 1$ and $S = \frac{3}{2}$ VBS states in order to better understand the practical cost of implementing the proposed methods. Concretely, we decompose the abstract controlled-$e^{-i \pi {\cal S}}$ gates, for local spins $S=1,\frac{3}{2}$, in terms of an elementary gate set $\left\{ \textsc{cnot}, \, U(\theta, \phi, \lambda) \right\}$, where $U(\theta, \phi, \lambda)$ is the general single-qubit operation
\begin{equation}
    U(\theta, \phi, \lambda) = \begin{pmatrix} 
    \cos(\theta/2) & -e^{i \lambda} \sin(\theta/2) \\
    e^{i\phi} \sin(\theta/2) & e^{i(\phi+\lambda)} \cos(\theta/2)
    \end{pmatrix}. 
    \label{eq:U-gate}
\end{equation}

To make the circuits as efficient as possible, we need to find their shallowest decompositions, as these enable faster execution times and, more importantly, less error accumulation in near-term noisy platforms. Decompositions with fewer \textsc{cnot} gates are prioritized because two-qubit operations 
are currently one to two orders of magnitude more error-prone than single-qubit ones \cite{Bruzewicz19, Kjaergaard20}.

We start with the decomposition of the controlled-$e^{-i \pi \mathcal{S}^{(\mathcal{N}=2)}}$ for local spin $S \equiv \frac{\mathcal{N}}{2} = 1$. Taking the exponential of the matrix representation of $\mathcal{S}^{(\mathcal{N}=2)}$ stated in Eq. (\ref{eqdefS_1}) in Appendix \ref{AppB}, one finds that
\begin{equation}
    e^{-i \pi \mathcal{S}^{(\mathcal{N}=2)}} = \begin{pmatrix}
     -1 & 0 & 0 & 0 \\
     0 & 0 & -1 & 0 \\
     0 & -1 & 0 & 0 \\
     0 & 0 & 0 & -1
    \end{pmatrix} = e^{i \pi} \textsc{swap}.
    \label{eq_SWAP_S}
\end{equation}
Hence, the controlled-$e^{-i \pi \mathcal{S}^{(\mathcal{N}=2)}}$ is just a Fredkin gate (i.e., a controlled-\textsc{swap}, or \textsc{cswap}, for short) preceded by a $\textsc{z}$ gate acting on the control-qubit to account for the global phase factor $e^{i \pi}$ in Eq. (\ref{eq_SWAP_S}) above, which becomes a relative phase factor upon controlling it. In reality, in the local Hadamard test, this $\textsc{z}$ gate can be skipped altogether; the only difference is that the desired outcome in the measurement of the site ancilla is $\ket{0}$ instead of $\ket{1}$. 

Optimized basis gate decompositions of the Fredkin gate with the above gate set are given in \cite{cruz2022optimized} for three different coupling and qubit assignment setups. Building upon the result assuming all-to-all qubit connectivity, with 7 \textsc{cnot}s and a depth of 10 operations, the circuit for the local Hadamard test can be straightforwardly compiled with the same depth and \textsc{cnot} count, since the Hadamard and \textsc{z} gates can be absorbed into the outer single-qubit operations of this Fredkin gate decomposition. Some processors, however, are restricted to nearest-neighbor couplings only. In that case, as will be described in Section \ref{Section9}, the natural placement of the ancilla qubit is between the two qubits that encode the spin-1 state at that site. The Hadamard test circuit can then be implemented with only 9 \textsc{cnot} gates and a depth of 19 operations by using the decomposition for the Fredkin gate in \cite{cruz2022optimized} where linear connectivity is assumed and the control qubit is placed at the center, dropping the unnecessary last three \textsc{cnot} gates that perform a \textsc{swap} of the control back to the central position.

Moving on to the case of the controlled-$e^{-i \pi \mathcal{S}^{(\mathcal{N}=3)}}$ for the scenario of local spin $S \equiv \frac{\mathcal{N}}{2} = \frac{3}{2}$ found in the honeycomb lattice, we first exponentiate the matrix representation of $\mathcal{S}^{(\mathcal{N}=3)}$, stated in Eq. (\ref{eqdefS3}), to obtain
\begin{equation}
e^{-i\pi\mathcal{S}^{(\mathcal{N}=3)}} =\left(\begin{array}{cccccccc}
    -1 & 0 & 0 & 0 & 0 & 0 & 0 & 0\\
    0 & \frac{1}{3} & -\frac{2}{3} & 0 & -\frac{2}{3} & 0 & 0 & 0\\
    0 & -\frac{2}{3} & \frac{1}{3} & 0 & -\frac{2}{3} & 0 & 0 & 0\\
    0 & 0 & 0 & \frac{1}{3} & 0 & -\frac{2}{3} & -\frac{2}{3} & 0\\
    0 & -\frac{2}{3} & -\frac{2}{3} & 0 & \frac{1}{3} & 0 & 0 & 0\\
    0 & 0 & 0 & -\frac{2}{3} & 0 & \frac{1}{3} & -\frac{2}{3} & 0\\
    0 & 0 & 0 & -\frac{2}{3} & 0 & -\frac{2}{3} & \frac{1}{3} & 0\\
    0 & 0 & 0 & 0 & 0 & 0 & 0 & -1
    \end{array}\right).
\label{eq:expSN3}
\end{equation}
Decomposing the controlled version of Eq. (\ref{eq:expSN3}) using the state-of-the-art optimized quantum Shannon decomposition \cite{Shende06, Krol21} takes $63$ $\textsc{cnot}$ gates \cite{iten2019introduction}. Starting from this method and then proceeding with a heuristic-based optimization approach, we have found a circuit that implements the controlled-$e^{-i \pi \mathcal{S}^{(\mathcal{N}=3)}}$ with just $26$ \textsc{cnot} gates.

First, the $63$-$\textsc{cnot}$ circuit yielded by the optimized quantum Shannon decomposition, assuming full connectivity, was recompiled with Qiskit's \texttt{transpile} function \cite{QiskitTranspile}, which can perform some basic simplifications. The \textsc{cnot} count was reduced to $54$. Then, subsequent simplification steps were carried out in the ZX-calculus language; after some effort, the number of $\textsc{cnot}$ gates was reduced by more than half, as described below.

To accomplish that, the 54-\textsc{cnot} circuit was converted into a ZX-diagram with the aid of the PyZX software library \cite{kissinger2020pyzx}, which contains a number of methods to simplify ZX-diagrams and convert them back into quantum circuits \cite{duncan2020graphtheoretic}. We first tested the full bundle of ZX-diagram simplification rules available, including spider-fusion, identity removal, pivoting, local complementation and removal of interior nodes, followed by the gadgetization technique and its simplification. All these procedures are conveniently combined into the \texttt{full\_reduce} function. The next step was extracting a new quantum circuit from the reduced ZX-graph \cite{backens2021there}. PyZX offers a few further methods to optimize the obtained circuit, which we employed before passing the result back through the Qiskit transpiler. However, in the end, the resulting quantum circuit was not simplified; rather, the \textsc{cnot} count increased to $64$.

Very different circuits are often extracted from equivalent ZX-graphs. Hence, one way to try to avoid generating a deeper circuit than the initial $54$-qubit one is to search for extracted circuits that minimize the gate count in the local space of ZX-diagrams that are equivalent to the previously simplified version. PyZX implements two optimization techniques to perform this local search: simulated annealing and genetic algorithms. The next leg of our circuit simplification pipeline built upon these methods with the intensive search and optimization procedure described in \cite{cruz2022optimized}, which often succeeds in escaping from local minima, thus optimizing decompositions further.

The final four-qubit circuit we have obtained for the controlled-$e^{-i \pi \mathcal{S}^{(\mathcal{N}=3)}}$ operation (up to a global phase factor) in terms of the aforementioned gateset and assuming full qubit connectivity comprises $26$ \textsc{cnot} gates and $42$ single-qubit gates, with depth $26$ $\textsc{cnot}$ gates (or depth $45$, if single-qubit layers are included). A similar optimization procedure was then repeated for the case of linear qubit connectivity, yielding a circuit with 39 \textsc{cnot} gates and 53 single-qubit gates, with depth $39$ $\textsc{cnot}$ gates (or depth $62$, including single-qubit layers). These circuits can be extracted from the QASM files 5 and 6 introduced in Appendix \ref{AppI}.

Finally, while the exact complexity of circuit extraction from a ZX-diagram is not yet known, a recent preprint shows it is at least \texttt{\#P}-hard \cite{de2022circuit}. Nevertheless, we were still able to dramatically reduce the number of operations necessary to build these circuits using a heuristic search. 

\section{Mitigating Repetition Overhead via Local Hadamard Test}\label{Section6}

The repetition overhead faced in the initialization of VBS states is common to other quantum many-body states, the preparation of which involves some non-unitary operation. This includes the Gutzwiller wave function \cite{Murta21,Seki22}, the ground state of the Kitaev honeycomb model \cite{Bespalova21}, or the Bethe ansatz for the XXZ model \cite{VanDyke21, VanDyke211}. In the latter case, for example, the number of repetitions was reduced via quantum amplitude amplification \cite{Brassard00} (a generalization of Grover's algorithm \cite{Grover97}), which we may also apply to the preparation of VBS states. In fact, the layer of local Hadamard tests introduced in the previous section allows for the construction of an appropriate oracle that can be implemented with, at most, $\mathcal{O}(N)$ circuit depth (cf. Appendix \ref{AppD}). However, the resulting quantum amplitude amplification scheme leads to a circuit with exponential depth in $N$, since the angle swept in the relevant two-dimensional subspace at each iteration is set by the exponentially vanishing overlap between the initial state $\ket{\psi_{\textrm{pre-VBS}}}$ and the final state $\ket{\psi_{\textrm{VBS}}}$. This would defeat our purpose of developing a low-depth preparation scheme for NISQ processors.

Returning to the probabilistic method introduced in the previous section, it is reasonable to expect to explore the locality of the Hadamard tests to reduce the repetition overhead. Indeed, measuring a single ancilla in state $\ket{1}$ is straightforward: it only takes $\frac{1}{p}$ repetitions on average. For example, for the spin-$1$ VBS state, $3$ out of every $4$ repetitions yield a site ancilla in state $\ket{1}$. Similarly, for the spin-$\frac{3}{2}$ VBS state, $1$ out of $2$ trials produces the desired outcome. The difficulty in preparing $\ket{\psi_{\textrm{VBS}}}$ lies in having to measure all $N$ site ancillas in state $\ket{1}$ simultaneously. Considering a coin toss analogy, having a large number $N$ of tossed coins land with ``heads'' facing up is an exponentially suppressed event, but obtaining ``heads'' from tossing a single coin is not.

If the input state happened to be a tensor product of sitewise states, one could repeat the local Hadamard test at each site as many times as required to obtain the successful outcome, resetting the respective qubits after an unsuccessful trial and before attempting a further one. Returning to the coin toss analogy, instead of having to obtain $N$ ``heads'' at once, one would toss each coin separately until yielding ``heads'', thus repeating only the toss of those coins that kept on yielding ``tails''. In principle, different sites would attain success after a different number of attempts, so the first ones to succeed would have to be held until all remaining sites were symmetrized. The crucial point is that, because of the absence of entanglement between different sites, resetting the qubits at one site would not affect the wave function at any other site.

However, the initial state $\ket{\psi_{\textrm{pre-VBS}}}$ involves valence bonds between neighboring sites, so resetting the qubits of one site after an unsuccessful local Hadamard test affects the remainder of the sites. In fact, it does not simply affect the nearest-neighboring sites. The effective action of the Hadamard test on the state $\ket{\psi}$ in the main $2S$-qubit register (i.e., $\mathcal{S} \ket{\psi}$ if the ancilla is measured in $\ket{1}$, and $(1-\mathcal{\mathcal{S}})\ket{\psi}$ otherwise) produces entanglement between the $2S$ qubits at that site. These qubits were already entangled to those from neighboring sites, and within each such neighboring site all $2S$ qubits become entangled to one another due to the respective Hadamard test. Hence, the entanglement spreads across the whole system, as expected from the hidden order captured by the string order parameter in VBS states \cite{Wei22}.

\begin{figure}[h]
\centering{\
\includegraphics[width=\linewidth]{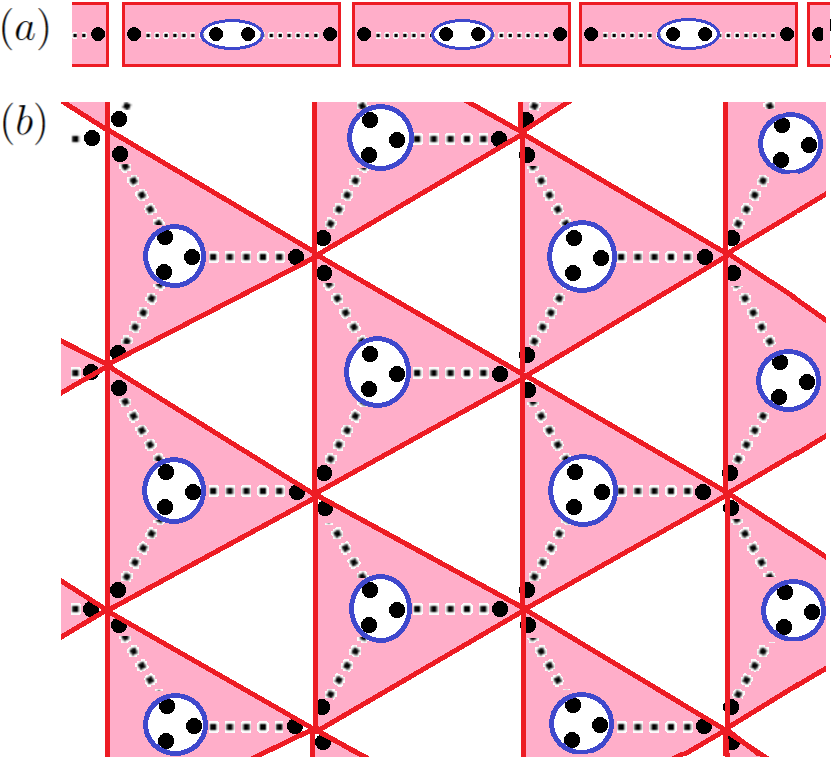}
\par}
\caption{Scheme of product state of valence bonds, $\ket{\psi_{\textrm{pre-VBS}}}$, in one-dimensional lattice (a) and honeycomb lattice (b) to support explanation of repetition overhead mitigation scheme. Dashed black lines represent valence bonds and black dots denote qubits. $4$-qubit islands within filled rectangles (a) and $6$-qubit islands within filled triangles (b) are not entangled with one another. Hence, application of Hadamard test at sites marked with blue circles, corresponding to a single sublattice, can be repeated (with reset of qubits between consecutive trials) as many times as required to achieve success.}
\label{fig_islands}
\end{figure}

If, however, one attempts to symmetrize the wave function \textit{only at the $\frac{N}{2}$ sites of one sublattice} --- assuming, of course, the lattice is bipartite ---, the entanglement resulting from the action, whether successful or unsuccessful, of each local Hadamard test is no longer spread across the whole system, being confined to a $4S$-qubit island (cf. Fig. \ref{fig_islands}) that only encompasses qubits from the respective site and its nearest neighbors (at which the local Hadamard test is \textit{not} applied). The trial-and-error approach outlined above is now viable because next-nearest-neighboring sites are not entangled to one another in the input state, so all qubits within each island can be reset without affecting the remaining qubits.

In brief, the repetition overhead mitigation strategy consists of applying the Hadamard test with $\mathcal{U} = e^{-i\pi \mathcal{S}}$, as in the probabilistic method described above, but only to the $\frac{N}{2}$ sites of one sublattice, provided that the lattice in question is bipartite. Some of these local Hadamard tests will result in a successful symmetrization of the corresponding site (those for which the ancilla is measured in $\ket{1}$) while others will fail (those for which the ancilla is measured in $\ket{0}$). In the latter, one simply resets the qubits, reprepares the valence bonds that emanate from the respective site, and tries the Hadamard test again, repeating this as many times as required to obtain a successful outcome. For clarity, Fig. \ref{fig_islands} details the independent $4S$-qubit islands for the $S = 1, \frac{3}{2}$ VBS states.

The cumulative probability of symmetrizing the $\frac{N}{2}$ sublattice sites after $n$ local Hadamard test rounds is
\begin{equation}
    P_n = (p R_n)^{N/2},
\end{equation}
where $R_n$ follows the recursive relation
\begin{equation}
    R_n = 1 + (1-p)R_{n-1}, \; R_1 = 1.
\end{equation}
As expected, for the first round, $P_1 = p^{\frac{N}{2}}$, which corresponds to the probability of measuring all $\frac{N}{2}$ ancillas in state $\ket{1}$ at once, as in the probabilistic method. As $n \to \infty$, $R_n$ converges to $\frac{1}{p}$, so that the cumulative probability $P_n$ approaches $1$ in that limit. 

The average number of rounds $\langle n \rangle$ required to symmetrize all $\frac{N}{2}$ sites of one sublattice can be computed numerically, yielding a logarithmic scaling with respect to the system size $N$. For the spin-$1$ VBS state ($p = \frac{3}{4}$),
\begin{equation}
    \langle n \rangle = 0.71 \log N + 0.49 \; \textrm{(2 s.f.),}
\end{equation}
and, for the spin-$\frac{3}{2}$ VBS state ($p = \frac{1}{2}$),
\begin{equation}
    \langle n \rangle = 1.4 \log N + 0.49 \; \textrm{(2 s.f.).}
\end{equation}
The logarithmic scaling with the system size $N$ could be anticipated from the fact that, on average, at each round, $\mathcal{S}$ is successfully applied at a fraction $p$ of the hitherto unsymmetrized sites. Since the circuit depth of a single layer of Hadamard tests (cf. Fig. \ref{Circ1}(c)) is independent of $N$, this strategy to reduce the repetition overhead translates into an additional circuit layer of $\mathcal{O}(\log N)$ depth. 

Once this step is complete, one is left with the symmetrization of only half of the sites (those belonging to the other sublattice), which requires, on average, only the square root of the original repetitions using the probabilistic method introduced in the previous section. For example, for the spin-$\frac{3}{2}$ VBS state on a honeycomb lattice with $N = 20$ sites, the unmitigated probabilistic method requires an average of $2^{20} = 1,048,576$ repetitions, while the repetition overhead mitigation strategy reduces this overhead to $2^{10} = 1024$ trials.

Two final notes about the integration of this repetition overhead mitigation scheme in the probabilistic method are in order. First, the repetition overhead mitigation strategy must be applied \textit{before} the probabilistic method, because the former exploits the entanglement structure of the initial state $\ket{\psi_{\textrm{pre-VBS}}}$, while the latter is agnostic to it, in that it applies the local symmetrization operator $\mathcal{S}$ at every site where the respective ancilla is measured in $\ket{1}$ regardless of the input state. Second, the $\frac{N}{2}$ ancillas used in the repetition overhead mitigation scheme can be reused in the probabilistic method, thus reducing the number of ancillas by half. 

\section{Constant-Depth Mitigation of Repetition Overhead}\label{Section7}

Despite the low circuit depth added by the repetition overhead mitigation layer, its implementation in the currently available quantum computers may be challenging due to the need to process the outcome of mid-circuit measurements in real time. Indeed, although the mid-circuit measurement and reset of qubits are already available in some state-of-the-art platforms \cite{Honeywell20,IBM21}, the delay resulting from performing the measurements in the quantum processor, deciding the next step on the conventional processor, and finally conveying this decision to the quantum processor may be great enough to the point of extending the execution time of the circuit beyond the limits set by the decoherence of the qubits.  

However, if the $4S$-qubit islands are sufficiently small, for each island one can simply prepare the respective state deterministically using generic quantum state preparation methods. Specifically, to prepare the $4$-qubit and $6$-qubit (normalized) states corresponding to each island in the spin-$1$ and spin-$\frac{3}{2}$ VBS states, we will make use of a scheme \cite{Plesch11,LosAlamos18} based on the Schmidt decomposition \cite{Nielsen00} of a general $2n$-qubit state.

Let us consider the preparation of a $2n$-qubit state
\begin{equation}
\ket{\phi} = \sum_{\substack{i_1, ..., i_n = 0 \\ j_1, ..., j_n = 0}}^{1} M_{j_1, ..., j_n}^{i_1, ..., i_n} \ket{i_1, ..., i_n; j_1, ..., j_n},
\end{equation}
where the separation of indices $\{i_k\}_{k=1}^{n}$ and $\{j_k\}_{k=1}^{n}$ reflects the symmetric bipartition we will adopt to perform the Schmidt decomposition of $\ket{\phi}$. Merging the indices $\{i_k\}_{k=1}^{n}$ and $\{j_k\}_{k=1}^{n}$ into a single index $\mathbf{i}$ and $\mathbf{j}$, respectively, we have
\begin{equation*}
\ket{\phi} = \sum_{\mathbf{i} = 0}^{2^{n}-1} \sum_{\mathbf{j} = 0}^{2^{n}-1} M_{\mathbf{i}; \mathbf{j}} \ket{\mathbf{i}} \otimes \ket{\mathbf{j}}.
\end{equation*}
The amplitudes of $\ket{\phi}$ are now cast in the form of a $2^{n} \times 2^{n}$ matrix $\mathbf{M}$. The singular value decomposition of $\mathbf{M}$ gives $\mathbf{M} = \mathbf{U} \mathbf{S} \mathbf{V}^{\dagger}$, in which case we can express $\ket{\phi}$ in terms of its Schmidt decomposition:
\begin{equation}
    \ket{\phi} = \sum_{k = 0}^{2^n-1} s_k (\mathbf{U} \ket{k}) \otimes (\mathbf{V} \ket{k}) \equiv \sum_{k = 0}^{2^n-1} s_k \ket{u_k} \otimes \ket{v_k}.
\label{phi_svd}
\end{equation}
$\{s_k\}_{k=0}^{2^n-1}$ are the singular values of $\mathbf{M}$, which are the real, non-negative entries of the diagonal matrix $\mathbf{S}$. $\{ u_k\}_{k=0}^{2^n-1}$ and $\{ v_k\}_{k=0}^{2^n-1}$ are the left and right singular vectors of $\mathbf{M}$, corresponding to the columns of the unitary matrices $\mathbf{U}$ and $\mathbf{V}$. Importantly, since $\mathbf{M}$ is itself unitary, $\sum_{k = 0}^{2^n-1} s_k^2 = 1$, so $(s_0, s_1, ..., s_{2^n-1})^T = \sum_{k = 0}^{2^n-1} s_k \ket{k}$ defines a properly normalized $n$-qubit state. 

\begin{figure}[h]
\centering{\
\includegraphics[width=\linewidth]{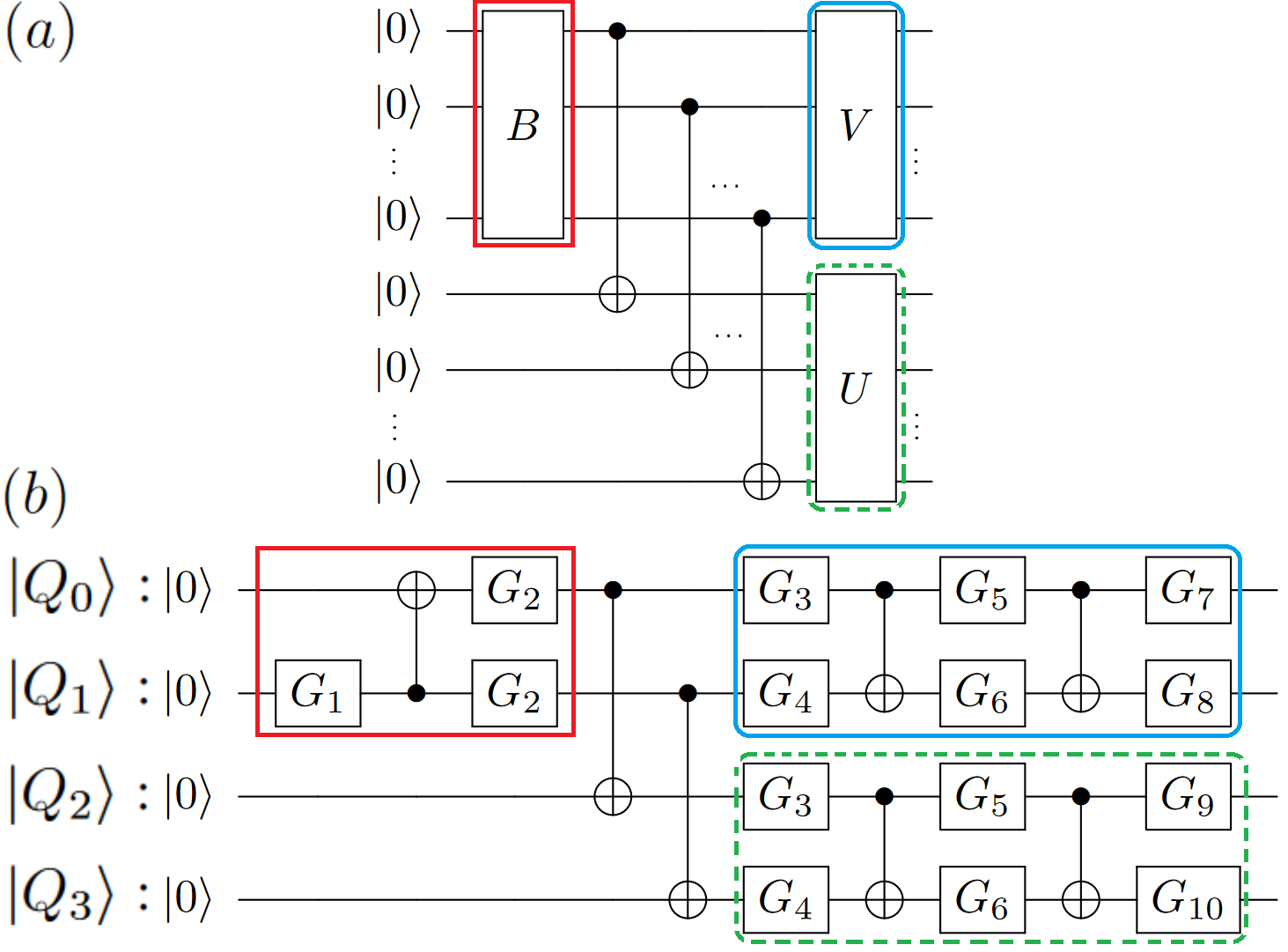}
\par}
\caption{(a) High-level scheme of general method \cite{Plesch11,LosAlamos18} to prepare $2n$-qubit state $\ket{\phi} = \sum_{\mathbf{i} = 0}^{2^{n}-1} \sum_{\mathbf{j} = 0}^{2^{n}-1} M_{\mathbf{i}; \mathbf{j}} \ket{\mathbf{i}; \mathbf{j}}$ by exploiting its Schmidt decomposition. $U$ and $V$ are found through the singular value decomposition of $\mathbf{M}$, i.e., $\mathbf{M} = \mathbf{U} \mathbf{S} \mathbf{V}^{\dagger}$. $B$ prepares state $\sum_{k=0}^{2^n-1} s_k \ket{k}$, where $\mathbf{S} = \textrm{diag}(s_0, ..., s_{2^n-1})$. (b) Application of this general method to preparation of four-qubit islands in spin-$1$ VBS state (cf. Eq. (\ref{psi_island_spin_1})). Valence bonds are associated with qubit pairs $(Q_0,Q_1)$ and $(Q_2,Q_3)$. Symmetrization operator $\mathcal{S}$ is applied at qubits $(Q_1,Q_2)$, which encode a single site. A total of $7$ $\textsc{cnot}$ gates and a depth of $4$ $\textsc{cnot}$ gates are required to prepare such state. Single-qubit gates $G_i$ are presented in compact form for the sake of clarity; details can be found in the QASM file 3 introduced in Appendix \ref{AppI}.}
\label{Schmidt_fig}
\end{figure}

Fig. \ref{Schmidt_fig}(a) shows a high-level scheme of the circuit that exploits the Schmidt decomposition to prepare $\ket{\phi}$ on a $2n$-qubit register, which is split into two $n$-qubit subregisters. It comprises three parts:
\begin{enumerate}
    \item{Initialize the state $\sum_{k = 0}^{2^n-1} s_k \ket{k}$ on one of the two subregisters via the $n$-qubit subcircuit $B$.}
    \item{Apply a network of $n$ $\textsc{cnot}$ gates, the i\textsuperscript{th} $\textsc{cnot}$ having as control the i\textsuperscript{th} most significant qubit of the subregister previously acted on by $B$ and as target the i\textsuperscript{th} most significant qubit of the other subregister. This prepares $\sum_{k=0}^{2^n-1} s_k \ket{k} \otimes \ket{k}$.}
    \item{Apply the $n$-qubit subcircuits $U$ and $V$ in parallel to one subregister each. This produces Eq. (\ref{phi_svd}).}
\end{enumerate}

In short, this method allows to turn the preparation of a $2n$-qubit state into the basis gate decomposition of $n$-qubit operations, thus simplifying the process considerably. For example, for $n = 1$, one can prepare an arbitrary entangled two-qubit state $\ket{\phi}$ with a single $\textsc{cnot}$ gate, since $B$, $U$ and $V$ are all single-qubit operations that can be decomposed into elementary operations trivially (e.g., via the ZYZ decomposition \cite{Nielsen00}). Of course, if $\ket{\phi}$ is separable, then $s_0 = 1$, $s_1 = 0$, in which case $\ket{\phi} = \ket{u_0} \otimes \ket{v_0}$ and the $\textsc{cnot}$ gate is redundant, as we only need to initialize two single-qubit states in parallel.

Let us now consider the application of this method to the preparation of the four-qubit state $\ket{\psi_{\textrm{island}}^{S=1}}$ corresponding to an island of a spin-$1$ VBS state (cf. Fig. \ref{fig_islands}(a)). Concretely, $\ket{\psi_{\textrm{island}}^{S=1}}$ results from preparing a product state of two valence bonds, applying the symmetrization operator $\mathcal{S}$ to the central pair of qubits, which encode a single site, and normalizing the resulting state:
\begin{equation}
    \begin{aligned}
    & \ket{\psi_{\textrm{island}}^{S=1}} = \frac{1}{\sqrt{3/4}} (\mathbb{1}_{2 \times 2} \otimes \mathcal{S} \otimes \mathbb{1}_{2 \times 2}) \left[  \begin{pmatrix}
     0 \\
     \frac{1}{\sqrt{2}} \\
     \frac{-1}{\sqrt{2}} \\
     0
    \end{pmatrix} \otimes \begin{pmatrix}
     0 \\
     \frac{1}{\sqrt{2}} \\
     \frac{-1}{\sqrt{2}} \\
     0
    \end{pmatrix} \right] \\
    & = \frac{1}{\sqrt{3}}\Big(0,0,0,\frac{1}{2},0,\frac{1}{2},-1,0,0,-1,\frac{1}{2},0,\frac{1}{2},0,0,0\Big)^T.
    \end{aligned}
\label{psi_island_spin_1}
\end{equation}
In general, preparing such a four-qubit state using the method from Fig. \ref{Schmidt_fig}(a) requires the preparation of a two-qubit state (via subcircuit B), which takes at most $1$ $\textsc{cnot}$ gate, and the application of two two-qubit subcircuits $U$ and $V$, each taking at most $3$ $\textsc{cnot}$ gates \cite{Vidal04}. Hence, a maximum of $9$ $\textsc{cnot}$ gates and a circuit depth of at most $5$ $\textsc{cnot}$ gates, ignoring connectivity constraints, are required to prepare an arbitrary four-qubit state. For the specific case of $\ket{\psi_{\textrm{island}}^{S=1}}$ (cf. Eq. (\ref{psi_island_spin_1})), its initialization involves a total of $7$ $\textsc{cnot}$ gates and a depth of $4$ $\textsc{cnot}$ gates, since $U$ and $V$ can both be decomposed into a circuit with only $2$ $\textsc{cnot}$ gates each using the Qiskit \texttt{two\_qubit\_cnot\_decompose} method \cite{Qiskit2QDecomp}, which implements the KAK decomposition \cite{Tucci05}. Figure \ref{Schmidt_fig}(b) presents a scheme of the corresponding quantum circuit. 

The application of the Schmidt-decomposition-based initialization method to a six-qubit state such as the spin-$\frac{3}{2}$ VBS state islands involves the preparation of a three-qubit state through subcircuit $B$ and the application of two three-qubit operations $U$ and $V$ in parallel. The latter take at most $20$ $\textsc{cnot}$ gates \cite{Shende06}. As for the former, even though the generic preparation method discussed in this section is only strictly applicable to states encoded in an even number of qubits, it is possible to consider an asymmetric bipartition, as discussed in Appendix \ref{AppE} for the three-qubit case, which yields a maximum of $4$ $\textsc{cnot}$ gates. Hence, ignoring connectivity constraints, a general six-qubit state can be prepared with at most $47$ $\textsc{cnot}$ gates and a circuit depth of $25$ $\textsc{cnot}$ gates. However, for the particular case of the six-qubit islands in the spin-$\frac{3}{2}$ VBS, the implementation of $B$ takes $3$ $\textsc{cnot}$ gates instead of the maximum $4$, and the $\textsc{cnot}$ count of the decomposition of $U$ and $V$ was reduced from the $20$ and $19$ $\textsc{cnot}$ gates yielded by the Cirq \texttt{three\_qubit\_matrix\_to\_operations} method to $14$ and $15$ $\textsc{cnot}$ gates, respectively, using the same procedure described in Section \ref{Section5} of converting the circuit to a ZX-diagram, simplifying it, and searching over the space of equivalent ZX-diagrams for optimal circuit extraction. To sum up, each spin-$\frac{3}{2}$ island can be prepared with a total of $35$ $\textsc{cnot}$ gates and a depth of $19$ $\textsc{cnot}$ gates. The respective QASM file is introduced in Appendix \ref{AppI}.

The use of this Schmidt-decomposition-based method to initialize the $4S$-qubit islands of the spin-$S$ VBSs as a strategy to reduce the repetition overhead of the probabilistic method is justified by the significant savings it generates relative to the default option of applying a standard state preparation method, such as that implemented via the Qiskit \texttt{initialize} function \cite{QiskitInitialize}. For the $S=1$ case, Qiskit \texttt{initialize} prepares the four-qubit island state (cf. Eq. (\ref{psi_island_spin_1})) with a total of $22$ $\textsc{cnot}$ gates, which compares with the $7$ $\textsc{cnot}$ gates yielded by the Schmidt-decomposition-based method. As for $S = \frac{3}{2}$, the number of $\textsc{cnot}$ gates is reduced from $114$ to $35$.

Although this constant-depth strategy is clearly the preferred option for the $S = 1, \frac{3}{2}$ cases, for larger local spins (i.e., larger coordination numbers of the lattice) it may be preferable to employ the logarithmic-depth scheme discussed in the previous section, as it only requires the basis gate decomposition of the controlled-$e^{-i \pi S}$ operation, for which a general implementation is discussed in the next section.

\section{Local Symmetrization via Linear Combination of Unitaries}\label{Section8}

We have proposed a NISQ-friendly probabilistic method to prepare any spin-$S$ VBS state on quantum hardware, with $S = 1, \frac{3}{2}, 2, \frac{5}{2}, 3, ...$ The implementation of this method (and of the logarithmic-depth repetition overhead mitigation strategy from Section \ref{Section6}) merely relies upon the determination of a $(2S + 1)$-qubit subcircuit that implements the controlled-$e^{-i \pi \mathcal{S}}$, where $\mathcal{S}$ is the local symmetrization operator acting on $2S$ spins-$\frac{1}{2}$. For the paradigmatic cases $S = 1, \frac{3}{2}$, we have performed an efficient basis gate decomposition of the respective $3$-qubit and $4$-qubit operations, as discussed in Section \ref{Section5}. 

However, for a larger local spin $S$, this process may become a bottleneck. Even for $S = 2$, corresponding to the local spin of a VBS state defined on a square lattice, the controlled-$e^{-i \pi \mathcal{S}}$ is a $5$-qubit operation, for which it is already particularly challenging to find a shallow decomposition via standard methods. 

This section presents a systematic method to find a circuit that, upon the measurement of ancillary qubits in the fiducial state, leads to the application of the local symmetrization operator $\mathcal{S}$ in the main register. The key idea is that, even though $\mathcal{S}$ is non-unitary (and therefore hard to implement on quantum hardware), $\mathcal{S}$ can be expressed as a linear combination of unitaries, each of which is easy to decompose into basis gates. This approach therefore allows to bypass the basis gate decomposition of a large subcircuit, particularly for high local spin $S$. The potential savings generated by this method for the spin-$2$ VBS state, a resource state for universal quantum computation \cite{Wei15}, will be discussed.

As a starting point, let us consider the simplest non-trivial case of the symmetrization operator $\mathcal{S}^{(\mathcal{N}=2)}$ acting on two spins-$\frac{1}{2}$, which is relevant for the preparation of the one-dimensional spin-$1$ VBS. Noting that the symmetrization operator is defined as the uniform linear combination of the elements of the symmetric group $S_n$ (cf. Eq. (\ref{eq_Symm_S_1}) for the particular case of two spins-$\frac{1}{2}$ and Eq. (\ref{eq_Symm}) for the general case), we have
\begin{equation*}
    \mathcal{S}^{(\mathcal{N}=2)} = \frac{1}{2!} 
    \left( \begin{pmatrix} 
    1 & 2 \\
    1 & 2  \end{pmatrix} + 
    \begin{pmatrix} 
    1 & 2 \\
    2 & 1  \end{pmatrix} \right),
\end{equation*}
where Cauchy's two-line notation is used to denote the permutations. In terms of quantum circuits, 
\begin{figure}[h]
\centering{\
\includegraphics[width=\linewidth]{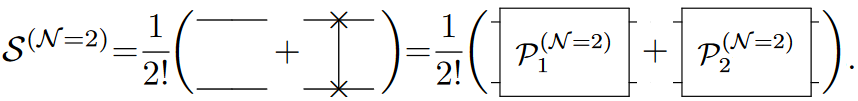}
\par}
\end{figure}

\noindent The notation used after the second equality will be relevant in the derivation of a systematic method to find the circuits for all permutations $\{ \mathcal{P}^{(\mathcal{N})}_{i}\}_{i=1}^{\mathcal{N}!}$ of an arbitrary number $\mathcal{N}$ of spins-$\frac{1}{2}$. As a sanity check, let us compute $e^{-i \pi \mathcal{S}^{(\mathcal{N}=2)}} = \mathbb{1} - 2 \mathcal{S}^{(\mathcal{N}=2)}$:
\begin{figure}[h]
\centering{\
\includegraphics[width=\linewidth]{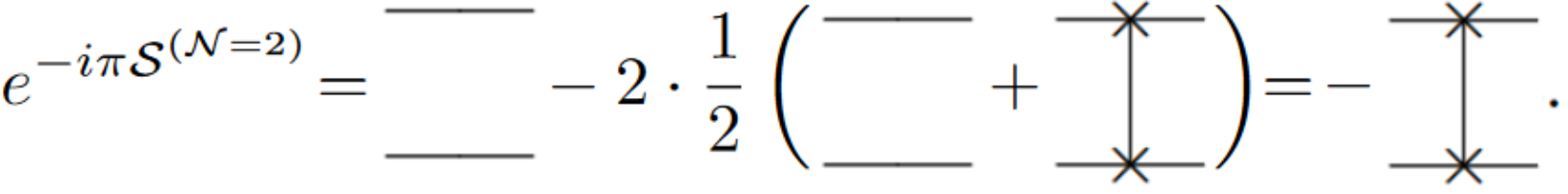}
\par}
\end{figure}

\noindent We arrive at Eq. (\ref{eq_SWAP_S}), derived earlier via matrix exponentiation in Section \ref{Section5}, effortlessly.

Let us now consider the next case, $\mathcal{N}=3$. We have
\begin{equation*}
    \begin{aligned}
    \mathcal{S}^{(\mathcal{N}=3)} = \frac{1}{3!} 
    \Bigg( & \begin{pmatrix} 
    1 & 2 & 3 \\
    1 & 2 & 3 \end{pmatrix} + 
    \begin{pmatrix} 
    1 & 2 & 3\\
    1 & 3 & 2 \end{pmatrix} + \begin{pmatrix} 
    1 & 2 & 3\\
    2 & 1 & 3 \end{pmatrix} + \\
    & \begin{pmatrix} 
    1 & 2 & 3\\
    2 & 3 & 1 \end{pmatrix} + \begin{pmatrix} 
    1 & 2 & 3\\
    3 & 2 & 1 \end{pmatrix} + 
    \begin{pmatrix} 
    1 & 2 & 3\\
    3 & 1 & 2 \end{pmatrix} \Bigg),
    \end{aligned}
\end{equation*}
or, in terms of quantum circuits,
\begin{figure}[h]
\centering{\
\includegraphics[width=\linewidth]{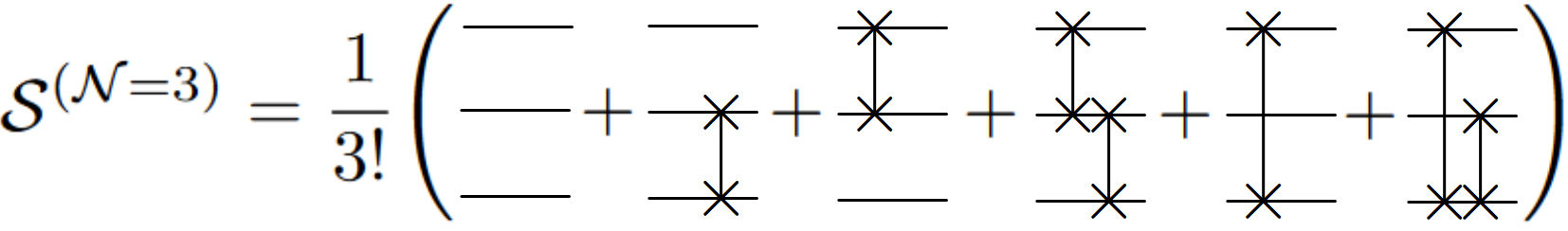}
\par}
\end{figure}

\noindent As in the $\mathcal{N}=2$ case, not only is each term in the linear combination unitary but the corresponding quantum circuit is simple. More importantly, a systematic, inductive way of finding these circuits for arbitrary $\mathcal{N}$ can be derived. We begin with the trivial case $\mathcal{N} = 1$, for which the symmetric group $S_1$ is the one-element group: 
\begin{figure}[h]
\centering{\
\includegraphics[width=0.8\linewidth]{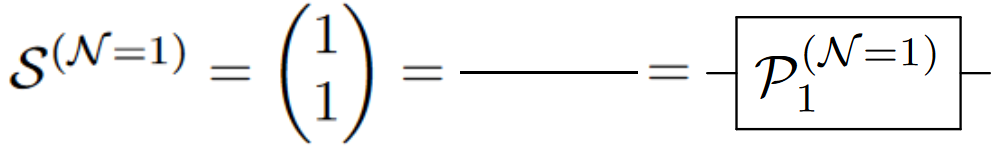}
\par}
\end{figure}

\noindent The expression for $\mathcal{S}^{(\mathcal{N}=2)}$ can be derived as follows:
\begin{figure}[h]
\centering{\
\includegraphics[width=0.8\linewidth]{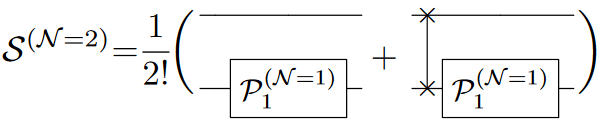}
\par}
\end{figure}

\noindent Similarly, we can obtain $\mathcal{S}^{(\mathcal{N}=3)}$ as
\begin{figure}[h!]
\centering{\
\includegraphics[width=\linewidth]{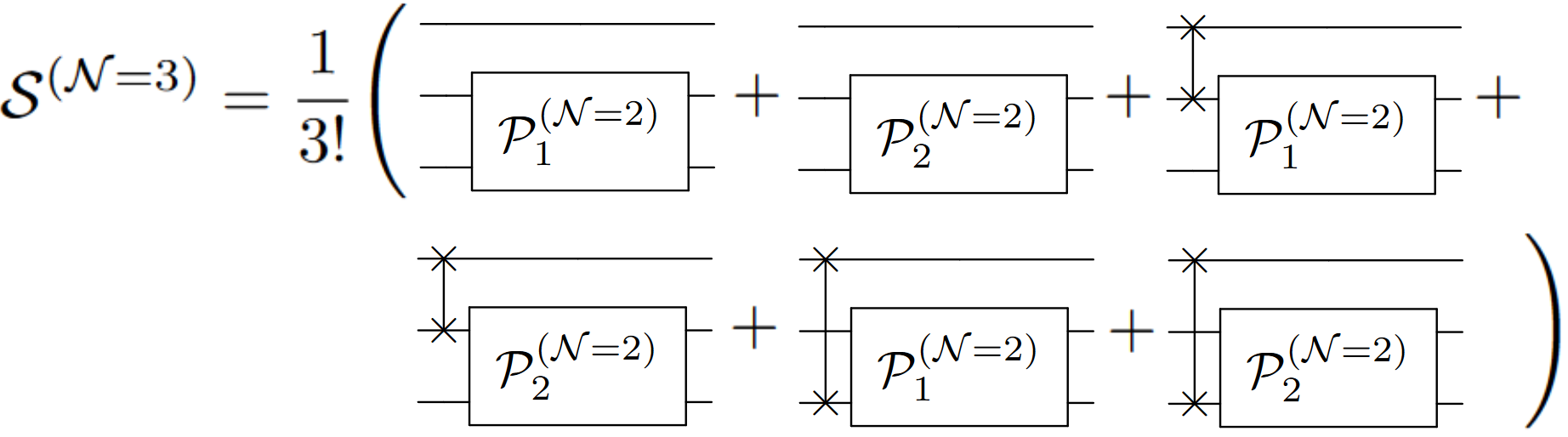}
\par}
\end{figure}

\noindent The general induction relation to derive the linear expansion of $\mathcal{S}^{(\mathcal{N})}$ in terms of the $\mathcal{N}!$ circuits for all permutations of $\mathcal{N}$ spins-$\frac{1}{2}$ is
\begin{figure}[h!]
\centering{\
\includegraphics[width=0.8\linewidth]{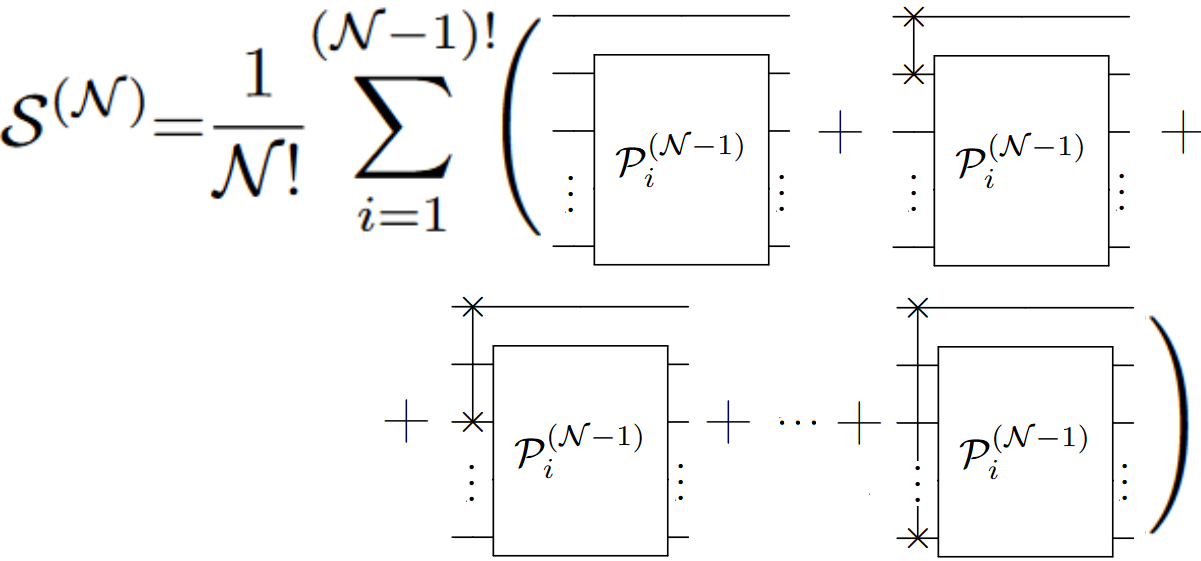}
\par}
\end{figure}

This iterative method generates the circuits with the lowest number of $\textsc{swap}$ gates for each permutation operation $\mathcal{P}^{(\mathcal{N})}_i$, but it does ignore qubit connectivity constraints. If one is restricted to linear qubit connectivity, the ``Amida lottery'' method devised by Seki et al. \cite{Seki20} could be considered instead.

\begin{figure}[h]
\centering{\
\includegraphics[width=0.75\linewidth]{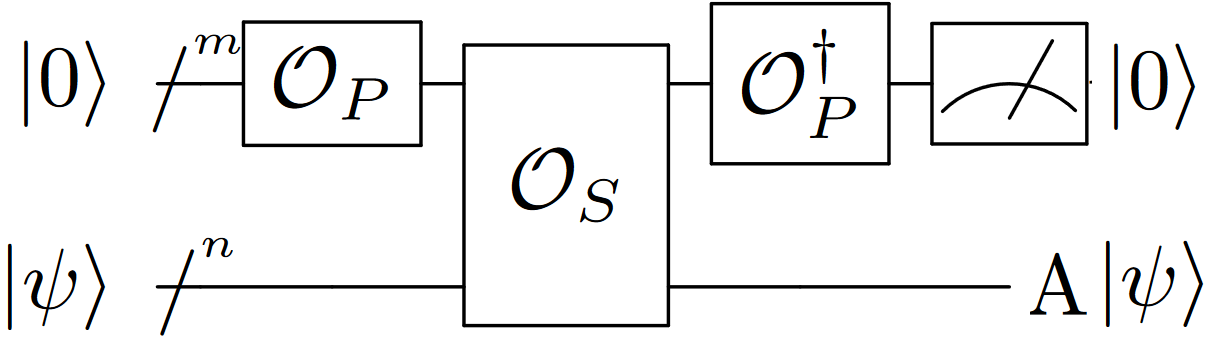}
\par}
\caption{High-level scheme of quantum circuit corresponding to Linear Combination of Unitaries (LCU) method \cite{Childs12} to prepare $n$-qubit state $A\ket{\psi} = \sum_{i = 0}^{m-1} |\alpha_i| e^{i \theta_i} V_i \ket{\psi}$, given circuits to prepare $\ket{\psi}$ and implement each $V_i$. The Prepare oracle $\mathcal{O}_P$ turns the fiducial state $\ket{0}^{\otimes m}$ into $\ket{\alpha} \coloneqq \frac{1}{\sqrt{s}} \sum_{i = 0}^{m-1} \sqrt{|\alpha_i|} \ket{2^i}$, where $s \coloneqq \sum_{i = 0}^{m-1} |\alpha_i|$. The Select oracle $\mathcal{O}_S$ corresponds to $\sum_{i = 0}^{m-1} \ket{2^i}\bra{2^i} \otimes e^{i \theta_i} V_i$. Measuring all $m$ ancillas in $\ket{0}$ ensures main $n$-qubit register is found in $A \ket{\psi}$.}
\label{LCU_fig}
\end{figure}

Having expressed the local symmetrization operator $\mathcal{S}^{(\mathcal{N})}$ as a sum of simple quantum circuits, we shall make use of the Linear Combination of Unitaries (LCU) method \cite{Childs12}, originally developed by Childs and Wiebe within the context of the Hamiltonian simulation problem. Given some operator $A = \sum_{i = 0}^{m-1} \alpha_i V_i$, where $A$ may or may not be unitary but each of the $m$ $V_i$ must be a $n$-qubit unitary for which we can find the corresponding quantum circuit, the LCU method applies $A$ to some input $n$-qubit state $\ket{\psi}$ by following the steps below:
\begin{enumerate}
    \item{Set up a $m$-qubit ancillary register initialized in the fiducial state $\ket{0}^{\otimes m}$. Initialize the state $\ket{\psi}$ in the main $n$-qubit register.}
    \item{Apply the Prepare oracle, $\mathcal{O}_{P}$, to the $m$-qubit ancillary register, thus preparing the state
    \begin{equation}
        \ket{\alpha} \coloneqq \frac{1}{\sqrt{s}} \sum_{i = 0}^{m-1} \sqrt{|\alpha_i|} \ket{2^i},
        \label{eq_alpha_state}
    \end{equation}
    with $s \coloneqq \sum_{i = 0}^{m-1} |\alpha_i|$ \footnote{For $\alpha_i \equiv |\alpha_i|e^{i \theta_i}$ with $\theta_i \neq 0$ (i.e., a negative real number or a complex number), the nontrivial phase factor is absorbed in the definition of the corresponding unitary $V_i$, thus carrying into the definition of the Select oracle $\mathcal{O}_S$ (cf. step 3).}.
    }
    \item{Apply the Select oracle, $\mathcal{O}_{S}$, defined as
    \begin{equation}
        \mathcal{O}_{S} \coloneqq \sum_{i = 0}^{m-1} \ket{2^i}\bra{2^i} \otimes e^{i \theta_i} V_i,
    \end{equation}
    where the phase factors arise from $\alpha_i = |\alpha_i| e^{i \theta_i}$.
    }
    \item{Apply the inverse of the Prepare oracle, $\mathcal{O}^{\dagger}_P$, to the $m$-qubit ancillary register.}
    \item{Measure the $m$ ancillary qubits in the computational basis. If all $m$ ancillas are measured in $\ket{0}$, the main register is found in the (normalized) state $A\ket{\psi}$. The probability of success is given by
    \begin{equation}
        P_{00...0} = \frac{1}{s^2} |\braket{\psi|A^{\dagger} A | \psi}|^2,
        \label{eq_prob_LCU}
    \end{equation}
    which reduces to $\frac{1}{s^2}$ if $A$ is unitary.
    }
\end{enumerate}
Fig. \ref{LCU_fig} shows a high-level scheme of the quantum circuit corresponding to the LCU method.

Applying the LCU method to the symmetrization operator $\mathcal{S}^{(\mathcal{N})}$ --- i.e., setting $A = \mathcal{S}^{(\mathcal{N})} = \sum_{i=1}^{\mathcal{N}!} \frac{1}{\mathcal{N}!} \mathcal{P}^{(\mathcal{N})}_i$, so that $\alpha_i = \frac{1}{\mathcal{N}!}, \forall i$ and $V_i = \mathcal{P}^{(\mathcal{N})}_i$ --- demands that the implementation of the corresponding Prepare and Select oracles $\mathcal{O}_P$ and $\mathcal{O}_S$ is considered. Regarding $\mathcal{O}_S$, its construction is simple: for every permutation circuit $\mathcal{P}^{(\mathcal{N})}_i$ derived above, all $\textsc{swap}$ gates are controlled by the i\textsuperscript{th} ancillary qubit, with $i = 1, 2, ..., m \equiv \mathcal{N}!$. The $\textsc{cswap}$ gate was previously discussed in Section \ref{Section5} and its basis gate decomposition can be found in \cite{cruz2022optimized}.

As for $\mathcal{O}_P$, since the symmetrization operator is given as a uniform linear combination of all $\mathcal{N}!$ permutation operators, the state $\ket{\alpha}$ (cf. Eq. (\ref{eq_alpha_state})) that $\mathcal{O}_P$ prepares starting from the fiducial state $\ket{0}^{\otimes \mathcal{N}!}$ corresponds to the $\mathcal{N}!$-qubit Dicke state of Hamming weight $1$ $\ket{D^{\mathcal{N}!}_{1}}$ \cite{Baerstschi19,Mukherjee20,Aktar21} (also referred to in the literature as $\ket{W_{\mathcal{N}!}}$ state \cite{Cruz19})
\begin{equation}
    \begin{aligned}
    & \quad \; \ket{W_{\mathcal{N}!}} \equiv \ket{D^{\mathcal{N}!}_1} \coloneqq \frac{1}{\sqrt{\mathcal{N}!}} \sum_{i = 0}^{{\mathcal{N}!}-1} \ket{2^i} = \\
    & = \frac{1}{\sqrt{\mathcal{N}!}} \big( \ket{10...0} + \ket{01...0} + ... + \ket{00...1} \big).
    \end{aligned}
\end{equation}
As detailed in Appendix \ref{AppG}, which follows \cite{Cruz19}, in general, the $\ket{W_m}$ state can be prepared in $\mathcal{O}(\log m)$ depth ignoring qubit connectivity constraints, and in $\mathcal{O}(m)$ depth with only nearest-neighbor couplings.

In summary, the second step of the preparation of a spin-$S = \frac{\mathcal{N}}{2}$ VBS state, corresponding to the symmetrization of $\mathcal{N}$ spins-$\frac{1}{2}$ at every lattice site of coordination number $\mathcal{N}$, can be accomplished by applying the LCU method with $A = \mathcal{S}^{(\mathcal{N})}$ instead of the local Hadamard test introduced in Section \ref{Section4}. The probability of success (and hence the average number of repetitions) of both methods is the same, since $s = \sum_{i=0}^{\mathcal{N}!-1} |\frac{1}{\mathcal{N}!}| = 1$, in which case Eq. (\ref{eq_prob_LCU}) coincides with Eq. (\ref{eq_prob_Hadamard}). The LCU-based approach is likely to yield shallower circuits for high local spins $S$, as it forgoes the basis gate decomposition of the $(2S+1)$-qubit controlled-$e^{-i \pi \mathcal{S}^{(\mathcal{N})}}$ required to implement the local Hadamard test.

The advantage of the LCU-based scheme may, in fact, be already attained for a local spin as low as $S = 2$, corresponding to a coordination number $\mathcal{N}=4$, as in the square lattice. The spin-$2$ VBS state is known to be a resource state for universal quantum computation \cite{Wei15}, so its implementation on quantum hardware, though more demanding than that of the spin-$\frac{3}{2}$ case, may be just as relevant. Ignoring qubit connectivity constraints, the Prepare oracle $\mathcal{O}_P$, which initializes the $\ket{W_{4!}}$ state, takes $46$ $\textsc{cnot}$ gates (cf. Appendix \ref{AppG} for details). As for the Select oracle $\mathcal{O}_S$, there are $4!-1 = 23$ terms (ignoring the identity), $6$ with just $1$ $\textsc{cswap}$ gate, $11$ with $2$ $\textsc{cswap}$ gates and the remaining $6$ with $3$ $\textsc{cswap}$ gates. Hence, $\mathcal{O}_S$ takes $6 \times 1 + 11 \times 2 + 6 \times 3 = 46$ $\textsc{cswap}$ gates, which amounts to $46 \times 7 = 322$ $\textsc{cnot}$ gates. Noting that the Prepare oracle must be reversed, the total number of $\textsc{cnot}$ gates is thus $2 \times 46 + 322 = 414$. This number is below the $469$ $\textsc{cnot}$ gates yielded by Qiskit \texttt{transpile} \cite{QiskitTranspile} for the decomposition of the $5$-qubit controlled-$e^{-i \pi \mathcal{S}^{(4)}}$. The $414$ $\textsc{cnot}$ gates required by the LCU method are also below the $444$ $\textsc{cnot}$ gates \cite{Krol21} that the optimized quantum Shannon decomposition \cite{Shende06} requires, in general, to decompose a $5$-qubit gate. 

The main limitation of the application of the LCU method to the implementation of $\mathcal{S}^{(\mathcal{N})}$ is arguably the potentially large number $\mathcal{N}!$ of ancillary qubits per site (e.g., $4! = 24$ for $S = 2$), which compares with a single ancilla for the local Hadamard test. Nevertheless, the LCU method introduced above corresponds to a \textit{sparse} version, for which $\mathcal{O}_P \ket{0}^{\otimes m} \sim \sum_{i = 0}^{m-1} \sqrt{|\alpha_i|} \ket{2^i}$ and $\mathcal{O}_S = \sum_{i = 0}^{m-1} \ket{2^i}\bra{2^i} \otimes e^{i \theta_i} V_i$. It is, however, possible to consider a \textit{dense} version that uses only $a = \lceil \log_2(m) \rceil$ ancillas, with $\mathcal{O}_P \ket{0}^{\otimes a} \sim \sum_{i = 0}^{m-1} \sqrt{|\alpha_i|} \ket{i}$ and $\mathcal{O}_S = \sum_{i = 0}^{m-1} \ket{i}\bra{i} \otimes e^{i \theta_i} V_i$. This leads to a more complex implementation of the Select and Prepare oracles, though. For the particular case of the symmetrization operator, each $\textsc{swap}$ gate in the Select oracle will be controlled by all $a$ ancillas instead of just one. Moreover, the initialization of $\sum_{i = 0}^{m-1} = \frac{1}{\sqrt{m}} \ket{i}$ is generally more difficult than that of $\ket{W_m}$. The one relevant exception corresponds to the case where $m = \mathcal{N}!$ is a power of $2$, in which case $\mathcal{O}_P$ is just the Walsh-Hadamard transform $H^{\otimes \log_2 m}$. This happens to be case for $\mathcal{N}=2$ (i.e., spin-$1$), for which the implementation of $\mathcal{S}^{(2)}$ via this dense LCU method is entirely equivalent to the local Hadamard test. In general, however, the dense LCU method gives rise to a deeper circuit than the sparse version, which is just a manifestation of the ubiquitous width-depth trade-off (cf. \cite{Zhang22} and references therein).

Finally, we note that East et al. \cite{East22} devised an implementation in the language of the ZXH-calculus of the symmetrization operator acting on $\mathcal{N}$ spins-$\frac{1}{2}$ to generate a single spin-$\frac{\mathcal{N}}{2}$ that is identical to the LCU-based method introduced in this section. East et al. \cite{East22} considered ZXH-diagrams instead of quantum circuits to represent the VBS states and compute their properties in a fully diagrammatic way.

\section{Discussion}\label{Section9}

\begin{figure*}
\includegraphics[width=\linewidth]{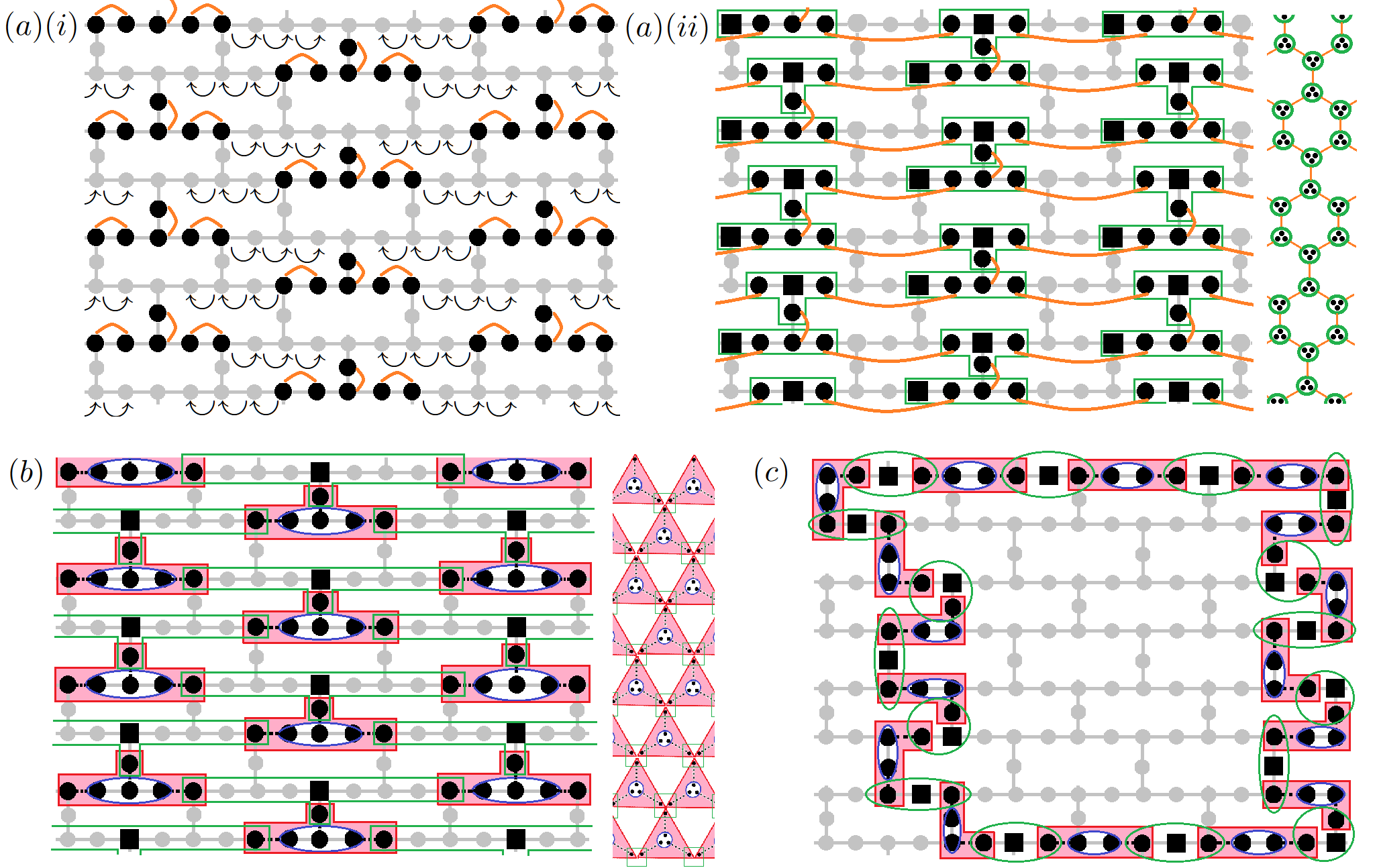}
\caption{Schemes of implementation of preparation of Valence-Bond-Solid (VBS) states on IBM Quantum heavy-hex lattice \cite{IBMQheavy_hex}. (a) Two-step description of preparation of spin-$\frac{3}{2}$ VBS state via bare probabilistic method. (i) In each set of six adjacent qubits shown as black dots, valence bonds are prepared at pairs of qubits linked by orange curved lines by applying the two-qubit circuit from Fig. \ref{Circ1}(a). Then, from each such six-qubit set, two qubits are moved past three redundant qubits (shown as gray dots), one towards the right and the other towards the left, as depicted by the curly arrows. This ensures that all three qubits that encode each site are nearby at the start of the second step. (ii) The local Hadamard test with $\mathcal{U} = e^{-i \pi \mathcal{S}}$ (cf. Fig. \ref{Circ1}(b)) is applied at the four-qubit sets enclosed by green boxes. Site ancillas, which were redundant qubits in (i), are now represented by black squares. T-shaped green boxes are associated with sites of one sublattice, and linear green boxes with those of the other. For the sake of clarity, a scheme of the segment of a honeycomb lattice corresponding to this circuit is shown on the right-hand side. (b) A single scheme describing the preparation of the spin-$\frac{3}{2}$ VBS state via the probabilistic method combined with the deterministic preparation of the six-qubit islands (cf. Section \ref{Section7}) to mitigate the repetition overhead. The main difference relative to (a) is the fact that the first step already symmetrizes the sites of one sublattice (as depicted by the blue loops), in addition to preparing all valence bonds (here represented as dashed black lines, as in Fig. \ref{fig_islands}). Hence, in second step only T-shaped green boxes require the application of the local Hadamard test. (c) Scheme of preparation of spin-$1$ VBS state for $32$-site ring via the probabilistic method preceded by the initialization of the four-qubit islands. As in (b), islands are illustrated with the color scheme adopted in Fig. \ref{fig_islands}.       
\label{fig_heavy_hex}}
\end{figure*}

In the previous sections we have tackled the generally challenging problem \cite{Deutsch95, Knill95, Shende06} of preparing quantum many-body states on quantum hardware, namely for the physically relevant class of Valence-Bond-Solid (VBS) states of given local spin $S = 1, \frac{3}{2}, 2, \frac{5}{2}, 3, ...$ (cf. Section \ref{Section2}). In particular, we have developed a probabilistic quantum scheme (cf. Section \ref{Section4}), inspired by the construction of the parent AKLT Hamiltonians \cite{AKLT87, AKLT88}, that gives rise to a circuit with depth independent of the lattice size $N$. In Section \ref{Section5} the detailed circuits to prepare the spin-$1$ and spin-$\frac{3}{2}$ VBS states were derived, achieving a depth of just $8$ and $27$ $\textsc{cnot}$ gates, respectively. Such low depths, valid for arbitrarily large lattices, render this method especially suitable for NISQ hardware.

However, thus far the issue of qubit connectivity constraints has been ignored in the determination of the circuit depth for the preparation of the VBS states. Although all-to-all connectivity has been achieved in trapped-ion quantum computers with as many as $20$ qubits \cite{Quantinuum20Q}, it remains unclear if such degree of connectivity can be maintained as the number of qubits increases \cite{Bruzewicz19}, despite ongoing efforts towards this end \cite{Wan20, Ramette22}. In superconducting-circuit-based quantum computers, restrictions in the connections between qubits are inevitable, with entangling gates between widely separated qubits being possible only through networks of $\textsc{swap}$ gates \cite{OGorman19}. Determining how the depth associated with preparing the spin-$1$ and spin-$\frac{3}{2}$ VBSs is affected by these qubit connectivity constraints is thus relevant to confirm its feasibility in near-term quantum hardware.

To this end, we will consider the heavy-hex lattice \cite{Chamberland20} adopted by IBM Quantum in their current and forthcoming devices \cite{IBMQheavy_hex0,IBMQheavy_hex,IBMQroadmap}. This architecture would be perfectly suited for the preparation of the spin-$\frac{3}{2}$ VBS state, were it not for the fact that there is only one qubit per lattice link instead of the two required to create a valence bond. A conceptually simple, yet technically challenging solution to this problem would be to replace every qubit along a link by a ququart (i.e., a four-dimensional qudit), in the spirit of previous proposals \cite{Ralph07, Lanyon09, Baekkegaard19} that exploit higher-dimensional local Hilbert spaces to simplify quantum circuits. A more feasible approach is to make use of only a fraction of the qubits available in the heavy-hex lattice, distributing them spatially in a way that suits the initialization of the spin-$\frac{3}{2}$ VBS. The resulting layout of the qubits across the heavy-hex lattice is illustrated in Figs. \ref{fig_heavy_hex}(a) and \ref{fig_heavy_hex}(b) for the cases where the bare probabilistic method and the combination of the preparation of the six-qubit islands and the probabilistic method at only one sublattice are employed, respectively. The circuit depth in $\textsc{cnot}$ gates for both cases is presented in Table \ref{TableA}; the details of the calculations can be found in Appendix \ref{AppH}. Even though, as expected, the depth does increase when qubit connectivity constraints are accounted for, the predicted depths of $51$ $\textsc{cnot}$ gates (for the bare probabilistic method) and $105$ $\textsc{cnot}$ gates (with mitigation of the repetition overhead) are still very likely to be within reach of noisy intermediate-scale quantum computers in the next few years.

\begin{table}[h]   
\caption{\label{TableA} Circuit depth in $\textsc{cnot}$ gates for preparation of spin-$1$ and spin-$\frac{3}{2}$ VBS states assuming different cases of qubit connectivity constraints, with and without mitigation of repetition overhead. Cf. Appendix \ref{AppH} for details of calculations.}
\begin{tabular}{| c | c | c |}
    \hline
    \thead{Local Spin, \\ Qubit Connectivity} & \thead{Probabilistic \\ Method at All \\ Lattice Sites} & \thead{4S-Qubit Islands +\\ Probabilistic Method \\ At One Sublattice} \\ \hline
    $S = 1$, all-to-all & $8$  & $11$  \\ \hline
    $S = 1$, linear & $10$  & $17$  \\ \hline
    $S = \frac{3}{2}$, all-to-all & $27$  & $45$  \\ \hline
    $S = \frac{3}{2}$, heavy-hex & $51$  & $105$ \\ \hline
\end{tabular}
\end{table}

\begin{table}[h]   
\caption{\label{TableB} Average number of repetitions required to successfully prepare spin-$1$ and spin-$\frac{3}{2}$ VBS states via probabilistic method introduced in Section \ref{Section4} with and without employing the repetition overhead mitigation strategies proposed in Sections \ref{Section6} and \ref{Section7}. Scaling of average number of repetitions for spin-$1$ and spin-$\frac{3}{2}$ is $(\frac{4}{3})^N$ and $2^N$ in unmitigated case, and $(\frac{4}{3})^{N/2}$ and $2^{N/2}$ if mitigation is applied.}
\begin{tabular}{| c | c | c | c | c | c |}
    \hline
    $N$ & 10 & 20 & 30 & 40 & 50 \\ \hline
    $S = 1$, unmitigated & $18$ & $315$ & $5,600$ & $99,000$ & $1.8 \times 10^6$ \\ \hline
    $S = 1$, mitigated & $4$ & $18$ & $75$ & $315$ & $1,300$  \\ \hline
    $S = \frac{3}{2}$, unmitigated & $1,000$ & $10^6$ & $10^{9}$ & $10^{12}$ & $10^{15}$ \\ \hline
    $S = \frac{3}{2}$, mitigated & $32$ & $1,000$ & $33,000$ & $10^6$ & $3.3 \times 10^7$ \\ \hline
\end{tabular}
\end{table}

For the spin-$1$ VBS state in one dimension, the restrictions related to the couplings between qubits are even less impactful: the circuit depth for the probabilistic method with and without $\mathcal{O}(1)$-depth mitigation of the repetition overhead increases from $10$ to $17$ $\textsc{cnot}$ gates and from $8$ to $11$ $\textsc{cnot}$ gates, respectively. For concreteness, Fig. \ref{fig_heavy_hex}(c) shows a scheme of the layout adopted in the IBM Quantum heavy-hex lattice \cite{IBMQheavy_hex} to prepare the spin-$1$ VBS state for a $32$-site ring. 

The downside of this low circuit depth is the probabilistic nature of the method, which means that the circuit must be repeated multiple times to successfully prepare the VBS state. In a sense, this reduction of the circuit depth by increasing the number of repetitions follows the spirit of hybrid variational algorithms \cite{Cerezo21}, which are naturally suited for NISQ hardware. Two strategies that achieve a quadratic reduction of this repetition overhead were devised by exploiting the entanglement structure of the initial state $\ket{\psi_{\textrm{pre-VBS}}}$. The latter (cf. Section \ref{Section7}) produces a $\mathcal{O}(1)$-depth overhead but requires the generic preparation of $4S$-qubit states, thus being more suitable for low spin-$S$. The former (cf. Section \ref{Section6}) results in a $\mathcal{O}(\log N)$-depth layer but only involves applying the controlled-$e^{-i \pi \mathcal{S}}$ used in the probabilistic method, for which a systematic scheme that avoids the basis gate decomposition of the corresponding $(2S + 1)$-qubit unitary operation was developed in Section \ref{Section8}, thus making it a more suitable option for high spin-$S$. It should be noted that the real-time processing of mid-circuit measurements, which is a key element of the former method, is an active field of research \cite{Riste20}, particularly due to its importance within the context of quantum error correction \cite{Lidar13}. In fact, a number of quantum routine proposals that make use of mid-circuit measurements have been put forth recently \cite{Botelho22,Yalovetzky21,Zhu21,Koh22}.

Table \ref{TableB} presents the average number of repetitions required to prepare the spin-$1$ and spin-$\frac{3}{2}$ VBS states with and without employing a mitigation overhead repetition strategy. Although the scaling of the average number of repetitions is exponential in the number of lattice sites $N$ in both cases, the mitigation of the repetition overhead can be decisive to make the probabilistic preparation of VBS states feasible for the intermediate values of $N$ at which quantum advantage can be achieved with near-term quantum computers.

An important application with a plausible prospect of achievable quantum advantage to which the preparation of the spin-$\frac{3}{2}$ VBS state on quantum hardware makes a relevant contribution is the simulation of a naturally-occurring gapped Hamiltonian such as the spin-$\frac{3}{2}$ AKLT model, or possibly a nearby but non-integrable model, to aid the experimental realization of a resource state for measurement-based quantum computation (MBQC) \cite{Raussendorf01, Raussendorf03, Briegel09}. This could be framed within the wider effort of using near-term quantum hardware to support the development of fault-tolerant quantum computers \cite{Iyer18,Kyaw21}.

The idea of employing the ground state of a naturally-occurring gapped Hamiltonian with an appropriate entanglement structure \cite{Gross09} as the resource state for MBQC is appealing for the flexible state preparation via cooling, on the one hand, and the stability against local perturbations, on the other \cite{Cai10}. Having prepared such a resource state, no entangling operations are required to perform the computations; single-qubit gates and local measurements suffice. Cluster states \cite{Briegel01} are the canonical example of a resource state for MBQC, but they cannot occur as the exact ground state of any naturally occurring physical system \cite{Nielsen06}. Given the proof that the spin-$\frac{3}{2}$ VBS state is a resource state for MBQC \cite{Wei11, Miyake11}, its experimental realization in solid-state platforms has been considered \cite{Miyake11, Koch-Janusz15,Mishra21} to develop a measurement-based quantum computer. 

Computing the expected properties of the ground state of such a naturally-occurring parent Hamiltonian, calculating its spectral gap, and determining the nature of its low-lying excitations could provide valuable assistance in the experimental realization of the desired robust resource state. However, computing these properties using conventional numerical methods is a formidable challenge, as demonstrated by the recent confirmation of the nonzero spectral gap of the spin-$\frac{3}{2}$ AKLT model \cite{Pomata20, Lemm20}. The semi-analytical proof developed by Lemm et al. \cite{Lemm20} involved a DMRG calculation on a $36$-site cluster, although the convergence to the excited states in all spin sectors would not have been possible without making use of the AKLT construction \cite{AKLT87,AKLT88}, which is only strictly possible at the integrable point. The challenge is certainly even more daunting away from the integrable point, where experimental systems are likely to be found \cite{Mishra21}. Ganesh et al. \cite{Ganesh11} investigated the phase diagram of a spin-$\frac{3}{2}$ model that interpolates between the Heisenberg and AKLT models using exact diagonalization on clusters with up to $N = 18$ spins-$\frac{3}{2}$.

Digital quantum simulation methods, in turn, may allow for a more viable calculation of these quantities. In the NISQ era, a hybrid variational algorithm such as the Variational Quantum Eigensolver (VQE) \cite{Peruzzo14} may be used to compute the ground state and low-lying excited states \cite{McClean17, Jones19, Higgott19, Nakanishi19} of such non-integrable models close to the spin-$\frac{3}{2}$ AKLT model. In this paper we have addressed one of the key challenges involved in this VQE simulation: The preparation of an initial state that overlaps significantly with the target ground state of the nearby non-integrable model. Such educated guess is essential to simplify the optimization process by reducing the number of layers of the ansatz applied on top of it and by avoiding barren plateaus \cite{McClean18, Holmes22}. An additional challenge that remains unaddressed in the literature is how to adapt VQE to the simulation of spin-$\frac{3}{2}$ (or, more generally, higher-spin) models, as all applications of VQE to the study of quantum magnetism have been restricted to local spin-$\frac{1}{2}$ degrees of freedom \cite{Seki20, Bespalova21, Liu19, Mazzola19, Li21, You21, Huerga22, Jattana22, Bosse22, Feulner22}. Improvements in the quantum hardware are also necessary, although the consistent rise in the quantum volume achieved by the leading manufacturers \cite{QVolume_IBM, QVolume_Quantinuum} and a recent simulation involving a two-qubit gate depth of up to 159 on 20 qubits \cite{Niroula22} suggest the required hardware may already be at our disposal in the next few years.

An alternative application where quantum advantage can be attained potentially sooner is the simulation of the quench dynamics \cite{Mitra18} of the spin-$1$ VBS state in one dimension. Even though DMRG \cite{White92, Rommer97, Schollwock05, Schollwock11} and time-dependent variants thereof (e.g., tDMRG \cite{White05, Daley04} and tangent-space methods \cite{Haegeman13,Haegeman16}) have been extremely successful at simulating static properties and short-time dynamics of one-dimensional systems (cf. \cite{Schollwock05, Schollwock11, Orus19} and references therein), for a sufficiently long duration of the time evolution the truncation error due to the bounded bond dimension of the underlying MPS becomes exceedingly large, signalling the ``runaway'' time at which the numerical simulation ceases to be reliable \cite{Gobert05}. Digital quantum computation offers an exponential speed-up in the simulation of quantum dynamics \cite{Lloyd96}, so it could become the leading method to validate experiments performed with cold atoms in optical lattices \cite{Lewenstein12}. Schemes to realize the spin-$1$ VBS state with cold atoms have been put forth \cite{Sharma21, Zaletel21}. By making use of the probabilistic method herein proposed to prepare the spin-$1$ VBS state on quantum hardware instead of the deterministic MPS-based method discussed in Section \ref{Section3}, considerable savings of the circuit depth devoted to the state initialization can be achieved, leaving more circuit depth available to implement the time evolution.

\section{Conclusions}\label{Section10}

In summary, we have proposed a method to prepare Valence-Bond-Solid (VBS) states of arbitrary local spin $S$ on quantum hardware. Inspired by the construction of the parent AKLT Hamiltonians, this method consists of initializing a product state of valence bonds, after which the local symmetrization operator is applied at all sites in parallel. As a result, the depth of the resulting quantum circuit is independent of the lattice size $N$, at the cost of requiring multiple independent repetitions to achieve success. Two schemes to reduce the average number of repetitions were developed by exploiting the entanglement structure of the initial state, one introducing a constant depth overhead but requiring the initialization of $4S$-qubit states and another leading to a $\mathcal{O}(\log N)$-depth overhead but involving the same building block as the main probabilistic method. The former is therefore appropriate for low $S$, while the latter is the method of choice for larger $S$. An alternative approach to implement the local symmetrization operator was also devised, bypassing the basis gate decomposition of the building block of the probabilistic method, which can be too onerous for high $S$.

These general methods were then applied to the particular cases of $S = 1, \frac{3}{2}$. Shallow circuits to implement the probabilistic method were devised using state-of-the-art basis gate decomposition methods, yielding circuits with depth $8$ and $27$ $\textsc{cnot}$ gates for the preparation of the spin-$1$ and spin-$\frac{3}{2}$ VBS states, respectively. Two applications with prospective quantum advantage were identified, one for each case. The impact of qubit connectivity constraints and the size of the repetition overhead required to prepare these Valence Bond States at the intermediate lattice sizes $N$ for which such quantum advantage is foreseeable were considered.

More broadly, in comparison with the tensor-network-based methods to prepare VBS states on quantum hardware, the novel quantum routine herein proposed trades circuit depth for the repetition of the same shallow circuit, which follows the spirit of hybrid variational algorithms. Extending this depth-repetitions trade-off to the preparation of other classes of quantum many-body states could be a prolific strategy. 

\vspace{0.5cm}

{\em Acknowledgements.} B.M. acknowledges financial support from the FCT PhD scholarship No. SFRH/BD/08444/2020. P.M.Q.C. acknowledges financial support from FCT Grant No. SFRH/BD/150708/2020. J.F.R. acknowledges financial support from the Ministry of Science and Innovation of Spain (grant No. PID2019-109539GB-41), from Generalitat Valenciana (grant No. Prometeo2021/017), and from FCT (grant No. PTDC/FIS-MAC/2045/2021).

\appendix

\section{Derivation of AKLT Hamiltonians}\label{AppA}

The AKLT Hamiltonian is a sum of local projectors acting on pairs of neighboring spins $\vec{S}_n$ and $\vec{S}_{n'}$ that restrict the total spin quantum number to its maximum value $S$. We therefore need to find an expression for
\begin{equation}
    P_{n,n'}^{S} = \mathcal{N} \prod_{S' \neq S} \big((\vec{S}_n + \vec{S}_{n'})^2 - S'(S'+1)\big),
\end{equation}
where $\mathcal{N}$ is a constant that ensures $P_{n,n+1}^{S}$ acts trivially on states with $S_{\textrm{total}}^{n,n+1} = S$, where $\vec{S}_{\textrm{total}}^{n,n+1} \equiv \vec{S}_n + \vec{S}_{n+1}$. 

\subsection{Spin-$1$ AKLT Model}

For the spin-$1$ AKLT model, $S = 2$, therefore $P_{n,n+1}^{S=2}$ projects out any states with $S_{\textrm{total}}^{n,n+1} = 0, 1$. The normalization constant $\mathcal{N}$ satisfies
\begin{equation}
    \mathcal{N} \big(2(2+1) - 0(0+1)\big) \big(2(2+1) - 1(1+1)\big) = 1,
\end{equation}
so $\mathcal{N} = \frac{1}{24}$. Using $(\vec{S}_n + \vec{S}_{n+1})^2 = 4 + 2\vec{S}_n \cdot \vec{S}_{n+1}$,
\begin{equation}
\begin{aligned}
    P_{n,n+1}^{S=2} & = \frac{1}{24} (4 + 2 \vec{S}_n \cdot \vec{S}_{n+1} - 0) (4 + 2 \vec{S}_n \cdot \vec{S}_{n+1} - 2) \\
    & = \frac{1}{3} + \frac{1}{2} \vec{S}_n \cdot \vec{S}_{n+1} + \frac{1}{6} (\vec{S}_n \cdot \vec{S}_{n+1})^2.
\end{aligned}
\label{eq_Proj_S_1}
\end{equation}
Comparing Eq. (\ref{eq_Proj_S_1}) to Eq. (\ref{eq_BLBQ_Ham}) gives Eq. (\ref{eq_AKLT_S_1}). 

\subsection{Spin-$\frac{3}{2}$ AKLT Model}

For the spin-$\frac{3}{2}$ AKLT model, $S = 3$ and $P_{n,n'}^{S=3}$ projects out states with $S_{\textrm{total}}^{n,n'} = 0, 1, 2$. $\mathcal{N}$ is given by
\begin{equation}
    \mathcal{N} \big(3(3+1) - 0\big) \big(3(3+1) - 2\big) \big(3(3+1) - 6\big) = 1,
\end{equation}
so $\mathcal{N} = \frac{1}{720}$. Using $(\vec{S}_n + \vec{S}_{n'})^2 = \frac{15}{2} + 2\vec{S}_n \cdot \vec{S}_{n'}$,
\begin{equation}
    P_{n,n'}^{S=3} = \frac{11}{192} + \frac{27}{160} \vec{S}_n \cdot \vec{S}_{n'} + \frac{29}{360} (\vec{S}_n \cdot \vec{S}_{n'})^2 + \frac{1}{90} (\vec{S}_n \cdot \vec{S}_{n'})^3.
\label{eq_Proj_S_3_2}
\end{equation}
Ignoring the additive constant and setting prefactor of the bilinear term to unity in Eq. (\ref{eq_Proj_S_3_2}) yields Eq. (\ref{eq_AKLT_S_3_2}). 

\section{Matrix Representation of Symmetrization $\mathcal{S}$}\label{AppB}

Although the matrix representation of the local symmetrization operator $\mathcal{S}$ can be obtained via its general definition (cf. Eq. (\ref{eq_Symm})), a leaner approach similar to the derivation of the expression for $P_{n,n'}^S$ in Appendix A can be adopted instead. Concretely, at site $n \in \Lambda$,
\begin{equation}
    \mathcal{S}_n = \mathcal{N} \prod_{S' \neq S} \big( (\vec{S}_{\textrm{total}}^n)^2 - S'(S'+1) \big),
\end{equation}
where, in this case, $\vec{S}_{\textrm{total}}^{n} = \sum_{i = 1}^{\mathcal{N}_n} \vec{S}_{i}$ is the total spin resulting from the sum of all $\mathcal{N}_n$ spins-$\frac{1}{2}$ and $S = \frac{\mathcal{N}_n}{2}$ is the spin that we wish to associate with the local degree of freedom. $\mathcal{N}$ is just a normalization constant that ensures that spin-$S$ states are acted on trivially by $\mathcal{S}_n$.

\subsection{Spin-$1$}

For the spin-$1$ case, we have two spins-$\frac{1}{2}$ per site, in which case the matrix representations of the three Cartesian components of $\vec{S}_{\textrm{total}}$ are of the form $S_{\textrm{total}}^i = S^i \otimes \mathbb{1} + \mathbb{1} \otimes S^i$, with
\begin{equation*}
    S^x = \frac{1}{2} \begin{pmatrix}
     0 & 1 \\
     1 & 0
    \end{pmatrix}, \;
    S^y = \frac{1}{2} \begin{pmatrix}
     0 & -i \\
     i & 0
    \end{pmatrix}, \;
    S^z = \frac{1}{2} \begin{pmatrix}
     1 & 0 \\
     0 & -1
    \end{pmatrix}.
\end{equation*}
The matrix representation of $(\vec{S}_{\textrm{total}})^2 = (S_{\textrm{total}}^x)^2 + (S_{\textrm{total}}^y)^2 + (S_{\textrm{total}}^z)^2$ is then
\begin{equation}
    (\vec{S}_{\textrm{total}})^2 = \begin{pmatrix}
     2 & 0 & 0 & 0 \\
     0 & 1 & 1 & 0 \\
     0 & 1 & 1 & 0 \\
     0 & 0 & 0 & 2
    \end{pmatrix}.
\label{eq:sym_S=1}
\end{equation}
The symmetrization operator removes $S_{\textrm{total}} = 0$ states, so $\mathcal{S} = \mathcal{N} (\vec{S}_{\textrm{total}}^2 - 0(0+1) \mathbb{1})$. Setting $\mathcal{N} = \frac{1}{2}$ gives
\begin{equation}
    \mathcal{S} = \begin{pmatrix}
     1 & 0 & 0 & 0 \\
     0 & \frac{1}{2} & \frac{1}{2} & 0 \\
     0 & \frac{1}{2} & \frac{1}{2} & 0 \\
     0 & 0 & 0 & 1
    \end{pmatrix}.
\label{eqdefS_1}
\end{equation}

\subsection{Spin-$\frac{3}{2}$}

For the spin-$\frac{3}{2}$ case, there are three spins-$\frac{1}{2}$ per site, so the matrix representations of the three Cartesian components of $\vec{S}_{\textrm{total}}$ are of the form $S_{\textrm{total}}^i = S^i \otimes \mathbb{1} \otimes \mathbb{1} + \mathbb{1} \otimes S^i \otimes \mathbb{1} + \mathbb{1} \otimes \mathbb{1} \otimes S^i $, with
\begin{equation*}
    \scriptstyle{
    S^x = \frac{1}{\sqrt{2}} \begin{pmatrix}
     0 & 1 & 0 \\
     1 & 0 & 1 \\
     0 & 1 & 0
    \end{pmatrix}, \;
    S^y = \frac{1}{\sqrt{2}i} \begin{pmatrix}
     0 & 1 & 0 \\
     -1 & 0 & 1 \\
     0 & -1 & 0
    \end{pmatrix}, \;
    S^z = \begin{pmatrix}
     1 & 0 & 0 \\
     0 & 0 & 0 \\
     0 & 0 & -1
    \end{pmatrix}.
    }
\end{equation*}
The matrix representation of $(\vec{S}_{\textrm{total}})^2 = (S_{\textrm{total}}^x)^2 + (S_{\textrm{total}}^y)^2 + (S_{\textrm{total}}^z)^2$ is therefore
\begin{equation}
    (\vec{S}_{\textrm{total}})^2 = \frac{1}{4}\begin{pmatrix}
     15 & 0 & 0 & 0 & 0 & 0 & 0 & 0 \\
     0 & 7 & 4 & 0 & 4 & 0 & 0 & 0 \\
     0 & 4 & 7 & 0 & 4 & 0 & 0 & 0 \\
     0 & 0 & 0 & 7 & 0 & 4 & 4 & 0 \\
     0 & 4 & 4 & 0 & 7 & 0 & 0 & 0 \\
     0 & 0 & 0 & 4 & 0 & 7 & 4 & 0 \\
     0 & 0 & 0 & 4 & 0 & 4 & 7 & 0 \\
     0 & 0 & 0 & 0 & 0 & 0 & 0 & 15 \\
    \end{pmatrix}.
\end{equation}

Symmetrizing removes $S_{\textrm{total}} = \frac{1}{2}$ states, so $\mathcal{S} = \mathcal{N} (\vec{S}_{\textrm{total}}^2 - \frac{1}{2}(\frac{1}{2}+1) \mathbb{1})$. Setting $\mathcal{N} = \frac{1}{3}$ gives
\begin{equation}
    \mathcal{S} = \begin{pmatrix}
     1 & 0 & 0 & 0 & 0 & 0 & 0 & 0 \\
     0 & \frac{1}{3} & \frac{1}{3} & 0 & \frac{1}{3} & 0 & 0 & 0 \\
     0 & \frac{1}{3} & \frac{1}{3} & 0 & \frac{1}{3} & 0 & 0 & 0 \\
     0 & 0 & 0 & \frac{1}{3} & 0 & \frac{1}{3} & \frac{1}{3} & 0 \\
     0 & \frac{1}{3} & \frac{1}{3} & 0 & \frac{1}{3} & 0 & 0 & 0 \\
     0 & 0 & 0 & \frac{1}{3} & 0 & \frac{1}{3} & \frac{1}{3} & 0 \\
     0 & 0 & 0 & \frac{1}{3} & 0 & \frac{1}{3} & \frac{1}{3} & 0 \\
     0 & 0 & 0 & 0 & 0 & 0 & 0 & 1 \\
    \end{pmatrix}.
\label{eqdefS3}
\end{equation}

\section{Norm Reduction due to Local Symmetrization $\mathcal{S}$ and Overlap $\braket{\psi_{\textrm{VBS}} | \psi_{\textrm{pre-VBS}}}$}\label{AppC}

First, we show that $\braket{\psi_{\textrm{VBS}} | \psi_{\textrm{pre-VBS}}} = \braket{\psi_{\textrm{VBS}} | \psi_{\textrm{VBS}}}$, where $\ket{\psi_{\textrm{VBS}}}$ is the unnormalized VBS state. Noting that the symmetrization operator $\mathcal{S}$ is Hermitian ($\mathcal{S} = \mathcal{S}^{\dagger}$) and idempotent ($\mathcal{S}^{2} = \mathcal{S}$), it follows that
\begin{equation}
\begin{aligned}
     \braket{\psi_{\textrm{VBS}} | \psi_{\textrm{VBS}}} & = \Big( \braket{\psi_{\textrm{pre-VBS}}
     | \otimes_{n} \mathcal{S}_n^{\dagger} \Big) \Big( \otimes_{n} \mathcal{S}_n | \psi_{\textrm{pre-VBS}}} \Big) \\
     & = \braket{\psi_{\textrm{pre-VBS}} | \Big( \otimes_{n} \mathcal{S}_n^2 \Big) | \psi_{\textrm{pre-VBS}}} \\
     & = \Big( \braket{\psi_{\textrm{pre-VBS}} | \otimes_{n} \mathcal{S}_n^{\dagger} \Big) | \psi_{\textrm{pre-VBS}}} \\
     & = \braket{\psi_{\textrm{VBS}} | \psi_{\textrm{pre-VBS}}}.
\end{aligned}
\label{eqcomp}
\end{equation}
In terms of the normalized VBS state, $\ket{\tilde{\psi}_{\textrm{VBS}}} \equiv \frac{\ket{\psi_{\textrm{VBS}}}}{\sqrt{\braket{\psi_{\textrm{VBS}} | \psi_{\textrm{VBS}}}}}$, Eq. (\ref{eqcomp}) can be expressed as
\begin{equation}
    \braket{\tilde{\psi}_{\textrm{VBS}} | \psi_{\textrm{pre-VBS}}} = \sqrt{\braket{\psi_{\textrm{VBS}} | \psi_{\textrm{VBS}}}}.
\end{equation}

Now, to arrive at Eq. (\ref{eq_overlap_main}) in the main text, we need to show that, at each application of the local symmetrization operator $\mathcal{S}$, the norm of $\ket{\psi_{\textrm{pre-VBS}}}$ is reduced by a constant factor $p$, ultimately leading to an exponentially vanishing overlap in the number of sites $N$. We will derive this result explicitly for the spin-$1$ and spin-$\frac{3}{2}$ VBS first, and then we will present the general case.

\subsection{Spin-$1$ Valence-Bond-Solid State}

The matrix representation of $\mathcal{S}$ for the spin-$1$ case is given by (\ref{eqdefS_1}). The eigenstates of $\mathcal{S}$ are
\begin{equation}
\begin{aligned}
    & \ket{1_1} = \ket{\uparrow \uparrow}, \\
    & \ket{1_0} = \frac{1}{\sqrt{2}}(\ket{\uparrow \downarrow} + \ket{\downarrow \uparrow}), \\
    & \ket{1_{-1}} = \ket{\downarrow \downarrow}, \\
    & \ket{0_0} = \frac{1}{\sqrt{2}}(\ket{\uparrow \downarrow} - \ket{\downarrow \uparrow}),
\end{aligned}
\end{equation}
where the respective eigenvalue corresponds to $s$ in $\ket{s_{m_s}}$. Focusing only on the valence bonds involving site $n$, $\ket{\psi_{\textrm{pre-VBS}}}$ can be written as
\begin{equation}
\begin{aligned}
    & \ket{\psi_{\textrm{pre-VBS}}} = \frac{1}{\sqrt{2}} \big( \ket{\uparrow}_{n-1,R} \ket{\downarrow}_{n,L} - \ket{\downarrow}_{n-1,R} \ket{\uparrow}_{n,L} \big) \; \otimes \\ & \otimes \frac{1}{\sqrt{2}} \big( \ket{\uparrow}_{n,R} \ket{\downarrow}_{n+1,L} - \ket{\downarrow}_{n,R} \ket{\uparrow}_{n+1,L} \big) \otimes ...,
\end{aligned}
\end{equation}
where the ellipsis includes all remaining valence bonds, which do not involve any of the two spins-$\frac{1}{2}$ from site $n$. Expanding the tensor product of the two highlighted valence bonds in the eigenbasis of $\mathcal{S}$ at site $n$ gives
\begin{equation}
    \begin{aligned}
    & \ket{\psi_{\textrm{pre-VBS}}} = \frac{1}{2}\Big[- \ket{1_1}_n \otimes (\ket{\downarrow}_{n-1,R} \ket{\downarrow}_{n+1,L}) \; - \\
    & - \ket{1_{-1}}_n \otimes (\ket{\uparrow}_{n-1,R} \ket{\uparrow}_{n+1,L}) \; + \\
    & + \ket{1_0}_n \otimes \frac{1}{\sqrt{2}}(\ket{\uparrow}_{n-1,R} \ket{\downarrow}_{n+1,L} + \ket{\downarrow}_{n-1,R} \ket{\uparrow}_{n+1,L}) \; + \\
    & + \ket{0_0}_n \otimes \frac{1}{\sqrt{2}}(\ket{\uparrow}_{n-1,R} \ket{\downarrow}_{n+1,L} + \ket{\downarrow}_{n-1,R} \ket{\uparrow}_{n+1,L}) \Big] \otimes ...
\end{aligned}
\label{eqexp}
\end{equation}
Upon application of $\mathcal{S}$ at site $n$, the first three terms on the right-hand side of Eq. (\ref{eqexp}) remain unchanged (since $\ket{1_{m_s}}, m_s \in \{1, 0, -1\}$, are eigenstates of $\mathcal{S}$ with eigenvalue $1$), while the fourth term vanishes (since $\ket{0_0}$ is an eigenstate of $\mathcal{S}$ with eigenvalue $0$). As a result, assuming $\ket{\psi_{\textrm{pre-VBS}}}$ is normalized, the norm is reduced from $1$ to $\frac{3}{4}$. In general, applying the local symmetrization operator $\mathcal{S}$ at one site of a spin-$1$ $\ket{\psi_{\textrm{pre-VBS}}}$ state reduces the norm by a factor $\frac{3}{4}$. 

Starting from a normalized spin-$1$ $\ket{\psi_{\textrm{pre-VBS}}}$, the application of the $N$ symmetrization operators in parallel leads to an exponential decrease of the norm of the resulting $\ket{\psi_{\textrm{VBS}}} = \bigotimes_{n = 1}^N \mathcal{S}_n \ket{\psi_{\textrm{pre-VBS}}}$. Asymptotically, 
\begin{equation}
    \braket{\psi_{\textrm{VBS}} | \psi_{\textrm{VBS}}} =  \Big(\frac{3}{4}\Big)^N \quad [\textrm{Spin-1}].
\end{equation}
However, for finite one-dimensional spin-$1$ AKLT models, there is a deviation from the asymptotic limit that depends on the boundary conditions:
\begin{itemize}
    \item{For periodic boundary conditions (i.e. N-site ring), 
    \begin{equation}
        \braket{\psi_{\textrm{VBS}} | \psi_{\textrm{VBS}}} =  (3/4)^N +3 (-1/4)^N.
    \label{eqfse1}
    \end{equation}}
    \item{For open boundary conditions (i.e. a $N$-site chain where the outermost spins-$\frac{1}{2}$ are not connected):
    \begin{itemize}
        \item{If outermost spins are aligned (e.g., both $\ket{\uparrow}$), 
        \begin{equation}
            \braket{\psi_{\textrm{VBS}} | \psi_{\textrm{VBS}}} =  (3/4)^N - (-1/4)^N.
        \label{eqfse2}
        \end{equation}}
        \item{If the outermost spins are anti-aligned (e.g., one $\ket{\uparrow}$ and another $\ket{\downarrow}$),
        \begin{equation}
            \braket{\psi_{\textrm{VBS}} | \psi_{\textrm{VBS}}} =  (3/4)^N + (-1/4)^N.
        \label{eqfse3}
        \end{equation}}
        \item{If the outermost spins are neither aligned nor anti-aligned, the finite-size correction takes values between Eqs. (\ref{eqfse2}) and (\ref{eqfse3}).}
    \end{itemize}}
\end{itemize}
Naturally, as $N \to \infty$, in all cases the finite-size effects vanish and the boundary conditions become redundant.

\subsection{Spin-$\frac{3}{2}$ Valence-Bond-Solid State}

The local symmetrization operator for the spin-$\frac{3}{2}$ case is given by Eq. (\ref{eqdefS3}). Its eigenstates are
\begin{equation}
\begin{aligned}
    & \ket{1_{\frac{3}{2}}} = \ket{\uparrow \uparrow \uparrow}, \\
    & \ket{1_{\frac{1}{2}}} = \frac{1}{\sqrt{3}} (\ket{\uparrow \uparrow \downarrow} + \ket{\uparrow \downarrow \uparrow} + \ket{\downarrow \uparrow \uparrow}), \\
    & \ket{1_{-\frac{1}{2}}} = \frac{1}{\sqrt{3}} (\ket{\uparrow \downarrow \downarrow} + \ket{\downarrow \uparrow \downarrow} + \ket{\downarrow \downarrow \uparrow}), \\
    & \ket{1_{-\frac{3}{2}}} = \ket{\downarrow \downarrow \downarrow}, \\
    & \ket{0_{\frac{1}{2}, 1}} = \frac{1}{\sqrt{2}}(\ket{\uparrow \uparrow \downarrow} - \ket{\downarrow \uparrow \uparrow}), \\
    & \ket{0_{\frac{1}{2}, 2}} = \frac{1}{\sqrt{6}} (\ket{\uparrow \uparrow \downarrow} -2 \ket{\uparrow \downarrow \uparrow} + \ket{\downarrow \uparrow \uparrow}), \\
    & \ket{0_{-\frac{1}{2}, 1}} = \frac{1}{\sqrt{2}}(\ket{\uparrow \downarrow \downarrow} - \ket{\downarrow \downarrow \uparrow}), \\
    & \ket{0_{-\frac{1}{2}, 2}} = \frac{1}{\sqrt{6}} (\ket{\uparrow \downarrow \downarrow} -2 \ket{\downarrow \uparrow \downarrow} + \ket{\downarrow \downarrow \uparrow}).
\end{aligned}
\end{equation}
Highlighting only the valence bonds involving site $n$, and denoting the three nearest neighbors by $n'_s, s \in \{ 1, 2, 3\}$, $\ket{\psi_{\textrm{pre-VBS}}}$ can be written as
\begin{equation}
\begin{aligned}
    \ket{\psi_{\textrm{pre-VBS}}}&  = \frac{1}{\sqrt{2}} \big( \ket{\uparrow}_{(n;n'_1)} \ket{\downarrow}_{(n'_1;n)} - \ket{\downarrow}_{(n;n'_1)} \ket{\uparrow}_{(n'_1;n)} \big) \; \otimes \\ & \otimes \frac{1}{\sqrt{2}} \big( \ket{\uparrow}_{(n;n'_2)} \ket{\downarrow}_{(n'_2;n)} - \ket{\downarrow}_{(n;n'_2)} \ket{\uparrow}_{(n'_2;n)} \big) \; \otimes \\
    & \otimes \frac{1}{\sqrt{2}} \big( \ket{\uparrow}_{(n;n'_3)} \ket{\downarrow}_{(n'_3;n)} - \ket{\downarrow}_{(n;n'_3)} \ket{\uparrow}_{(n'_3;n)} \big) \otimes ...
\end{aligned}
\end{equation}
Expanding the tensor product of the three highlighted valence bonds in the eigenbasis of $\mathcal{S}$ at site $n$ gives
\begin{equation}
    \begin{aligned}
    & \ket{\psi_{\textrm{pre-VBS}}} = \frac{1}{2\sqrt{2}} \Big[ \ket{1_{\frac{3}{2}}}_{n} \otimes \ket{\downarrow \downarrow \downarrow}_{n'} - \ket{1_{-\frac{3}{2}}}_{n} \otimes \ket{\uparrow \uparrow \uparrow}_{n'} \\
    & - \ket{1_{\frac{1}{2}}}_{n} \otimes \frac{1}{\sqrt{3}} \big( \ket{\uparrow \downarrow \downarrow}_{n'} + \ket{\downarrow \uparrow \downarrow}_{n'} + \ket{\downarrow \downarrow \uparrow}_{n'} \big) \\
    & - \ket{1_{-\frac{1}{2}}}_{n} \otimes \frac{1}{\sqrt{3}} \big( \ket{\uparrow \uparrow \downarrow}_{n'} + \ket{\uparrow \downarrow \uparrow}_{n'} + \ket{\downarrow \uparrow \uparrow}_{n'} \big) \\
    & + \ket{0_{\frac{1}{2}, 1}}_{n} \otimes \frac{1}{\sqrt{2}} \big( \ket{\uparrow \downarrow \downarrow}_{n'} - \ket{\downarrow \downarrow \uparrow}_{n'} \big) \\ 
    & - \ket{0_{\frac{1}{2}, 2}}_{n} \otimes \big(\frac{1}{\sqrt{6}} \ket{\downarrow \downarrow \uparrow}_{n'} - \frac{\sqrt{2}}{\sqrt{3}} \ket{\downarrow \uparrow \downarrow}_{n'} + \frac{1}{\sqrt{6}} \ket{\uparrow \downarrow \downarrow}_{n'} \big) \\ 
    & + \ket{0_{-\frac{1}{2}, 1}}_{n} \otimes \frac{1}{\sqrt{2}} \big( \ket{\downarrow \uparrow \uparrow}_{n'} - \ket{\uparrow \uparrow \downarrow}_{n'} \big) \\ 
    & + \ket{0_{-\frac{1}{2}, 2}}_{n} \otimes \big(\frac{1}{\sqrt{6}} \ket{\downarrow \uparrow \uparrow}_{n'} - \frac{\sqrt{2}}{\sqrt{3}} \ket{\uparrow \downarrow \uparrow}_{n'} + \frac{1}{\sqrt{6}} \ket{\uparrow \uparrow \downarrow}_{n'} \big) \Big] \\
    & \otimes ...
\end{aligned}
\label{eqexp3}
\end{equation}
Upon application of $\mathcal{S}$ at site $n$, the last four terms on the right-hand side of Eq. (\ref{eqexp3}) vanish, which reduces the norm from $1$ to $\frac{1}{2}$. Hence, applying the local symmetrization operator $\mathcal{S}$ at one site of a spin-$\frac{3}{2}$ $\ket{\psi_{\textrm{pre-VBS}}}$ state decreases the norm by a factor $\frac{1}{2}$. In the asymptotic limit, regardless of the boundary conditions, the norm of the Valence Bond State $\ket{\psi_{\textrm{VBS}}} = \bigotimes_{n = 1}^N \mathcal{S}_n \ket{\psi_{\textrm{pre-VBS}}}$ at a lattice with $N$ sites is
\begin{equation}
    \braket{\psi_{\textrm{VBS}} | \psi_{\textrm{VBS}}} =  \Big(\frac{1}{2}\Big)^N \quad \Big[\textrm{Spin-}\frac{3}{2}\Big].
\end{equation}

\subsection{General Valence-Bond-Solid State}

In general, ignoring finite-size effects, the norm of the VBS, $\ket{\psi_{\textrm{VBS}}} = \bigotimes_{n} \mathcal{S}_n \ket{\psi_{\textrm{pre-VBS}}}$, is given by
\begin{equation}
    \braket{\psi_{\textrm{VBS}} | \psi_{\textrm{VBS}}} =  p^N,
\end{equation}
where $p \in (0,1)$ corresponds to the fraction of symmetric spin states selected by $\mathcal{S}$ at each site. In other words, $p$ is the ratio between the number of states corresponding to the desired local spin-$\frac{\mathcal{N}_n}{2}$ and the total number of states resulting from adding the $\mathcal{N}_n$ spins-$\frac{1}{2}$.

For concreteness, let us consider the previous two examples plus the spin-$2$ VBS state on a square lattice. 
\begin{itemize}
    \item{Spin-$1$ VBS: the addition of two spins-$\frac{1}{2}$ gives $\frac{1}{2} \oplus \frac{1}{2} = 0, 1$, so we have $2 \times 1 + 1 = 3$ spin-1 states and $2 \times 0 + 1 = 1$ spin-0 state, yielding $p = \frac{3}{4}$.}
    \item{Spin-$\frac{3}{2}$ VBS: adding three spins-$\frac{1}{2}$ gives $\frac{1}{2} \oplus \frac{1}{2} \oplus \frac{1}{2} = \frac{1}{2}, \frac{1}{2}, \frac{3}{2}$, so there are $2 \times \frac{3}{2} + 1 = 4$ spin-$\frac{3}{2}$ states and $2 \times (2 \times \frac{1}{2} + 1) = 4$ spin-$\frac{1}{2}$ states, thus $p = \frac{4}{8} = \frac{1}{2}$.}
    \item{Spin-$2$ VBS: adding four spins-$\frac{1}{2}$ gives $\frac{1}{2} \oplus \frac{1}{2} \oplus \frac{1}{2} \oplus \frac{1}{2} = 0, 0, 1, 1, 1, 2$, so there are $2 \times (2 \times 0 + 1) = 2$ spin-$0$ states, $3 \times (2 \times 1 + 1) = 9$ spin-$1$ states and $2 \times 2 + 1 = 5$ spin-$2$ states, in which case $p = \frac{5}{16}$.}
\end{itemize}


\section{Initialization of Matrix Product States with Physical Index Dimension $d=4$ and Virtual Index Dimension $D=2$ on Quantum Hardware}\label{AppTrueD}

This section explains how the MPS form of the 1D spin-$1$ VBS state can be exploited to construct the circuits shown in Fig. \ref{fig_mps_vbs} of the main text. In fact, this discussion applies to an arbitrary left-canonical MPS with virtual and physical index dimensions $D = 2$ and $d = 4$, of which the spin-$1$ VBS state is one example. First, the case of open boundary conditions will be considered, resulting in a deterministic scheme with $N$ sequential operations for a chain of $N$ sites. Then, the case of periodic boundary conditions will be discussed, yielding a probabilistic method with $\mathcal{O}(1)$ success probability.

The general method for the preparation of an MPS with open boundary conditions was first introduced by Sch\"{o}n et al. \cite{Schoen05}, and later adapted to digital quantum computing by Ran \cite{Ran20}, who explicitly discussed its application to a MPS with $d = D = 2$. Here, we adapt this general method to the case $d = 4, D = 2$. We also extend it to the case of periodic boundary conditions.

\subsection{Open Boundary Conditions}

\begin{figure}[h]
    \centering{
    \includegraphics[width=\linewidth]{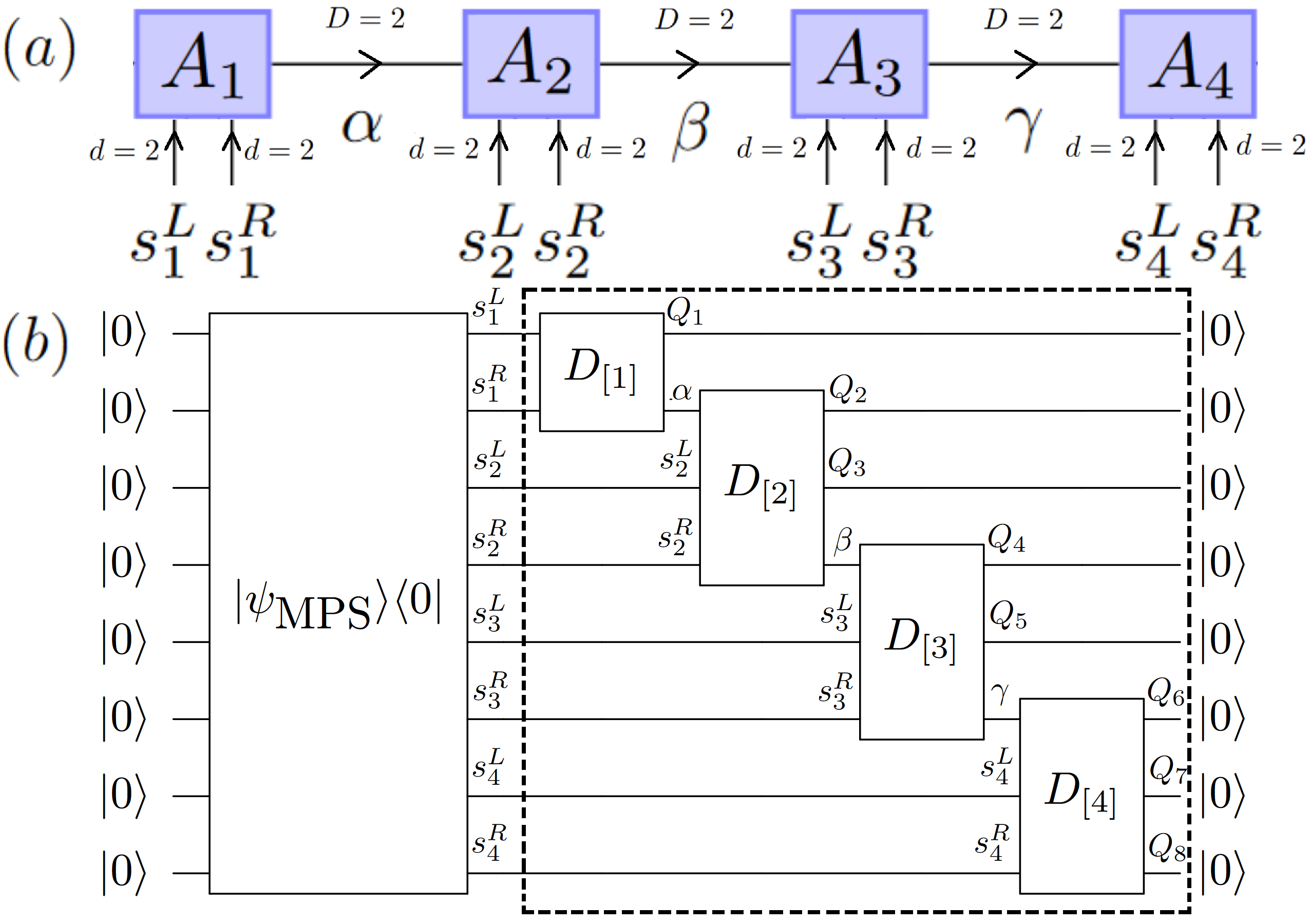}}
    \caption{(a) Diagram of MPS with open boundary conditions, virtual index dimension $D=2$ and physical index dimension $d =4$ for $N=4$ sites. (b) Scheme of retrosynthetic construction of quantum circuit to initialize MPS defined in (a). Sequential application of matrix product disentanglers (MPD) $\{ D_{[i]}\}_{i=1}^{4}$ reverses the black-box preparation of the MPS. Qubit labels serve to make connection with tensor network diagrams used in step-by-step explanation. Inverse of quantum circuit inside dashed-line box prepares MPS on $2N$-qubit register, in accordance with Fig. \ref{fig_mps_vbs}(c) in the main text.}
    \label{fig_MPS_app_circ}
\end{figure}

Fig. \ref{fig_MPS_app_circ}(a) shows the MPS we wish to prepare on a quantum computer. WELOG, the number of sites is $N = 4$. Moreover, the MPS is assumed to be in left-canonical form, hence the directions of the arrows. Any MPS can be turned into left-canonical form via a sequential application of singular value decomposition at every site from left to right. We assume the singular value at the last site is discarded, so the MPS is normalized, as required to be initialized on a quantum computer. 
Every tensor satisfies the left-normalization condition,

\begin{figure}[h]
    \centering{
    \includegraphics[width=\linewidth]{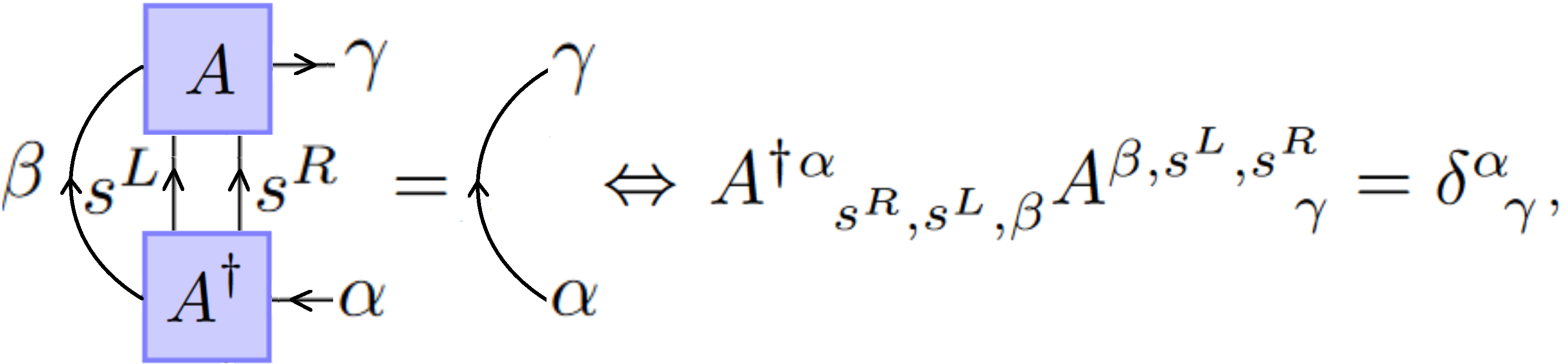}}
\end{figure}

\noindent where the Einstein summation convention is implied in the algebraic expression. Of course, for the tensor $A_1$ at the first site, $\beta$ is just a singleton index, so the sum over it is redundant. Likewise, for the tensor $A_{N}$ at the last site, $\alpha$ and $\gamma$ are also singleton indices, so the identity on the right-hand-side of this equality is just the scalar $1$.

As illustrated in Fig. \ref{fig_MPS_app_circ}(b), the derivation of the circuit that initializes such a MPS will follow a retrosynthetic approach, whereby we assume there is a black-box circuit that prepares this MPS and our goal is to invert its effect, thereby retrieving the fiducial state $\ket{0}^{\otimes2N}$. Each of the $N$ steps of this process involves the application of a unitary matrix product disentangler (MPD) $D_{[i]}$, which disentangles site $i$ from the remainder of the MPS. 

Let us begin with the first site. We wish to apply a MPD, $D_{[1]}$, that cancels the action of $A_1$. Making use of the aforestated left-normalization condition, $D_{[1]}$ would simply correspond to the adjoint of such local tensor, $A_1^{\dagger}$. However, $A_1$ is not unitary, which is clear from the fact that the dimensions of the incoming legs $s^{L}_1$ and $s^{R}_1$ ($2 \times 2 = 4$) do not match that of the outgoing leg $\alpha$ ($2$). To address this issue, we define an enlarged tensor $A'^{\; s^{L}_{1}, s^{R}_{1}}_{1 \quad \; \; \; \;  \alpha, Q_1}$ that has an extra outgoing leg $Q_1$ of dimension $2$ relative to $A^{\; s^{L}_{1}, s^{R}_{1}}_{1 \quad \; \; \; \alpha}$, so that the incoming and outgoing legs dimensions are balanced. The original and enlarged tensors are related by
\begin{equation}
    A^{\; s^{L}_{1}, s^{R}_{1}}_{1 \quad \; \; \; \alpha} = A'^{\; s^{L}_{1}, s^{R}_{1}}_{1 \quad \; \; \; \;  \alpha, 0}.
\end{equation}
In words, the action of the enlarged tensor $A'_1$ is equivalent to that of the original tensor $A_1$ provided that the dummy leg $Q_1$ is set to $0$ (or, in the language of gate-based quantum computing, the corresponding qubit is in state $\ket{0}$). Replacing $A_1$ by $A'_1$ in the MPS diagram gives 

\begin{figure}[h]
    \centering{
    \includegraphics[width=\linewidth]{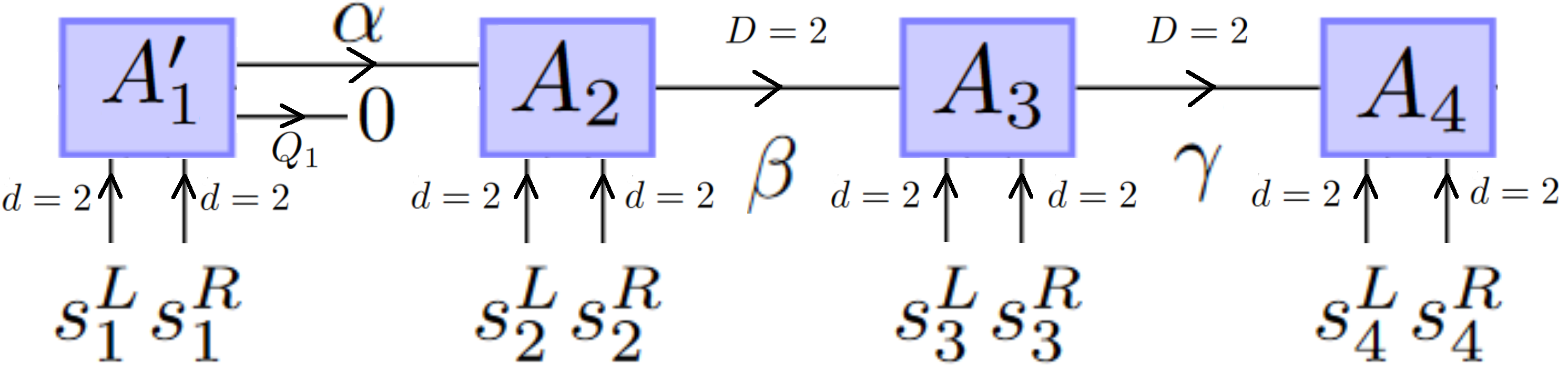}}
\end{figure}

\noindent As explained below, $A'_1$ can be constructed from $A_1$ so that its $4 \times 4$ matrix representation --- after fusing $(s^{L}_1,s^{R}_1)$ and $(\alpha,Q_1)$ --- is unitary. The MPD $D_{[1]}$ is therefore given by $A_{1}^{' \dagger}$, which, being unitary, can be implemented as a two-qubit operation on quantum hardware. The action of the MPD $D_{[1]}$ on the MPS yields

\begin{figure}[h]
    \centering{
    \includegraphics[width=\linewidth]{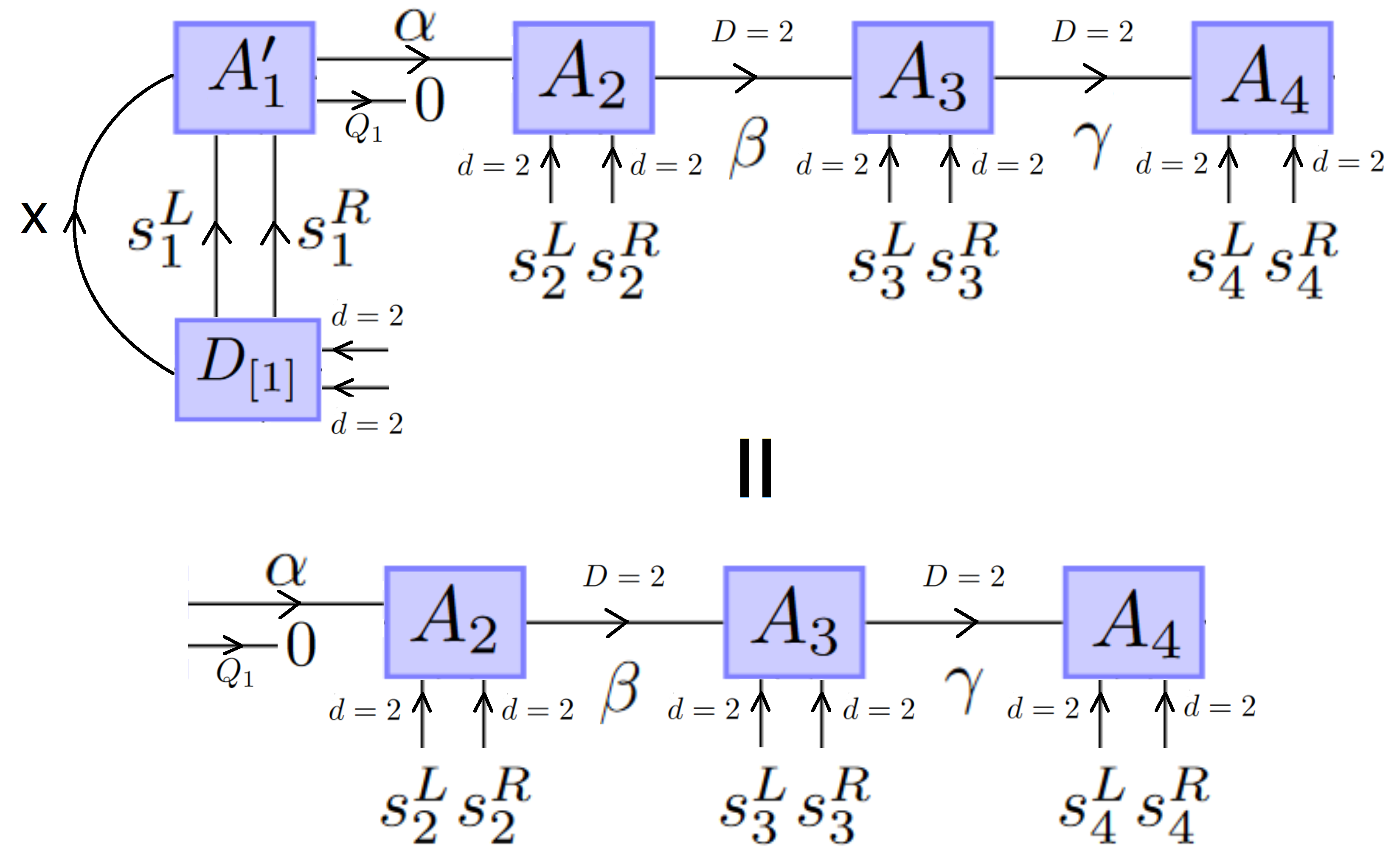}}
\end{figure}

\noindent where a dummy singleton index x was explicitly added in order to make the connection to the left-normalization condition clearer. Naturally, the resulting sum over x is redundant. Prior to the application of $D_{[1]}$, there were two open legs $s^{L}_1$ and $s^{R}_1$ encoding the local degree of freedom at site $1$. After applying $D_{[1]}$, one of those open legs, now termed $Q_1$, is separated from the MPS, as desired. The other encodes the virtual index $\alpha$ that connects sites $1$ and $2$. This will be an input for the next MPD, $D_{[2]}$.

The only missing element is how to construct the enlarged tensor $A'_1$ from the original tensor $A_1$. Merging all incoming/outgoing legs into a single incoming/outgoing leg, $A_1$ and $A'_1$ are, respectively, $4 \times 2$ and $4 \times 4$ matrices. Because $A_1$ and $A'_1$ share the same incoming legs, the columns of $A_1$ can be set as the first two columns of $A'_1$:
\begin{equation}
    A'_1 = \begin{pmatrix} 
    \uparrow & \uparrow & \uparrow & \uparrow \\
    A^{\; s^{L}_1,s^{R}_1}_{1 \quad \quad 0} & A^{\; s^{L}_1,s^{R}_1}_{1 \quad \quad 1} & \vec{u} & \vec{v} \\
    \downarrow & \downarrow & \downarrow & \downarrow \end{pmatrix}.
\label{eq_enlarged_tensor}
\end{equation}
The fact that $A_1$ is left-normalized ensures that $A^{\; s^{L}_1,s^{R}_1}_{1 \quad \quad 0}$ and $A^{\; s^{L}_1,s^{R}_1}_{1 \quad \quad 1}$ are orthogonal to each other and normalized. For $A'_1$ to be unitary, its columns (and rows) must form an orthonormal set. Hence, to complete the construction of $A'_1$, we just need to find the two remaining columns $\vec{u}$ and $\vec{v}$, which must be normalized four-dimensional vectors orthogonal to each other and to the first two columns. Concretely, ($\vec{u}$, $\vec{v}$) span the kernel of the $2 \times 4$ matrix
\begin{equation}
    \begin{pmatrix} 
    \leftarrow & (A^{\; s^{L}_1,s^{R}_1}_{1 \quad \quad 0})^{\dagger} & \rightarrow \\
    \leftarrow & (A^{\; s^{L}_1,s^{R}_1}_{1 \quad \quad 1})^{\dagger} & \rightarrow
    \end{pmatrix}.
\end{equation}
The MPD $D_{[1]}$ is simply given by the adjoint of Eq. (\ref{eq_enlarged_tensor}).

For the remaining sites, the construction of the respective MPDs proceeds in a similar spirit to what was described for the first site. The main difference is that more dummy outgoing legs need to be introduced to ensure the enlarged tensor, when cast in matrix form, is unitary. The resulting MPDs are three-qubit operations.  For the second site, 

\begin{figure}[h]
    \centering{
    \includegraphics[width=0.8\linewidth]{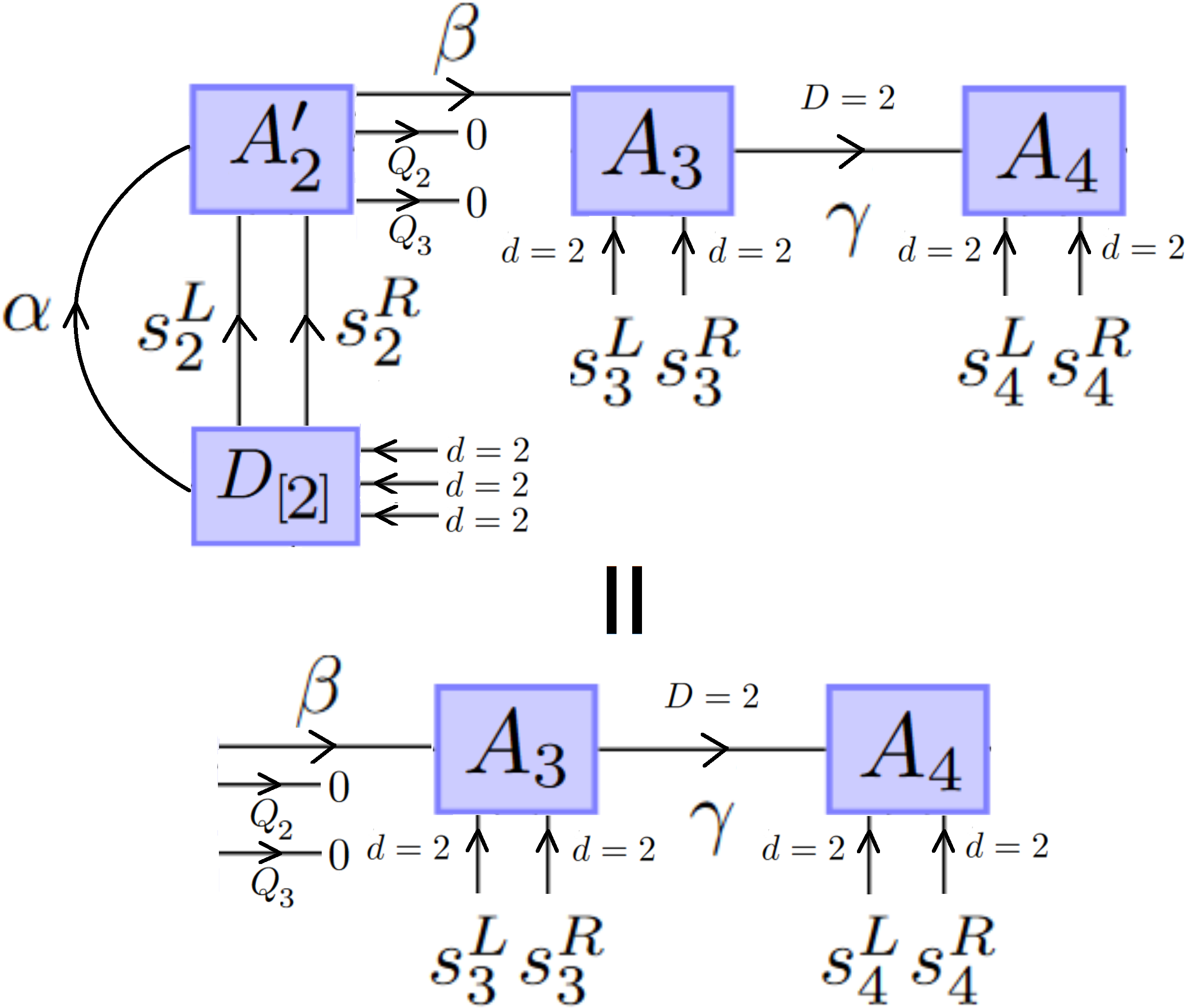}}
\end{figure}

\noindent where $A^{\; \alpha, s^{L}_{2}, s^{R}_{2}}_{2 \quad \; \; \; \; \; \beta} = A'^{\; \alpha, s^{L}_{2}, s^{R}_{2}}_{2 \quad \; \; \; \; \; \; \beta, 0, 0} $ and $D_{[2]} = A'^{\dagger}_{2}$. Similarly, for the third site, we have

\begin{figure}[h]
    \centering{
    \includegraphics[width=0.8\linewidth]{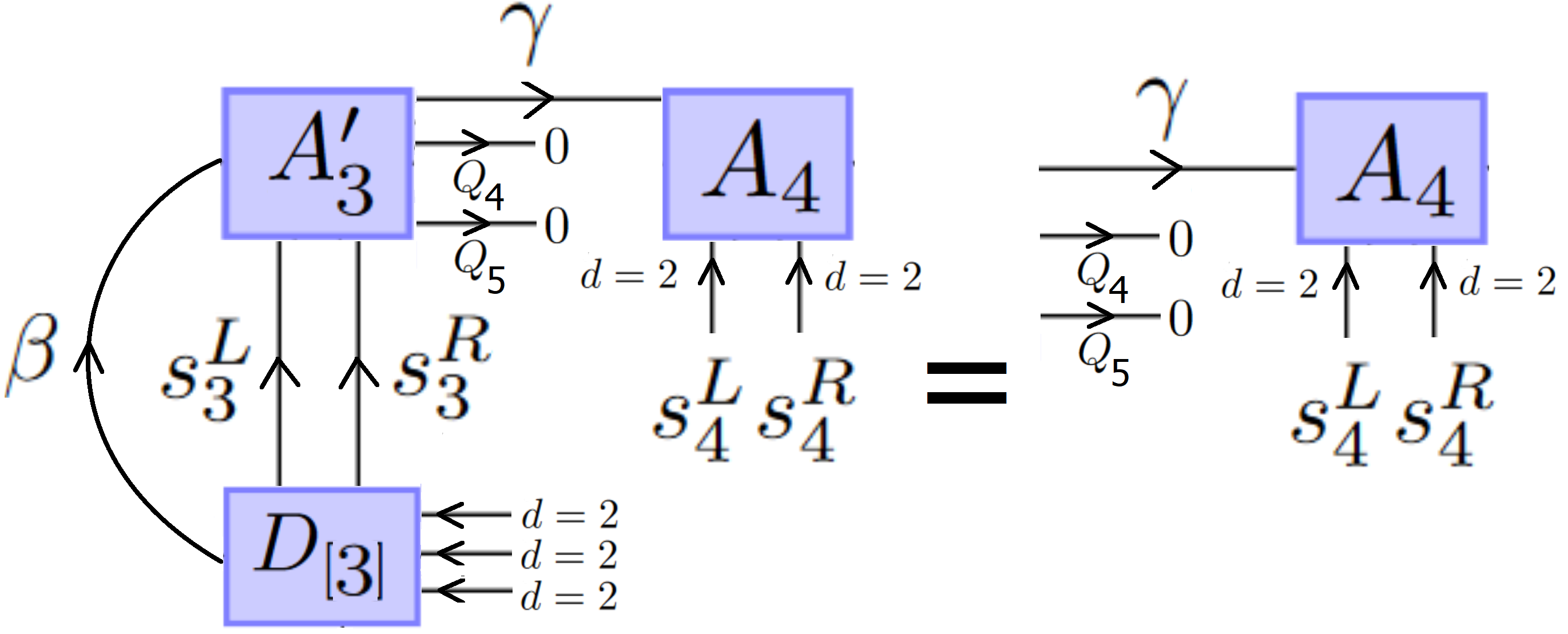}}
\end{figure}

\noindent where $A^{\; \beta, s^{L}_{3}, s^{R}_{3}}_{3 \quad \; \; \; \; \; \gamma} = A'^{\; \beta, s^{L}_{3}, s^{R}_{3}}_{3 \quad \; \; \; \; \; \; \gamma, 0, 0} $ and $D_{[3]} = A'^{\dagger}_{3}$. Of the three incoming legs, two encode the local degree of freedom at site $i = 2, 3$, and the other is associated with the virtual index that links sites $i-1$ and $i$. After the application of the MPD $D_{[i]}$, two of the legs become disentangled from the remainder of the MPS, while the third contains the virtual index that connects sites $i$ and $i+1$.

Finally, at the last site, since there are three incoming legs and no outgoing ones, three dummy legs must be added, with $A^{\; \gamma, s^{L}_{4}, s^{R}_{4}}_{4} = A'^{\; \gamma, s^{L}_{4}, s^{R}_{4}}_{4 \quad \; \; \; \; \; \; \; 0, 0, 0} $. After applying the respective MPD $D_{[4]} = A'^{\dagger}_{4}$, all three remaining legs are disentangled.

\begin{figure}[h]
    \centering{
    \includegraphics[width=0.6\linewidth]{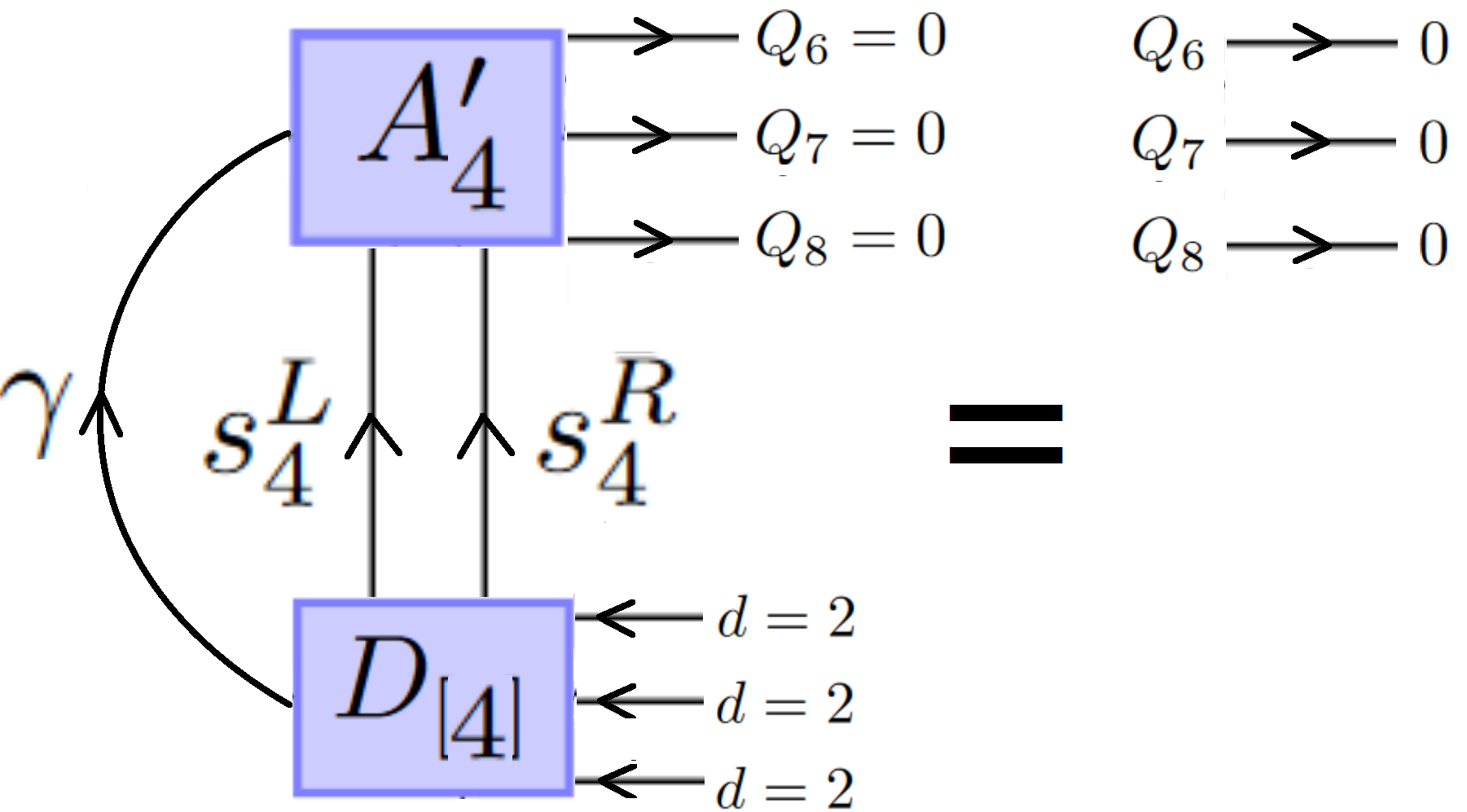}}
\end{figure}

Putting all pieces together, the sequential application of the $N$ MPDs $\{D_{[i]}\}_{i=1}^{N}$ reverses the black-box initialization of the MPS (cf. Fig. \ref{fig_MPS_app_circ}(b)). Hence, the quantum circuit that prepares such MPS is given by the inverse of the circuit within the dashed-line box in Fig. \ref{fig_MPS_app_circ}(b), in agreement with Fig. \ref{fig_mps_vbs}(c) in the main text.

\subsection{Periodic Boundary Conditions}

The adaptation of the method discussed for a MPS with open boundary conditions to the case of periodic boundary conditions appears challenging at first glance. In the former case, starting from one of the two ends of the chain (the left end, by hypothesis, since we assumed the MPS is in left-canonical form) meant that only the physical indices $s^{L}_1$ and $s^{R}_1$ appeared as inputs for the first MPD $D_{[1]}$, which was just as well, since, given the black-box initialization of the MPS, we had access to the physical indices but not to the virtual ones. However, for a MPS with periodic boundary conditions, regardless of the site from which we start, there will always be an incoming virtual leg that we cannot access. For example, let us consider the following translationally-invariant MPS for which the local rank-$4$ ($2 \times 2 \times 2 \times 2$) tensor $A$ is assumed to be left-normalized WLOG (hence the arrows defining the directions of the legs in the diagram). 

\begin{figure}[h]
    \centering{
    \includegraphics[width=\linewidth]{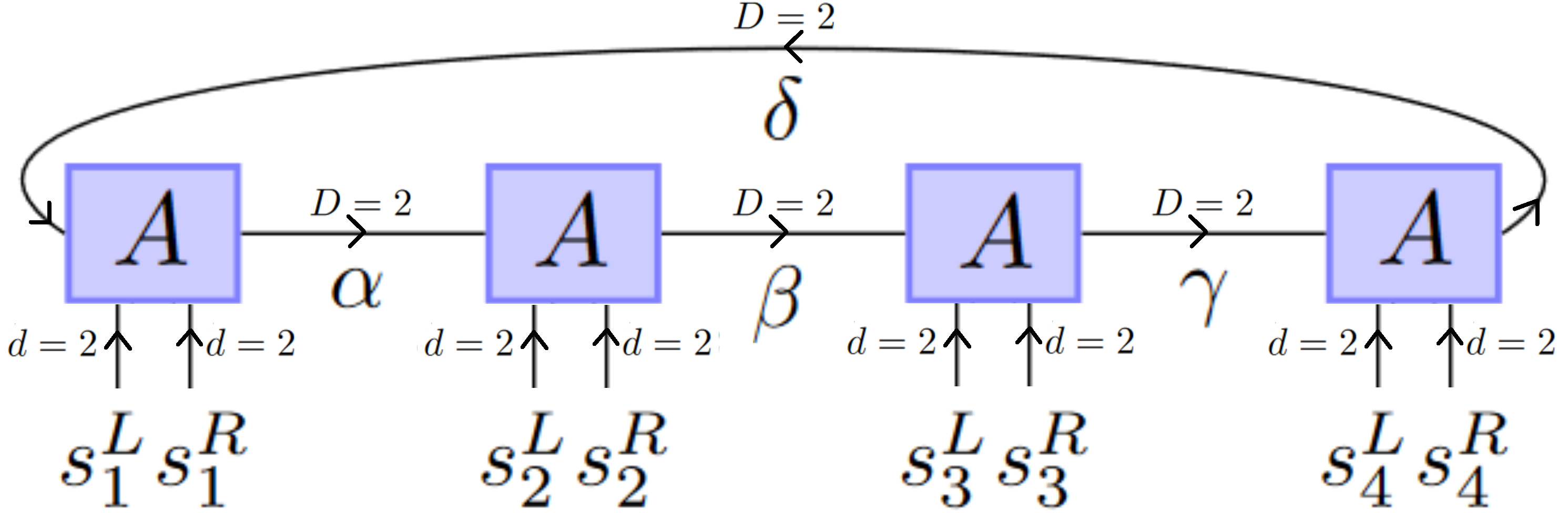}}
\end{figure}

\noindent If we start from site $1$ (any site will do, since they are all equivalent), the first MPD, $D_{[1]}$, will take as inputs $s^{L}_{1}$, $s^{R}_{1}$ \textit{and} $\delta$, outputting the virtual index $\alpha$. We do have access to the physical indices $s^{L}_{1}$, $s^{R}_{1}$ in our retrosynthetic approach (cf. Fig. \ref{fig_MPS_app_circ}(b)), but not to the virtual index $\delta$.

We therefore need a different approach to the first site. Specifically, we will cast the local tensor $A^{\delta, s^{L}_1, s^{R}_1}_{\quad \quad \; \; \alpha}$ at site $1$ into a $4 \times 4$ matrix $A^{(s^{L}_1, s^{R}_1)}_{\quad \quad \; \; (\alpha, \delta)} \equiv \tilde{A}$ by fusing the two physical indices $(s^{L}_1, s^{R}_1)$ into a single index labelling the rows and the two virtual indices $(\alpha, \delta)$ into a single index labelling the columns. Notice that, in this case, the dimensions of the inputs and outputs coincide, so there seems to be no need to add dummy legs to construct the MPD $D_{[1]}$. The problem, however, is that $\tilde{A}$ is, in general, non-unitary, because we are fusing one incoming index, $\delta$, with an outgoing index, $\alpha$. Hence, although the MPD defined as $D_{[1]} = \tilde{A}^{\dagger}$ does produce the desired effect of turning the two physical indices from site $1$, $s^{L}_1$ and $s^{R}_1$, into the two virtual indices $\alpha$ and $\delta$, implementing such $4 \times 4$ non-unitary matrix on quantum hardware is not immediately obvious.

To address this problem, we will embed this $4 \times 4$ non-unitary matrix $\tilde{A}$ in an $8 \times 8$ unitary matrix $U_{\tilde{A}}$. $\tilde{A}$ will correspond to the top-left quadrant of $U_{\tilde{A}}$, in which case the most significant qubit on which the three-qubit operation $U_{\tilde{A}}$ acts is an ancilla prepared in $\ket{0}$ and projected onto $\ket{0}$ by post-selection, thus ensuring that the effective action of $U_{\tilde{A}}$ on the remaining two qubits is the desired $\tilde{A}$. Then, the three-qubit $D'_{[1]}$ that appears in Fig. \ref{fig_mps_vbs}(d) in the main text is simply given by $D'_{[1]} = U_{\tilde{A}}^{\dagger}$. The probability of success is, in general, $\mathcal{O}(1)$; for the spin-$1$ VBS state, $50\%$ of trials are successful regardless of $N$.

A general scheme to accomplish this embedding of a $N$-qubit non-unitary operation in a $(N+1)$-qubit gate was presented by Lin et al. \cite{Lin21} (cf. Appendix C). Applying it to the specific case considered here, the goal is to find the $8 \times 8$ unitary matrix 
\begin{equation}
    U_{\tilde{A}} = \begin{pmatrix} 
    n \tilde{A} &   \\
        & B \\
    C   &   \end{pmatrix},
\end{equation}
where $C$ is a $4 \times 4$ matrix, $B$ is a $8 \times 4$ matrix, and $n$ is a scalar that ensures that no column of $\tilde{A}$ has norm greater than $1$, otherwise it is impossible to ensure that all columns of $U_{\tilde{A}}$ are orthonormal, which is a defining property of unitary matrices. $C$ can be derived from $\tilde{A}$ by performing its singular value decomposition, $\tilde{A} = U S V^{\dagger}$, so that
\begin{equation}
    C = U (\mathbb{1} - n^2 S)^{\frac{1}{2}} V^{\dagger},
\end{equation}
where  $n$ is any scalar that satisfies the requirement
\begin{equation}
    0 < n < 1/\sqrt{\textrm{max}(\{ s^2_i \}_{i=1}^{4})},
\end{equation}
with $\{ s_i \}_{i=1}^{4}$ the singular values of $\tilde{A}$ (i.e., $S = \textrm{diag}(s_1,s_2,s_3,s_4)$). Having derived $n$ and $C$ from $\tilde{A}$, we already have the first four columns of $U_{\tilde{A}}$, which are orthonormal. We just need to find the remaining four columns (i.e., the matrix $B$), which can be accomplished similarly to the way we completed the enlarged tensors for the MPS with open boundary conditions. That is, the four columns of $B$ are the orthonormal basis that spans the kernel of the $4 \times 8$ matrix 
\begin{equation}
    \begin{pmatrix} 
    n \tilde{A} \\
    C  \end{pmatrix}^{\dagger}.
\end{equation}

For sites $i = 2, 3, ..., N-1$, we can proceed exactly as for the bulk sites in the MPS with open boundary conditions. In this case, however, because of the translational invariance, all MPDs $D_{[2]}$, $D_{[3]}$, ..., $D_{[N-1]}$ are equal. For the particular case of the spin-$1$ VBS state, using the definition of the local tensor $A$ stated in Eq. (\ref{local_tensor_two_qubits}) in the main text and following the steps described in the previous section of this appendix, we obtain
\begin{equation}
    D^{\dagger}_{[i]} = \begin{pmatrix}
    0 & \sqrt{\frac{2}{3}} & 0 & 0 & 0 & 0 & 0 & -\frac{1}{\sqrt{3}} \\
    -\frac{1}{\sqrt{6}} & 0 & 0 & a & a & c & 0 & 0 \\
    -\frac{1}{\sqrt{6}} & 0 & 0 & -\frac{1}{\sqrt{12}} & -\frac{1}{\sqrt{12}} & d & 0 & 0 \\
    0 & 0 & 0 & 0 & 0 & 0 & 1 & 0 \\
    0 & 0 & 1 & 0 & 0 & 0 & 0 & 0 \\
    0 & \frac{1}{\sqrt{6}} & 0 & \frac{1}{2} & -\frac{1}{2} & 0 & 0 & \frac{1}{\sqrt{3}} \\
    0 & \frac{1}{\sqrt{6}} & 0 & -\frac{1}{2} & \frac{1}{2} & 0 & 0 & \frac{1}{\sqrt{3}} \\
    -\sqrt{\frac{2}{3}} & 0 & 0 & b & b & e & 0 & 0
    \end{pmatrix},
\end{equation}
with $a = \frac{2\sqrt{2}}{3} + \frac{\sqrt{3}}{30}$, $b = -\sqrt{\frac{5}{12} - a^2}$, $c = -(1 + f^2 + \frac{1}{4}(1+f)^2)^{-\frac{1}{2}}$, where $f = \frac{a-\frac{1}{2}b}{\frac{1}{\sqrt{12}} + \frac{1}{2}b}$, $d = f c$, and $e = -\frac{c + d}{2}$. The circuit for this three-qubit operation can be obtained via the Cirq \texttt{three\_qubit\_matrix\_to\_operations} method \cite{Cirq3QDecomp}, resulting in a count of $20$ $\textsc{cnot}$ gates.

For the final site, the rank-$4$ tensor $A^{\gamma, s^{L}_{N=4}, s^{R}_{N=4}}_{\quad \quad \quad \quad \; \; \delta}$ is turned into a $16$-dimensional vector by fusing all four indices into a single one. This corresponds to the first column of the enlarged $A'_{N=4}$, and the remaining columns have to be found as in the case of open boundary conditions. There is, however, one important difference: Because we fused three incoming legs ($\gamma, s^{L}_{N=4}, s^{R}_{N=4}$) with an outgoing leg ($\delta$), the vectorized form of $A_{N=4}$ is not guaranteed to be normalized, so we have to normalize it explicitly. Of course, due to the extra bond linking sites $1$ and $N$, $D_{[N=4]}$ is a four-qubit gate as opposed to the three-qubit gate found for a chain.

In any case, in the initialization of the MPS, $D^{\dagger}_{[N=4]}$ is the first operation, so only its action on the fiducial state $\ket{0}^{\otimes 4}$ matters. As a result, instead of having to find the basis gate decomposition of a four-qubit gate $D^{\dagger}_{[N=4]}$, we only have to find the circuit that prepares the (normalized) vectorized form of $A_{N=4}$, which is a four-qubit state. For example, for the spin-$1$ VBS state, the following state must be initialized:
\begin{equation*}
    \frac{\sqrt{3}}{6}\Big(0,2,-1, 0,-1, 0, 0, 0, 0, 0, 0, 1, 0,1,-2,0\Big)^T.
\end{equation*}
Using the Schimdt-decomposition-based method discussed in Section \ref{Section7} of the main text, this can be accomplished with a total of at most $9$ $\textsc{cnot}$ gates.

\section{Quantum Amplitude Amplification Applied to Preparation of VBS States}\label{AppD}

\begin{figure}[h]
    \centering{
    \includegraphics[width=\linewidth]{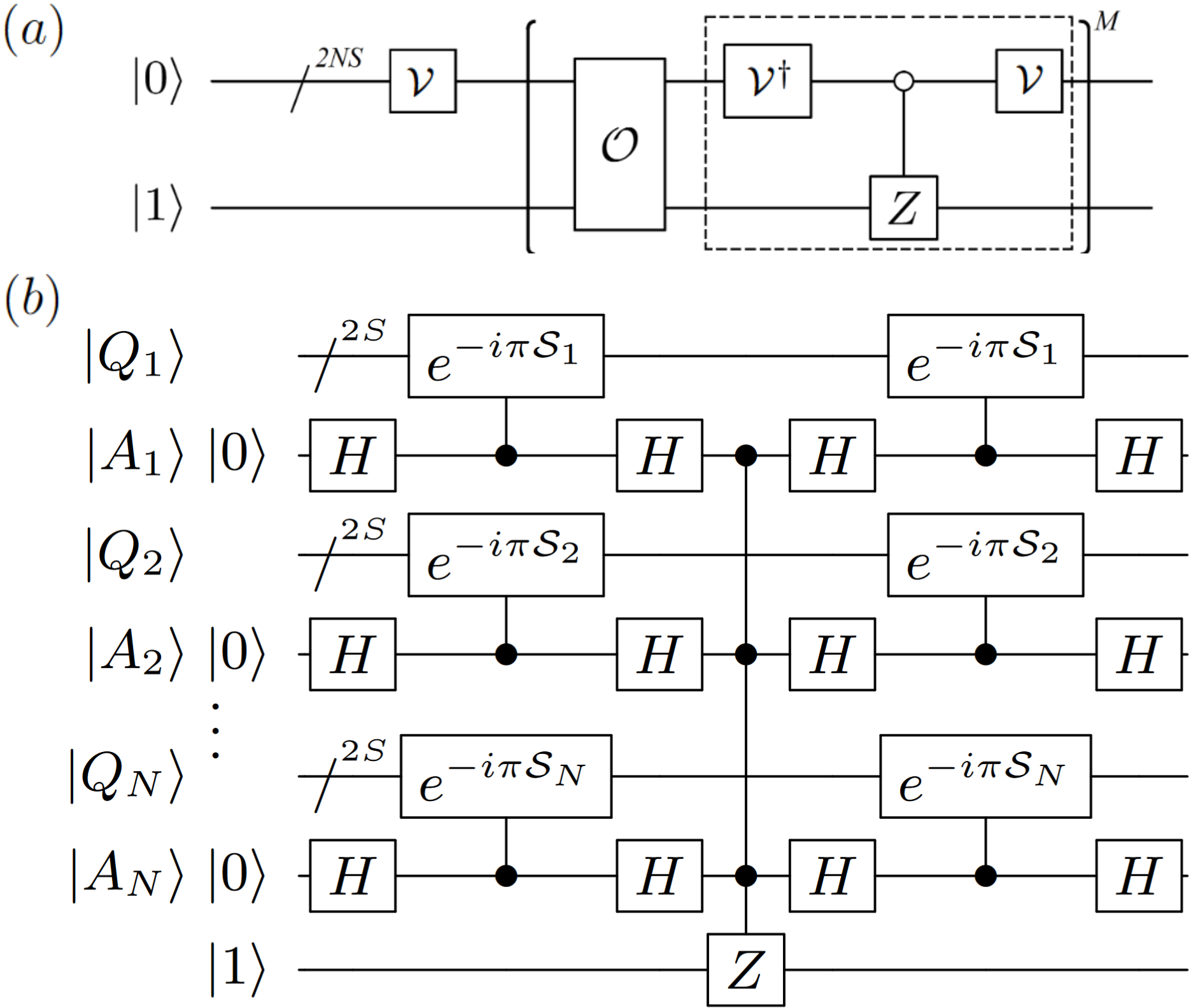}
    \caption{(a) High-level circuit for quantum amplitude amplification algorithm. (b) Quantum circuit for oracle $\mathcal{O}$ that is suitable for the application of quantum amplitude amplification to prepare $\ket{\psi_{\textrm{VBS}}}$ starting from $\ket{\psi_{\textrm{pre-VBS}}} = \mathcal{V} \ket{0}^{\otimes 2NS}$.}
    \label{QAA_Circuit}}
\end{figure}

Fig. \ref{QAA_Circuit}(a) shows a schematic representation of the circuit corresponding to the quantum amplitude amplification algorithm for the preparation of VBS states. The subcircuit $\mathcal{V}$ prepares the initial state, $\ket{\psi_{\textrm{pre-VBS}}}$, and its structure can be found in Fig. \ref{Circ1} in the main text. The only missing element is the oracle that marks the target state, $\ket{\psi_{\textrm{VBS}}} = \bigotimes_{n} \mathcal{S}_n \ket{\psi_{\textrm{pre-VBS}}}$, with a phase shift $e^{i \pi}$. An implementation of such oracle that results in a circuit depth of $\mathcal{O}(N)$ is presented in Fig. \ref{QAA_Circuit}(b). The first part corresponds to the layer of local Hadamard tests, of constant depth relative to $N$, used in the probabilistic method discussed in the main text. Then, the $N$ ancillas are used as control-qubits of a controlled-$Z$ gate acting on an additional ancilla initialized in $\ket{1}$. Such a gate can be implemented with $\mathcal{O}(N)$ depth (cf. Lemma 7.11 in \cite{Barenco95}). If all ancillas are in $\ket{1}$, the main $2NS$-qubit register is in the fully symmetrized state $\ket{\psi_{\textrm{VBS}}}$, and the controlled-$Z$ gate is triggered, applying a phase shift $e^{i \pi}$ to the main register per the phase kickback effect \cite{JonesJaksch12}. The final layer of local Hadamard tests resets the ancillas, so that they can be reused in the following iteration. 

Despite the $\mathcal{O}(N)$ depth of the circuit that realizes the oracle $\mathcal{O}$, the number of iterations $M$ corresponding to the repetition of the subcircuit shown inside brackets in Fig. \ref{QAA_Circuit}(a) is exponential in $N$. Indeed, quantum amplitude amplification amounts to a rotation in a two-dimensional subspace spanned by $\{ \ket{\tilde{\psi}_{\textrm{VBS}}}, \ket{\psi^{\perp}_{\textrm{VBS}}} \}$, where $\ket{\psi^{\perp}_{\textrm{VBS}}} \equiv \ket{\psi_{\textrm{pre-VBS}}} - \ket{\tilde{\psi}_{\textrm{VBS}}}$. Since $|\braket{\tilde{\psi}_{\textrm{VBS}} | \psi_{\textrm{pre-VBS}}}|^2 = p^{N}$ (cf. Eq. (\ref{eq_overlap_main})), for $N \gg 1$, $p^{N} \ll 1$, so the angle covered in each iteration is $\sim 2 p^{N/2}$ \cite{JonesJaksch12}, in which case maximizing the overlap with the target state $\ket{\psi_{\textrm{VBS}}}$ takes $\mathcal{O}(p^{-\frac{N}{2}})$ iterations. Since all iterations have to be implemented consecutively, this results in a total circuit depth that is far beyond the capabilities of near-term quantum hardware even for small $N$.

\section{Deterministic Preparation of General $3$-Qubit States via Schmidt Decomposition}\label{AppE}

As mentioned in Section \ref{Section7} of the main text, the application of the Schmidt-decomposition-based method to the initialization of the six-qubit islands $\ket{\psi_{\textrm{island}}^{S=3/2}} = \sum_{\mathbf{i}=0}^{7} \sum_{\mathbf{j}=0}^{7} \mathbf{M}_{\mathbf{i},\mathbf{j}} \ket{\mathbf{i}} \otimes \ket{\mathbf{j}}$ of the spin-$\frac{3}{2}$ VBS state involves the preparation of a three-qubit state $\sum_{k=0}^{7} s_k \ket{k}$, where $\{ s_k \}_{k = 0}^{7}$ are the singular values of $\mathbf{M}$. Even though the number of qubits is not even, the Schmidt-decomposition-based method can be adapted to this three-qubit case. Specifically, we can consider an asymmetric bipartition with two qubits on one side and one qubit on the other: $\sum_{k=0}^{7} s_k \ket{k} = \sum_{i = 0}^{3} \sum_{j = 0}^{1} \mathbf{M'}_{i,j} \ket{i} \otimes \ket{j}$. Performing the singular value decomposition of the $4 \times 2$ matrix $\mathbf{M'}$ whilst retaining the redundant all-zero rows of $\mathbf{S}$ (thus ensuring $\mathbf{U}$ is unitary and not just left-normalized) gives $\mathbf{M'} = \mathbf{U'} \mathbf{S'} \mathbf{V'}^{\dagger}$, with
\begin{equation}
    \begin{aligned}
    & \quad \; \mathbf{U'} = \begin{pmatrix} 
    \uparrow & \uparrow & \uparrow & \uparrow \\
    \ket{u'_0} & \ket{u'_1} & \ket{u'_2} & \ket{u'_3} \\
    \downarrow & \downarrow & \downarrow & \downarrow \end{pmatrix} \\
    & \mathbf{S'} = \begin{pmatrix} 
    s'_0 & 0 \\
    0 & s'_1 \\
    0 & 0 \\
    0 & 0
    \end{pmatrix} \quad
    \mathbf{V'} = \begin{pmatrix} 
    \uparrow & \uparrow \\
    \ket{v'_0} & \ket{v'_1} \\
    \downarrow & \downarrow
    \end{pmatrix}.
    \end{aligned}
\end{equation}
Hence, $\sum_{k=0}^{7} s_k \ket{k} = \sum_{l = 0}^{1} s'_l \ket{u'_l} \otimes \ket{v'_l}$, which can be prepared via the quantum circuit shown in Fig. \ref{SVD_method_3Q_state} below. $B'$ and $V'$ are both single-qubit operations, so their decomposition involves no $\textsc{cnot}$ gates. $U'$ is a two-qubit operation, so it can be decomposed in terms of, at most, $3$ $\textsc{cnot}$ gates \cite{Vidal04}, yielding a maximum of $4$ $\textsc{cnot}$ gates to prepare an arbitrary three-qubit state via this method.

\begin{figure}[h]
    \centering{\large \Qcircuit @C=0.8em @R=0.8em {
    \quad & \quad & \quad & \quad & \quad & \quad & \quad & \quad & \quad & \lstick{|0\rangle} & \gate{B'} & \ctrl{1} & \gate{V'} & \qw \\
    \quad & \quad & \quad & \quad & \quad & \quad & \quad & \quad & \quad & \lstick{|0\rangle} & \qw & \targ & \multigate{1}{U'} & \qw \\
    \quad & \quad & \quad & \quad & \quad & \quad & \quad & \quad & \quad & \lstick{|0\rangle} & \qw & \qw & \ghost{U'} & \qw
    }}
    \caption{Adaptation of Schmidt-decomposition-based method to preparation of arbitrary three-qubit state $\ket{\phi} = \sum_{i=0}^{3} \sum_{j=0}^{1} \mathbf{M'}_{i,j} \ket{i} \otimes \ket{j}$. $U'$ and $V'$ can be obtained from the singular value decomposition of $\mathbf{M'} = U' S' V'$. Single-qubit operation $B'$ prepares state $B' \ket{0} = \sum_{l=0}^{1} s'_l \ket{l}$, where $\{s'_0, s'_1 \}$ are the singular values of $\mathbf{M}'$.}
    \label{SVD_method_3Q_state}
\end{figure}
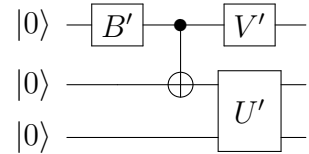

\section{Preparation of $\ket{W_m}$ states}\label{AppG}

The $\ket{W_m}$ state is a $m$-qubit state defined as
\begin{equation*}
    \ket{W_m} \coloneqq \sum_{i = 0}^{m-1} \frac{\ket{2^i}}{\sqrt{m}} = \left( \frac{\ket{10...0}}{\sqrt{m}} + \frac{\ket{01...0}}{\sqrt{m}} + ... + \frac{\ket{00...1}}{\sqrt{m}} \right).
\end{equation*}

\begin{figure}[h]
    \centering{
    \includegraphics[width=0.9\linewidth]{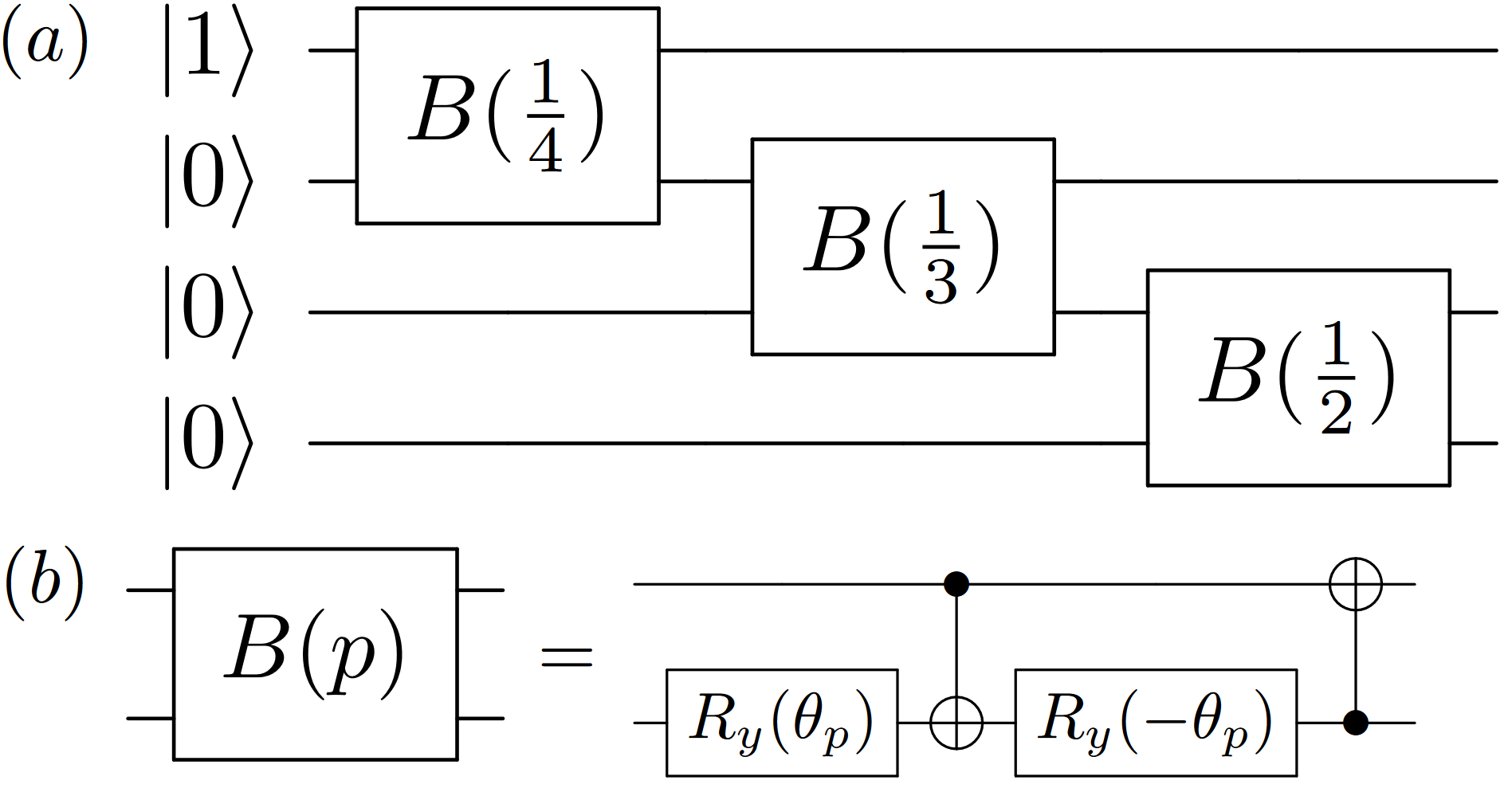}}
    \caption{(a) Scheme of quantum circuit that prepares $\ket{W_4}$ state. For the general case $\ket{W_m}$, a sequence of two-qubit blocks $B(\frac{1}{m})$, $B(\frac{1}{m-1})$, ..., $B(\frac{1}{3})$ and $B(\frac{1}{2})$ is applied in the staircase-like structure of Fig. 11(a), with the first qubit initialized in $\ket{1}$ and all remaining $m-1$ qubits in $\ket{0}$. (b) Basis gate decomposition of two-qubit block $B(p)$, with $0 < p < 1$. $\theta_p = \arcsin(\cos(\arctan(\sqrt{\frac{1-p}{p}})))$.}
    \label{fig_W_state}
\end{figure}

Fig. \ref{fig_W_state}(a) shows the circuit that prepares such W state for $m=4$, as first presented in \cite{Cruz19}. The key building block of this circuit is the parameterized two-qubit subcircuit $B(p)$, with $p \in (0,1)$, shown in Fig. \ref{fig_W_state}(b).

The generalization of this preparation scheme of $\ket{W_m}$ for arbitrary $m \in \mathbb{N}$ is self-evident. It consists of applying the sequence of two-qubit blocks $B(\frac{1}{m})$, $B(\frac{1}{m-1})$, ..., $B(\frac{1}{3})$ and $B(\frac{1}{2})$ in the staircase-like structure of Fig. \ref{fig_W_state}(a), with the first qubit initialized in $\ket{1}$ and all remaining $m-1$ qubits in $\ket{0}$. Hence, a total of $m-1$ executions of the $B(p)$ block are required to prepare $\ket{W_m}$. However, as alluded to in Section \ref{Section8} of the main text, the execution of such $m-1$ $B(p)$ blocks need not be sequential; it is possible to exploit the freedom to choose the target qubit of the final $\textsc{cnot}$ in the $B(p)$ building block to execute multiple such blocks in parallel, thus yielding a $\mathcal{O}(\log m)$-depth circuit, as opposed to the $\mathcal{O}(m)$ depth of the sequential approach. In any case, the total number of blocks is the same for both cases: $m-1$ $B(p)$.

Let us consider the case $m = 4! = 24$, which is relevant for the construction of the Prepare oracle $\mathcal{O}_P$ for the LCU implementation of the symmetrization operator acting on $\mathcal{N}=4$ spins-$\frac{1}{2}$, $\mathcal{S}^{(\mathcal{N}=4)}$. $24-1=23$ $B(p)$ blocks are involved in the preparation of $\ket{W_{24}}$. Each $B(p)$ takes $2$ $\textsc{cnot}$ gates, so the implementation of the Prepare oracle $\mathcal{O}_P$ for $\mathcal{S}^{(\mathcal{N}=4)}$ takes $23 \times 2 = 46$ $\textsc{cnot}$ gates.

\section{Circuit Depth in Preparation of Spin-$1$ and Spin-$\frac{3}{2}$ Valence-Bond-Solid States}\label{AppH}

In this appendix, the depth of the circuits that prepare the spin-$1$ and spin-$\frac{3}{2}$ VBS states is determined, taking into account the qubit connectivity constraints of the NISQ device. The $S=1$ and $S = \frac{3}{2}$ cases will be examined separately. All-to-all qubit connectivity will be assumed first, and then the qubit connectivity arrangement obtained by embedding these circuits on the IBM Quantum heavy-hex lattice \cite{IBMQheavy_hex}, as shown in Fig. \ref{fig_heavy_hex} in Section \ref{Section9}, will be considered. The results herein discussed are summarized in Table \ref{TableA} on Section \ref{Section9}.

\subsection{Spin-$1$ Valence-Bond-Solid State}

Regarding the spin-$1$ VBS state defined on a one-dimensional lattice, assuming all-to-all qubit connectivity for the moment, the probabilistic method without any mitigation of the repetition overhead results in a circuit of depth $8$ $\textsc{cnot}$ gates, $7$ of which are associated with the local Hadamard test (cf. decomposition of Fredkin gate in \cite{cruz2022optimized}) and $1$ with the initialization of the product state of valence bonds (cf. Fig. \ref{Circ1}(a)). The average number of repetitions in such case is $\left( \frac{4}{3} \right)^N$ (ignoring the finite-size effects discussed in Appendix \ref{AppC}).

A quadratic reduction of this repetition overhead can be achieved by preparing the four-qubit islands deterministically, as discussed in Section \ref{Section7}, before applying the probabilistic method to the remaining sublattice. This merely increases the circuit depth from $8$ to $11$ $\textsc{cnot}$ gates, as the repetition overhead mitigation layer takes $4$ $\textsc{cnot}$ gates of depth, $3$ more than the initialization of the product state of valence bonds.

Let us now assume the most restrictive case of linear qubit connectivity. If the probabilistic method alone is implemented, then all $N$ ancillary qubits are required, one for each site. These ancillas are placed between the two qubits that encode the spin-$1$ of the respective site, so that the two qubits forming a valence bond are connected to each other, thus allowing all valence bonds to be initialized with depth $1$ $\textsc{cnot}$. Then, in the implementation of the local Hadamard test, given that the ancilla is between the two site qubits, we make use of the circuit for the $\textsc{cswap}$ in \cite{cruz2022optimized} that assumes linear qubit connectivity and the placement of the control-qubit in the central position of the three-qubit register. This takes $9$ $\textsc{cnot}$ gates, ignoring the final network of $3$ $\textsc{cnot}$ gates that swap the ancilla with one of the site qubits, which is immaterial, since the ancilla is measured immediately afterwards. Having discarded this $\textsc{swap}$ gate, the qubit to be measured must be changed accordingly; this is already accounted for in the respective QASM file (cf. Appendix \ref{AppI}). In total, the depth is of $1+9=10$ $\textsc{cnot}$ gates.

Alternatively, the $\mathcal{O}(1)$-depth repetition overhead mitigation layer can be used to reduce the average number of repetitions. The Qiskit \texttt{transpile} function was applied to the original $4$-qubit island preparation circuit, imposing the linear connectivity constraint via the coupling\_map input. After a few trials, a circuit with a depth of $8$ $\textsc{cnot}$ gates was generated. The layer of local Hadamard tests at the sites of the other sublattice must then be added, taking the same $9$ $\textsc{cnot}$ gates stated in the previous paragraph. The only difference relative to the bare probabilistic method is that this scheme requires only $\frac{N}{2}$ ancillas, placed between consecutive $4$-qubit islands, as illustrated in Fig. \ref{fig_heavy_hex}(c) in Section \ref{Section9}. In total, with linear qubit connectivity, the probabilistic method preceded by the repetition overhead mitigation layer takes $8 + 9 = 17$ $\textsc{cnot}$ gates of depth.

\subsection{Spin-$\frac{3}{2}$ Valence-Bond-Solid State}

Ignoring qubit connectivity constraints, applying the probabilistic method without mitigating the repetition overhead to prepare the spin-$\frac{3}{2}$ VBS state defined on a honeycomb lattice takes $27$ $\textsc{cnot}$ gates of depth, $26$ of which due to the local Hadamard test at every site (cf. Section \ref{Section5}) and the remaining $\textsc{cnot}$ due to the initialization of the product state of valence bonds $\ket{\psi_{\textrm{pre-VBS}}}$.

If, instead, one initializes the six-qubit islands to symmetrize all sites of one sublattice deterministically before applying the probabilistic method to the other sublattice, the average repetition overhead is decreased from $2^N$ to $2^{\frac{N}{2}}$ at the cost of increasing the circuit depth. Concretely, the initialization of the six-qubit islands takes $19$ $\textsc{cnot}$ gates of depth, which adds to the $26$ $\textsc{cnot}$ gates from the local-Hadamard-test layer to yield a total circuit depth of $19 + 26 = 45$ $\textsc{cnot}$ gates.

\begin{figure}[h]
\centering{\
\includegraphics[width=0.8\linewidth]{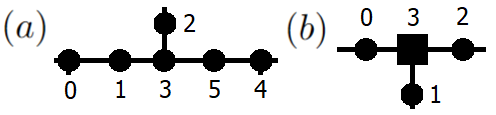}
\par}
\caption{Illustration of the qubit coupling maps considered in the implementation in the IBM Quantum heavy-hex lattice of the initialization of the six-qubit islands (a) and the local Hadamard test for the remaining sublattice (b) for the preparation of the spin-$\frac{3}{2}$ VBS state via the probabilistic method combined with the $\mathcal{O}(1)$-depth repetition overhead mitigation scheme. Numbering of qubits in (a) ensures that qubit pairs (1,2), (3,4), and (5,6) form valence bonds, and site encoded by qubit trio (1,3,5) is symmetrized, in agreement with Fig. \ref{fig_heavy_hex}(b). Black square in (b) corresponds to site ancilla.}
\label{heavy_hex_4Q_labelling}
\end{figure}

Let us now take realistic qubit connectivity constraints into account by considering the preparation of the spin-$\frac{3}{2}$ VBS state on the IBM Quantum heavy-hex lattice, as illustrated in Fig. \ref{fig_heavy_hex}(a,b). First, we will adopt the local Hadamard test at all lattice sites, which corresponds to the case depicted in Fig. \ref{fig_heavy_hex}(a). Then, we will assume the six-qubit islands are initialized deterministically before the probabilistic method is employed at half of the sites at which the wave function is yet to be locally symmetrized, which is the method contemplated in Fig. \ref{fig_heavy_hex}(b).

Regarding the bare probabilistic method, the first step amounts to the initialization of all valence bonds, which are represented by the orange curved lines in Fig. \ref{fig_heavy_hex}(a). This simply involves the execution of the two-qubit subcircuit shown in Fig. \ref{Circ1}(a) for each valence bond. All such circuits can still be executed in parallel, so the depth of this first step is of $1$ $\textsc{cnot}$ gate. At this point, we have the prepared the $\ket{\psi_{\textrm{pre-VBS}}}$ state, with sets of three adjacent valence bonds (the six qubits of which are represented as black dots in Fig. \ref{fig_heavy_hex}(a)(i)) separated from one another in the qubit heavy-hex lattice by redundant qubits (shown as gray dots). Before we can proceed to the local symmetrization at every site, we must ensure that all three qubits that encode a local spin-$\frac{3}{2}$, plus the respective site ancilla, are next to one another. To accomplish this, two qubits from each set of three adjacent valence bonds are displaced in the qubit lattice by swapping them with three consecutive redundant qubits, as depicted by the curly arrows included in Fig. \ref{fig_heavy_hex}(a)(i). One of these displaced qubits moves to the left, while the other moves to the right. As a result of these displacements, two out of every three valence bonds are now defined between pairs of separated qubits, as illustrated in Fig. \ref{fig_heavy_hex}(a)(ii) by the stretching of the orange curved lines.

We are now in a position to apply the local Hadamard test with $\mathcal{U} = e^{-i \pi \mathcal{S}}$ (cf. Fig. \ref{Circ1}(b)) at the sets of four qubits enclosed by the green boxes in Fig. \ref{fig_heavy_hex}(a)(ii). Notice that there are two types of arrangements of these sets of four qubits, one being T-shaped and the other having a linear shape. Each corresponds to a different sublattice. The ancillary qubits are identified as black squares. For the T-shaped arrangements, the ancillary qubit corresponds to the redundant qubit from Fig. \ref{fig_heavy_hex}(a)(i) between the two qubits that joined it from either side. For the linear arrangement, in turn, the site ancilla is the very first redundant qubit that is swapped with the left-moving qubit that departed from there. The qubit connectivity layout for the T-shaped set is shown in Fig. \ref{heavy_hex_4Q_labelling}(b), while that of the linear set can be identified with the $(0, 1, 3, 5)$ subset from Fig. \ref{heavy_hex_4Q_labelling}(a), with the ancilla being qubit $0$ in this case. Applying the Qiskit \texttt{transpile} \cite{QiskitTranspile} function to the controlled-$e^{-i \pi \mathcal{S}}$ and providing the coupling\_map input according to the layout of the T-shaped and linear sets, the circuit depth increases from the $26$ $\textsc{cnot}$ gates found without any qubit connectivity constraints to $39$ and $41$ $\textsc{cnot}$ gates, respectively. For the sake of reproducibility, the QASM files corresponding to these basis gate decompositions are provided (cf. Appendix \ref{AppI}), since the Qiskit \texttt{transpile} function does not always produce the same outcome, as this is a hard optimization problem. The circuits provided are the outcomes with lowest $\textsc{cnot}$ depth and total count yielded by Qiskit \texttt{transpile} over many trials. To sum up, the total circuit depth corresponding to the preparation of the spin-$\frac{3}{2}$ VBS state on the IBM Quantum heavy-hex lattice via the bare probabilistic method is $1 + 9 + 41 = 51$ $\textsc{cnot}$ gates, where the first term is due to the initialization of the product state of valence bonds, the second to the $3$ $\textsc{swap}$ gates required to place all qubits in the right positions for the final step, and the third to the local Hadamard test.

Alternatively, the spin-$\frac{3}{2}$ VBS state can be prepared by first initializing the six-qubit islands deterministically and then applying the probabilistic method at only half of the sites, resulting in a quadratic reduction of the average number of repetitions. Fig. \ref{fig_heavy_hex}(b) illustrates the execution of this method on the IBM Quantum heavy-hex lattice. The six-qubit islands can be prepared in the same sets of six qubits considered in the bare probabilistic method above for the initialization of the valence bonds. Considering the qubit coupling map shown in Fig. \ref{heavy_hex_4Q_labelling}(a), the circuit depth is found to increase from $19$ $\textsc{cnot}$ gates (assuming all-to-all connectivity) to $57$ $\textsc{cnot}$ gates. At this point, not only are all valence bonds already initialized but all sites of one sublattice are symmetrized as well. Before the probabilistic method is applied at the T-shaped sets of four qubits corresponding to the sites of the other sublattice, the same $3$ $\textsc{swap}$ gates with redundant qubits considered above need to be performed, even though they are not explicitly shown in Fig. \ref{fig_heavy_hex}(b). The total circuit depth of this method is therefore $57 + 9 + 39 = 105$ $\textsc{cnot}$ gates.

\section{OpenQASM Instruction Files}\label{AppI}

To foster the reproducibility of the methods herein presented, we have made available the following QASM files in the Supplemental Material section:
\begin{enumerate}
    \item{\texttt{S\_1\_Hadamard\_test\_all\_to\_all.qasm}}
    \item{\texttt{S\_1\_Hadamard\_test\_linear.qasm}}
    \item{\texttt{S\_1\_4Q\_island\_all\_to\_all.qasm}}
    \item{\texttt{S\_1\_4Q\_island\_linear.qasm}}
    \item{\texttt{S\_3\_2\_Hadamard\_test\_all\_to\_all.qasm}}
    \item{\texttt{S\_3\_2\_Hadamard\_test\_heavy\_hex\_linear\_box.qasm}}
    \item{\texttt{S\_3\_2\_Hadamard\_test\_heavy\_hex\_T\_box.qasm}}
    \item{\texttt{S\_3\_2\_6Q\_island\_all\_to\_all.qasm}}
    \item{\texttt{S\_3\_2\_6Q\_island\_heavy\_hex.qasm}}
\end{enumerate}
We note that the files were actually provided in \texttt{.txt} format, as the APS Submissions Server does not contemplate the upload of \texttt{.qasm} files. In any case, the \texttt{.txt} files can be resaved in \texttt{.qasm} format in any text editor. 

The use of these QASM files is described below.

\subsection{$S=1$ VBS State via Bare Probabilistic Method}

First, the product state of valence bonds, $\ket{\psi_{\textrm{pre-VBS}}}$, must be initialized by applying at every pair of qubits representing a lattice link the two-qubit subcircuit presented in Fig. \ref{Circ1}(a), for which no QASM file is provided, given its simple structure. 

Then, the local Hadamard test subcircuit must be applied at all sets of three qubits associated with one site (two encoding the local spin-$1$ and the remaining acting as the site ancilla). Depending on whether the qubit connectivity between these three qubits is all-to-all or linear, the three-qubit circuits corresponding to the QASM files 1 and 2 from the list above should be used, respectively. 

In both cases, the whole structure of the local Hadamard test is already included, namely the Hadamard gates acting on the ancilla before and after the controlled-$e^{-i \pi \mathcal{S}}$, the $\textsc{z}$ gate on the ancilla before the Fredkin gate to account for the global phase factor in Eq. (\ref{eq_SWAP_S}) (so that the desired outcome of the ancilla measurement is $\ket{1}$), and the actual measurement of the ancilla at the end. In QASM file 1, the ancilla is the most significant qubit, q[2]. In QASM file 2, the ancilla qubit is initially q[1], in agreement with the fact that the ancilla should be between the two site qubits to ease the initialization of the valence bonds (cf. Fig. \ref{fig_heavy_hex}(c)). However, at the end of the circuit of QASM file 2, the ancilla becomes encoded in q[0] because a $\textsc{swap}$ gate is skipped, as discussed in Appendix \ref{AppH}. As a result, the measured qubit is q[0], not q[1].

\subsection{$S=1$ VBS State via Preparation of $4$-Qubit Islands and Probabilistic Method at One Sublattice}

With respect to the bare probabilistic method described above, the initialization of the product state of valence bonds is replaced by the deterministic preparation of the $4$-qubit islands (cf. Fig. \ref{fig_islands}(a)) using the circuits from the QASM files 3 and 4 for all-to-all and linear qubit connectivity, respectively. 

In both cases the valence bonds are prepared at qubit pairs (q[0],q[1]) and (q[2],q[3]), and the symmetrized site is encoded by the qubit pair (q[1],q[2]), in agreement with the discussion from Section \ref{Section7}. A schematic diagram of the circuit corresponding to QASM file 3 is shown in Fig. \ref{Schmidt_fig}(b), and the details of the single-qubit operations can be inferred from the QASM file. 

Then, the local Hadamard test must be applied at the remaining half of the sites using the circuits from QASM files 1 or 2, as described in the previous case.

\subsection{$S=\frac{3}{2}$ VBS State via Bare Probabilistic Method}

First, all valence bonds are prepared in parallel by applying the two-qubit subcircuit in Fig. \ref{Circ1}(a) at every qubit pair representing a lattice link, just like in the spin-$1$ case. Then, if the IBM Q heavy-hex lattice \cite{IBMQheavy_hex} is considered, the three consecutive $\textsc{swap}$ gates represented by the curly arrows in Fig. \ref{fig_heavy_hex}(a)(i) must be applied to place all three qubits encoding each local spin-$\frac{3}{2}$ next to one another. 

The local Hadamard test is then applied to these three qubits, plus the respective ancilla, represented by the black squares in Fig. \ref{fig_heavy_hex}(a)(ii). In the heavy-hex lattice, there are two different circuits for this $4$-qubit operation, one for each sublattice. QASM file 6 applies to the sublattice associated with the linear-shaped green boxes in Fig. \ref{fig_heavy_hex}(a)(ii), and QASM file 7 to the other sublattice corresponding to the T-shaped green boxes (cf. numbered coupling scheme in Fig. \ref{heavy_hex_4Q_labelling}(b)). 

If instead a quantum processor with all-to-all connectivity is used, the circuit from QASM file 5 should be used to implement the local Hadamard test. In QASM files 5, 6, and 7, the site ancilla is always the most significant qubit, q[3]. As in the spin-$1$ case, all elements of the Hadamard test are already included, including the measurement of the ancilla.

\subsection{$S=\frac{3}{2}$ VBS State via Preparation of $6$-Qubit Islands and Probabilistic Method at One Sublattice}

The starting point corresponds to the deterministic initialization of the $6$-qubit islands (cf. Fig. \ref{fig_islands}(b)). For a quantum computer with all-to-all connectivity between the six qubits, the circuit to be used is provided in QASM file 8 from the list above. If, instead, the IBM Quantum heavy-hex \cite{IBMQheavy_hex} architecture is used, the circuit to prepare the $6$-qubit islands is given by QASM file 9, which adopts the coupling map shown in Fig. \ref{heavy_hex_4Q_labelling}(a). In both cases, the valence bonds are associated with the qubit pairs (q[0],q[1]), (q[2],q[3]), and (q[4],q[5]). The site at the center of this island that is already symmetrized in this step is encoded by (q[1],q[3],q[5]).

The remaining part corresponds to applying the local symmetrization operator at the sites of the other sublattice. For the case of all-to-all connectivity, the corresponding local Hadamard test is implemented via the $4$-qubit subcircuit given in QASM file 5. In the case of the IBM Quantum heavy-hex architecture \cite{IBMQheavy_hex}, first, the sequence of three $\textsc{swap}$ gates must be applied to place all three qubits associated with each site of this sublattice next to each other, and then the local Hadamard test is executed through the circuit in QASM file 7.

\bibliographystyle{apsrev4-1} 
\bibliography{main.bib}

\end{document}